\documentclass[notitlepage,apj,numberedappendix]{emulateapj}

\usepackage{verbatim}
\usepackage{amsmath}
\usepackage{hyperref}
\usepackage{breakurl}
\usepackage{float}

\newcommand{\simgt}{\,\hbox{\lower0.6ex\hbox{$\sim$}\llap{\raise0.6ex\hbox{$>$}}}\,}
\newcommand{\simlt}{\,\hbox{\lower0.6ex\hbox{$\sim$}\llap{\raise0.6ex\hbox{$<$}}}\,}

\newcommand{\code}[1]{\texttt{#1}}

\begin{document}

\title{Simulating {\it Astro-H} Observations of Sloshing Gas Motions in the Cores of Galaxy Clusters}

\author{J. A. ZuHone\altaffilmark{1}, E. D. Miller\altaffilmark{1}, A. Simionescu\altaffilmark{2}, M. W. Bautz\altaffilmark{1}}

\altaffiltext{1}{Kavli Institute for Astrophysics and Space Research, Massachusetts
Institute of Technology, 77 Massachusetts Avenue, Cambridge, MA 02139, USA}
\altaffiltext{2}{Institute of Space and Astronautical Science (ISAS), JAXA, 3-1-1 Yoshinodai, Chuo-ku, Sagamihara, Kanagawa, 252-5210 Japan}


\begin{abstract}
{\it Astro-H} will be the first X-ray observatory to employ a high-resolution microcalorimeter, capable of measuring the shift and width of individual spectral lines to the precision necessary for estimating the velocity of the diffuse plasma in galaxy clusters. This new capability is expected to bring significant progress in understanding the dynamics, and therefore the physics, of the intracluster medium. However, because this plasma is optically thin, projection effects will be an important complicating factor in interpreting future {\it Astro-H} measurements. To study these effects in detail, we performed an analysis of the velocity field from simulations of a galaxy cluster experiencing gas sloshing, and generated synthetic X-ray spectra, convolved with model {\it Astro-H} Soft X-ray Spectrometer (SXS) responses. We find that the sloshing motions produce velocity signatures that will be observable by {\it Astro-H} in nearby clusters: the shifting of the line centroid produced by the fast-moving cold gas underneath the front surface, and line broadening produced by the smooth variation of this motion along the line of sight. The line shapes arising from inviscid or strongly viscous simulations are very similar, indicating that placing constraints on the gas viscosity from these measurements will be difficult. Our spectroscopic analysis demonstrates that, for adequate exposures, {\it Astro-H} will be able to recover the first two moments of the velocity distribution of these motions accurately, and in some cases multiple velocity components may be discerned. The simulations also confirm the importance of accurate treatment of PSF scattering in the interpretation of {\it Astro-H}/SXS spectra of cluster plasmas. 
\end{abstract}

\section{Introduction}\label{sec:intro}

X-ray observatories have yielded a wealth of information about the thermodynamic and chemical properties of the intracluster medium (ICM) of galaxy clusters. However, measuring the kinematics of the ICM has been up to now beyond the capability of present instruments. Gas motions in the ICM can be detected by the Doppler shifting and broadening of spectral lines: the former caused by large-scale bulk motions, and the latter by turbulence or a complex projection of components with different bulk motions along a line of sight. The Reflection Grating Spectrometer (RGS) grating on {\it XMM-Newton} can provide upper limits on Doppler broadening of spectral lines in cool-core clusters \citep[][and references therein]{san11,san13,pin15}, but so far no direct velocity measurements have been made, mainly due to the inadequate spectral resolution of existing instruments.

{\it Astro-H} \citep{tak14}, a joint JAXA/NASA endeavor, will be launched in early 2016, and will be the first X-ray observatory capable of detecting motions in the intracluster medium of galaxy clusters using measurements of the shifting and broadening of spectral lines. {\it Astro-H} will possess a Soft X-ray Spectrometer (SXS) micro-calorimeter with an energy resolution of $\Delta{E} \leq 7$~eV (FWHM) within the energy range $E \sim 0.3-12.0$~keV, covering a 3'$\times$3' field. At the energy of the Fe-K${\alpha}$ line, $E \approx 6.7$~keV, this enables the measurement of velocities at resolutions of tens of km~s$^{-1}$.

Determining the properties of gas motions in clusters is important to studies of cluster astrophysics and cosmology. Turbulence and bulk motions can serve as a transport mechanism for heat, metals, and cosmic rays throughout the cluster \citep[e.g.,][]{fuj04,den05,reb06,vaz10,ens11}. Furthermore, these gas motions provide non-negligible pressure support against gravity, biasing mass estimates based on the assumption of hydrostatic equilibrium, and likely explain discrepancies between hydrostatic and weak lensing-derived masses \citep{zha10,mah13,vdl14,app14}. Simulations predict that up to $\sim$20-30\% of pressure support in even some relaxed clusters is due to non-thermal sources, most of which will be comprised of gas motions \citep{evr96,ras06,nag07,pif08,nel14}. Finally, ICM turbulence is likely a key ingredient for the origin of non-thermal phenomena such as radio halos and radio mini-halos \citep{bru07,don13,zuh13}. X-ray instruments with micro-calorimeters such as {\it Astro-H}, {\it Athena}\footnote{\url{http://www.the-athena-x-ray-observatory.eu/}}, and the mission concept {\it X-ray Surveyor} \citep{wei15}, are essential to reveal the details of the kinematics of the cluster gas, which will shed light on these questions.

One class of clusters, those with bright central ``cool'' cores, appear relatively relaxed. However, in these clusters there is one type of gas motion in cool-core galaxy clusters that can already be inferred from imaging studies: ``sloshing'' motions, evidenced by the presence of sharp discontinuities in surface brightness and temperature. In these features, the denser (brighter) side of the discontinuity is colder than the lighter (fainter) side, hence they have been dubbed ``cold fronts'' \citep[see][for a review]{MV07}. Typically, one or several cold fronts may appear in the core region, often laid out in a spiral pattern. Simulations have demonstrated that these features can be produced by gravitational perturbations of the cool core induced by encounters with subclusters or strong shocks \citep{chu03,tit05,AM06,zuh10,rod11}. The motions associated with cold fronts in these simulations are subsonic, with typical velocities of several hundred km~s$^{-1}$ (${\cal M} \sim 0.3-0.5$) and associated length scales of $\sim$100s of kpc, even up to $\sim$1~Mpc in at least a few clusters \citep{sim12,ros13,wal14}. In the simulations, these gas motions amplify magnetic fields to near-equipartition levels \citep{zuh11a}, generate turbulence \citep{vaz12,zuh13}, and advect and mix metals and entropy \citep{zuh10,rod12}. By some estimates, sloshing cold fronts appear in as many as half to two-thirds of relaxed galaxy clusters \citep{ghi10}, indicating that the associated bulk motions are common in the cores of massive, non-merging clusters. Therefore, it is important to determine if these motions can be detected by {\it Astro-H}, and, if so, how they affect the shape of the observed lines.

Previous works have used simulated high-resolution X-ray spectra from simulations of galaxy clusters to measure characteristics of the gas velocity. \citet{ino03} performed an early analysis of the effects on spectral lines from Doppler shifting and broadening by a turbulent medium. \citet{fuj05} showed that X-ray spectra with resolution of several eV could measure the turbulent gas velocity in a cluster cool core. \citet{reb08} looked at the radial dependence of the line width for isotropic and radially directed gas motions, proposing to use this radial dependence as a test of the directionality of the gas velocity. \citet{sha12} demonstrated that mixing model analyses of {\it Astro-H} spectra could distinguish between different components of the velocity distribution, provided the exposure time was adequate. \citet{nag13} and \citet{ota15} used adaptive mesh refinement cosmological simulations and mock {\it Astro-H} observations to measure the velocity dispersion profiles of relaxed galaxy clusters and line shifts and widths for a merging system with multiple components. \citet{bif13} used clusters produced in a smoothed particle hydrodynamics cosmological simulation and mock {\it Athena} observations to investigate the velocity structure of the ICM and its impact on the $L - T$ relation.

Our aim in this work is to examine the specific case of sloshing motions in a massive galaxy cluster, and ``observe'' them with {\it Astro-H}. For this, we use cluster merger simulations from \citep[][hereafter ZMJ10]{zuh10}, which included runs with varying viscosity. We also employ recently developed tools for producing synthetic X-ray images and spectra from simulations, and produce mock observations which are convolved with the appropriate instrumental responses to make them as realistic as possible \citep{zuh14}. This work will be relevant for a number of nearby clusters with indications of gas sloshing, including those that already have planned {\it Astro-H} observations and others that merit investigation (see Section \ref{sec:targets}).

This paper is organized as follows: in Section \ref{sec:methods} we describe the cluster merger simulations and our method for producing synthetic X-ray observations. In Section \ref{sec:results}, we first examine the characteristics of the velocity field from the simulations, and then generate synthetic X-ray spectra, performing standard spectral analyses to determine what aspects of the velocity field may be measured from its effect on the spectral lines. We then compare these results to those from the simulation. Finally, in Section \ref{sec:disc} we discuss implications of our results and limitations of our methods, and in Section \ref{sec:summary} we summarize our conclusions. We assume a $\Lambda$CDM cosmology with $h = 0.71$, $\Omega_m$ = 0.27, and $\Omega_\Lambda$ = 0.73.

\section{Methods}\label{sec:methods}

\subsection{Hydrodynamic Simulations of Cluster Mergers}\label{sec:hydro_sims}

For our cluster candidate, we examine a minor merger simulation from ZMJ10, an off-center collision between a large, cool-core cluster and a smaller gasless subcluster, where it is assumed that the smaller cluster has been stripped of its gas on a previous passage.\footnote{A gas-filled subcluster with the same orbital setup would produce a shock front and turbulence without smooth cold fronts, see \citet{zuh10}.} This configuration produces sloshing cold fronts in the large cluster's core. This simulation was performed using the \code{FLASH} code \citep{dub09}, a parallel hydrodynamics/$N$-body adaptive mesh refinement (AMR) astrophysical simulation code. Full details of the physics employed and the initial setup of the simulation can be found in ZMJ10, but we provide a short summary here.

\begin{figure*}
\begin{center}
\includegraphics[width=0.96\textwidth]{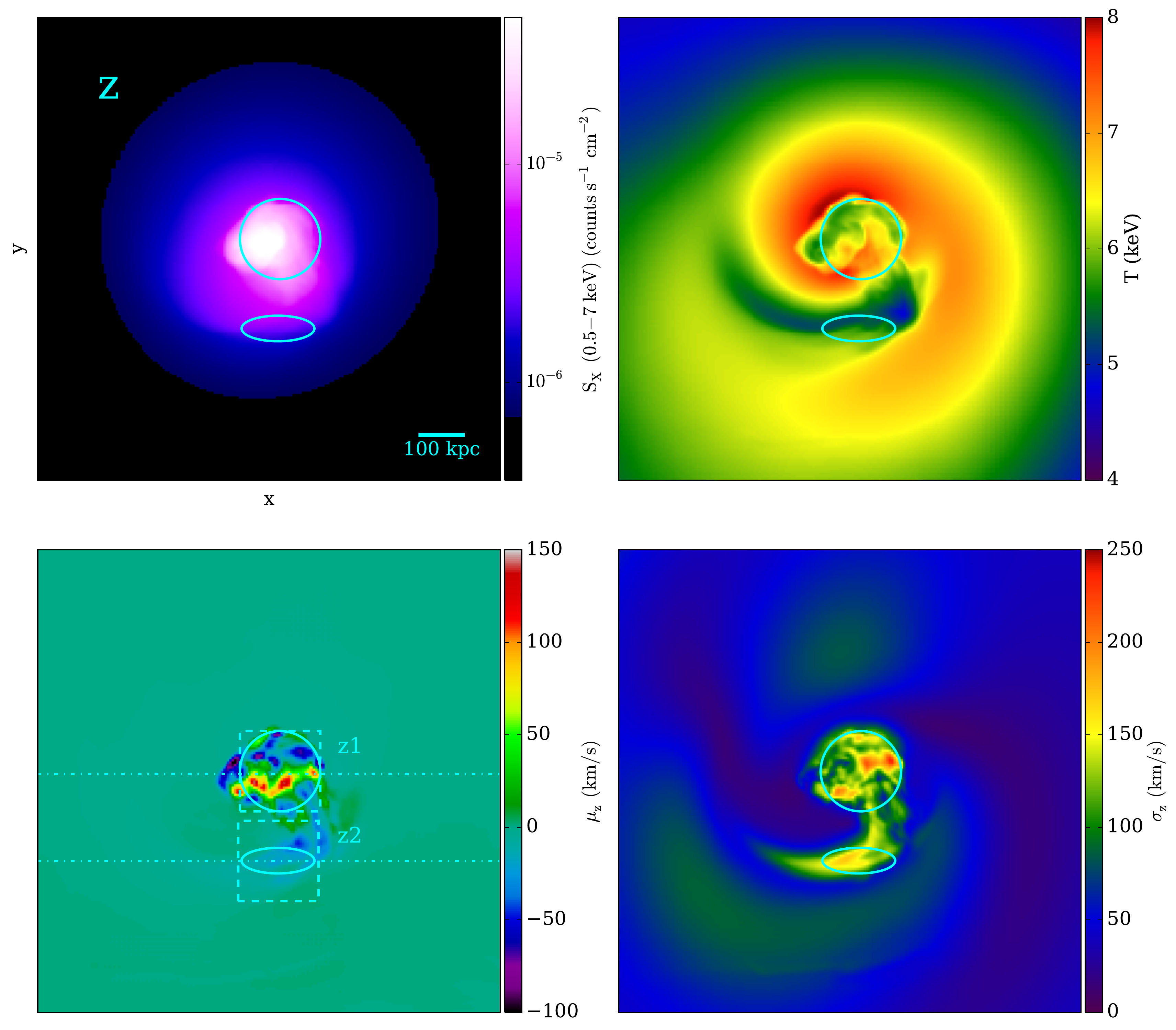}
\caption{Projections of various quantities along the $z$-axis of the inviscid simulation, $\sim$1.6~Gyr after core passage. Upper-left: X-ray surface brightness (0.5-7 keV band), upper-right: spectroscopic temperature, lower-left: line-of-sight velocity, lower-right: line-of-sight velocity dispersion. Square dashed regions indicate the locations of simulated {\it Astro-H} SXS pointings (FOV of 3’$\times$3’), and solid elliptical regions correspond to the locations of regions of interest from which we will examine velocity distributions and extract spectra. The cyan dot-dashed lines indicate the locations of slices shown in Figures \ref{fig:vz_dist1} and \ref{fig:vz_dist2}.\label{fig:map_z}}
\end{center}
\end{figure*}

The simulation used \code{FLASH}'s standard hydrodynamics module employing the Piecewise-Parabolic Method of \citet{col84} for treatment of the cluster plasma, under the assumption of an ideal gas equation of state with $\gamma = 5/3$, and a mean molecular weight of $\mu = 0.592$, appropriate for an ionized H/He gas with a hydrogen mass fraction of $X = 0.75$. We will examine two versions of this simulation: one inviscid, and another with a significant physical viscosity. In the latter, we used an isotropic Spitzer dynamic viscosity \citep{spi62,sar88}:
\begin{eqnarray}
\mu             &=& 0.960\,\frac{n_{\rm i}k_{\rm B}T}{\nu_{\rm ii}} \\
\nonumber  &\approx& 2.2 \times 10^{-15}\frac{T^{5/2}}{\ln\Lambda_{\rm i}}~{\rm g~cm^{-1}~s^{-1}},
\end{eqnarray}
where the temperature $T$ is in Kelvin and the ion Coulomb logarithm $\ln\Lambda_{\rm i} \approx 40$, appropriate for conditions in the ICM. It is unlikely that the ICM is this viscous, due to the anisotropic nature of the ion viscosity in a high-$\beta$ magnetized plasma \citep{bra65}, and also because microscale plasma instabilities may set an upper limit on the viscosity that is much lower than expected for a collisional plasma \citep{kun14}. However, this simulation still serves as a useful test case, since it allows us to examine the effects of the sloshing motions on the spectral lines in the limit that turbulence and instabilities are completely suppressed. The collisionless dark matter component of the two clusters is modeled as a set of gravitating particles, using an $N$-body module which uses the particle-mesh method to map accelerations from the AMR grid to the particle positions. The gravitational field due to both the gas and dark matter is computed using a multigrid solver \citep{ric08}.

Our merging clusters consist of a large, ``main'' cluster with a mass of $1.25 \times 10^{15}$~$M_\odot$, and a small infalling subcluster, with a mass of $2.5 \times 10^{14}$~$M_\odot$, for a mass ratio of 5. Our main cluster closely resembles A2029 \citep{vik05}, a hot, relatively relaxed cluster with sloshing in the cool core. Initially, the two clusters are set up in hydrostatic and virial equilibrium, at a distance of 3~Mpc away from each other, with an impact parameter of 0.5~Mpc on a bound orbit. Each simulation contains $\sim$10 million dark matter particles and has a finest cell size of $\Delta{x} \sim 5$~kpc.

\begin{figure*}
\begin{center}
\includegraphics[width=0.96\textwidth]{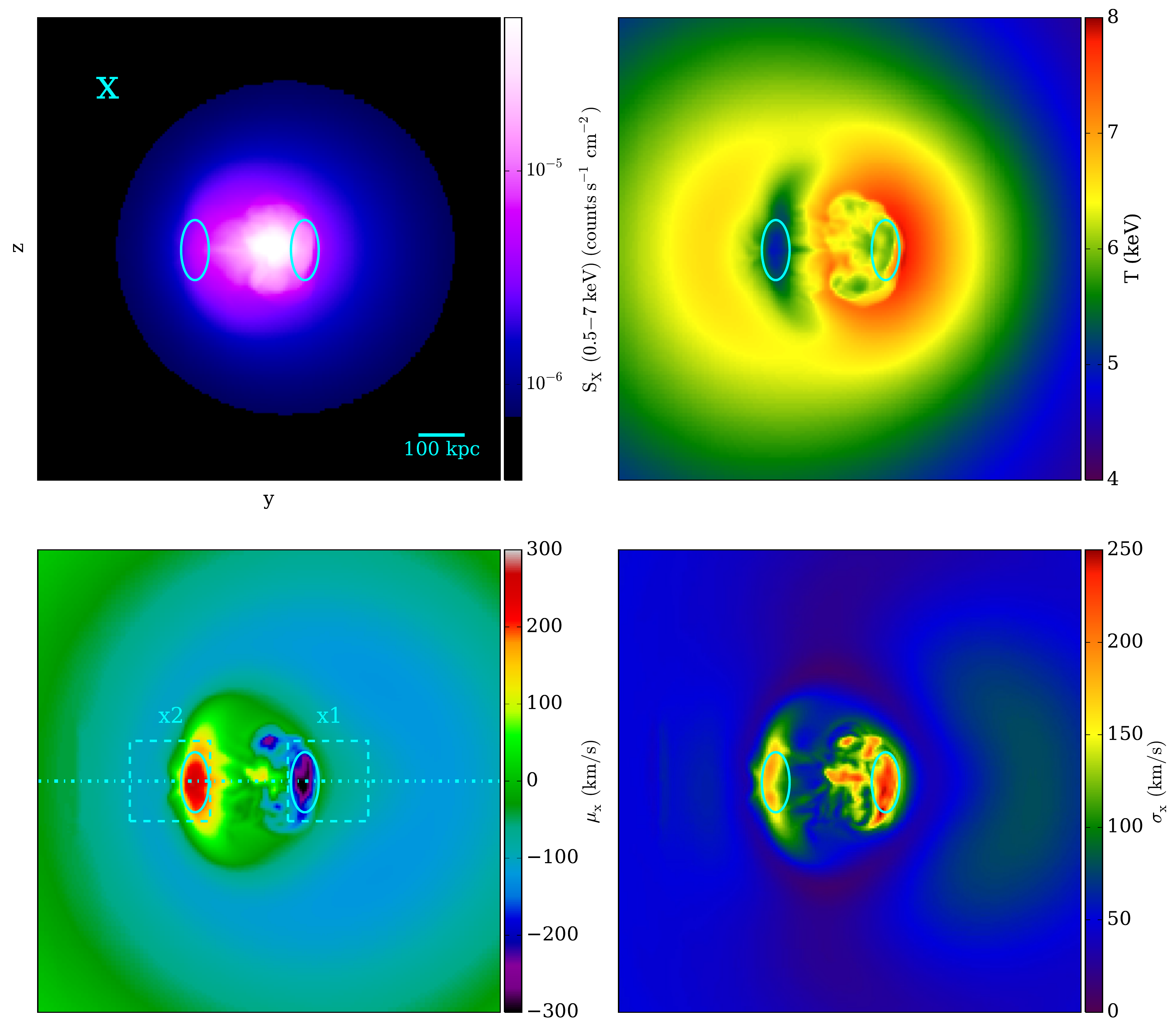}
\caption{Projections of various quantities along the $x$-axis of the inviscid simulation, $\sim$1.6~Gyr after core passage. Upper-left: X-ray surface brightness (0.5-7 keV band), upper-right: spectroscopic temperature, lower-left: line-of-sight velocity, lower-right: line-of-sight velocity dispersion. Square dashed regions indicate the locations of simulated {\it Astro-H} SXS pointings (FOV of 3’$\times$3’), and solid elliptical regions correspond to the locations of regions of interest from which we will examine velocity distributions and extract spectra. The cyan dot-dashed line indicates the location of the slices shown in Figure \ref{fig:vx_dist}.\label{fig:map_x}}
\end{center}
\end{figure*}

\begin{figure*}
\begin{center}
\includegraphics[width=0.96\textwidth]{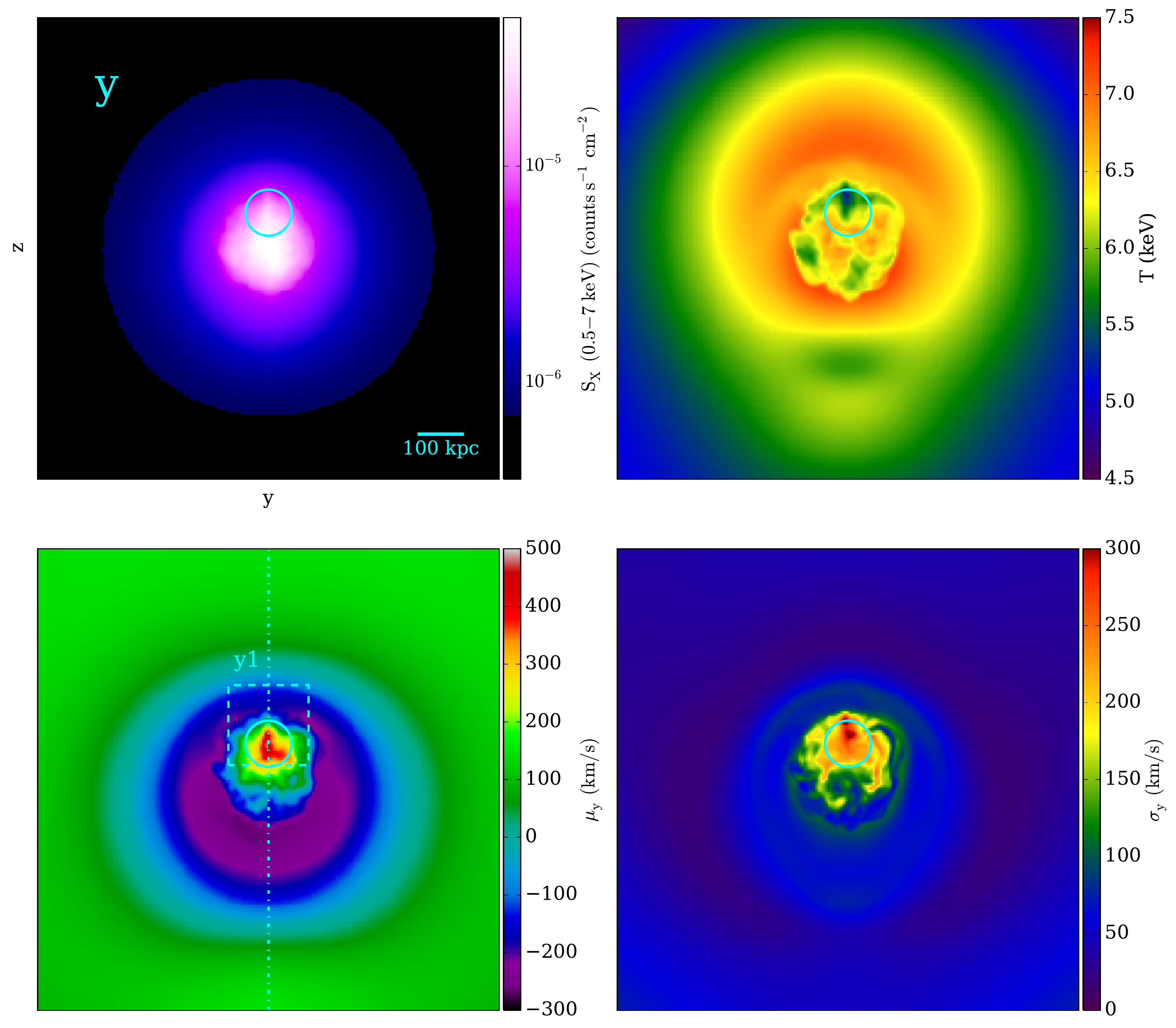}
\caption{Projections of various quantities along the $y$-axis of the inviscid simulation, $\sim$1.6~Gyr after core passage. Upper-left: X-ray surface brightness (0.5-7 keV band), upper-right: spectroscopic temperature, lower-left: line-of-sight velocity, lower-right: line-of-sight velocity dispersion. The square dashed region indicates the location of a simulated {\it Astro-H} SXS pointing (FOV of 3’$\times$3’), and the solid elliptical region corresponds to the location of a region of interest from which we will examine the velocity distribution and extract a spectrum. The cyan dot-dashed line indicates the location of the slices shown in Figure \ref{fig:vy_dist}.\label{fig:map_y}}
\end{center}
\end{figure*}

\begin{figure*}
\begin{center}
\includegraphics[width=0.96\textwidth]{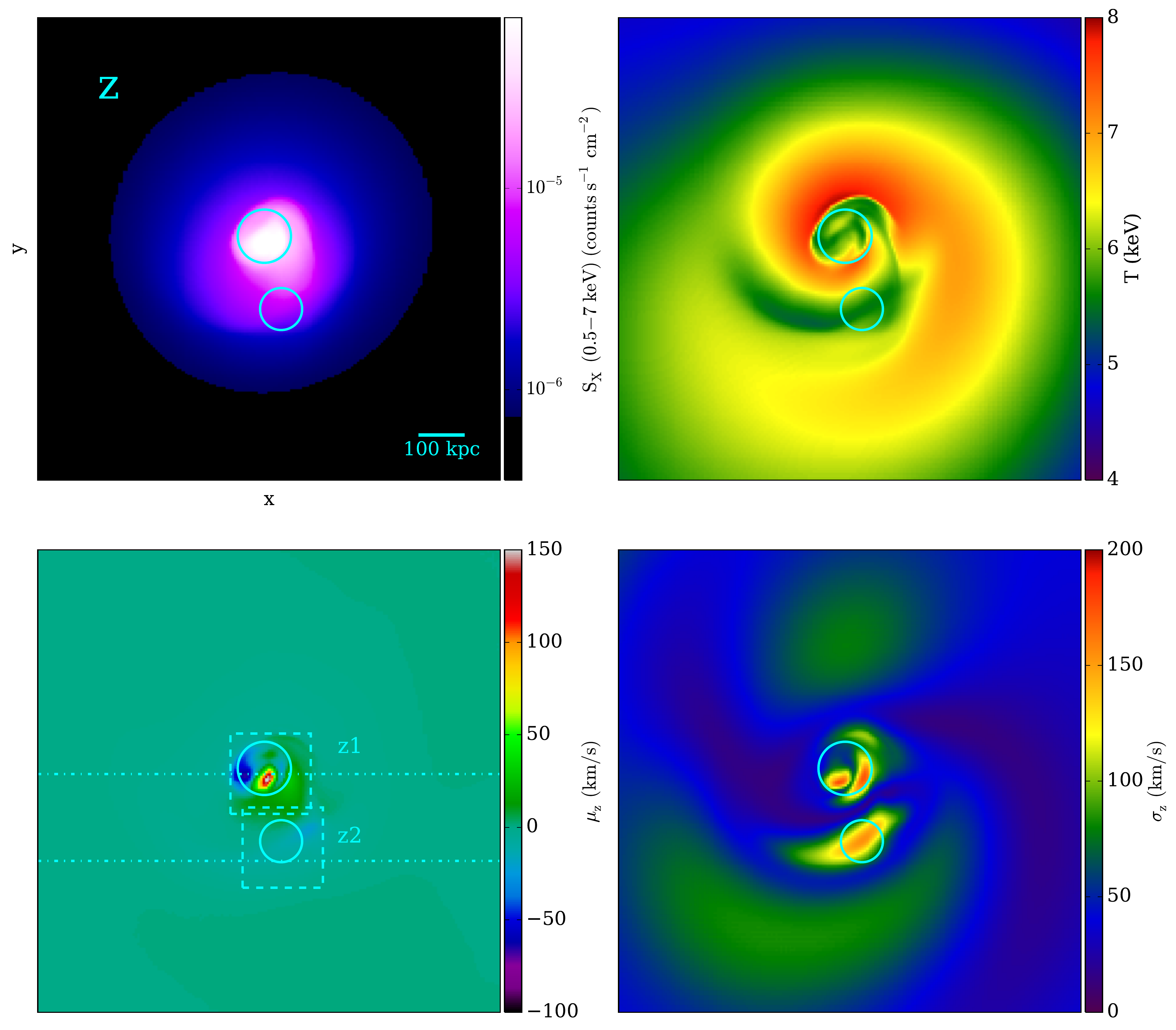}
\caption{Projections of various quantities along the $z$-axis of the viscous simulation, $\sim$1.6~Gyr after core passage. Upper-left: X-ray surface brightness (0.5-7 keV band), upper-right: spectroscopic temperature, lower-left: line-of-sight velocity, lower-right: line-of-sight velocity dispersion. Square dashed regions indicate the locations of simulated {\it Astro-H} SXS pointings (FOV of 3’$\times$3’), and solid elliptical regions correspond to the locations of regions of interest from which we will examine velocity distributions. The cyan dot-dashed lines indicate the locations of slices shown in Figures \ref{fig:vz_visc_dist1} and \ref{fig:vz_visc_dist2}.\label{fig:map_z_visc}}
\end{center}
\end{figure*}

\begin{figure*}
\begin{center}
\includegraphics[width=0.96\textwidth]{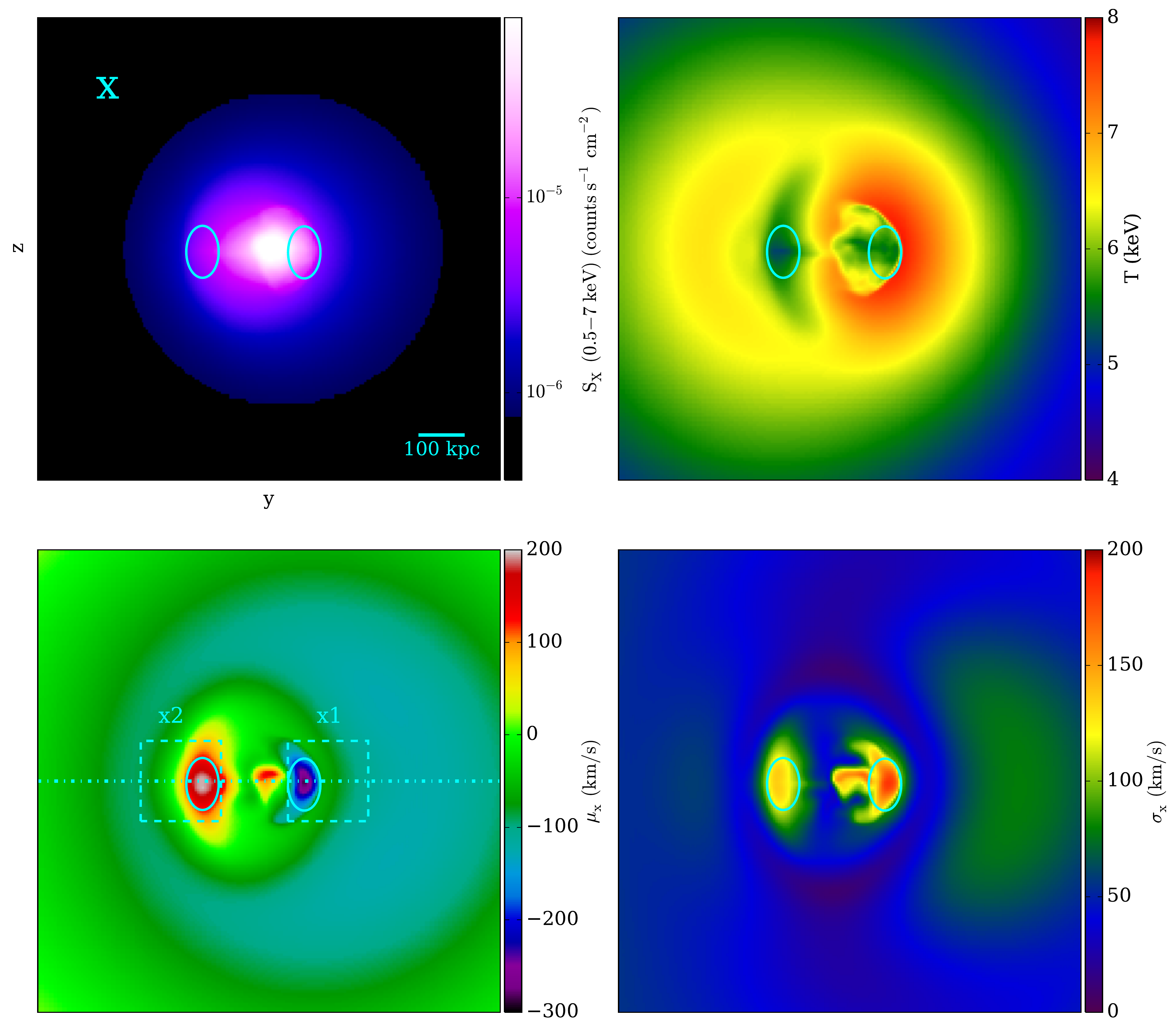}
\caption{Projections of various quantities along the $x$-axis of the viscous simulation, $\sim$1.6~Gyr after core passage. Upper-left: X-ray surface brightness (0.5-7 keV band), upper-right: spectroscopic temperature, lower-left: line-of-sight velocity, lower-right: line-of-sight velocity dispersion. Square dashed regions indicate the locations of simulated {\it Astro-H} SXS pointings (FOV of 3’$\times$3’), and solid elliptical regions correspond to the locations of regions of interest from which we will examine velocity distributions. The cyan dot-dashed line indicates the location of the slices shown in Figure \ref{fig:vx_visc_dist}.\label{fig:map_x_visc}}
\end{center}
\end{figure*}

\begin{figure*}
\begin{center}
\includegraphics[width=0.96\textwidth]{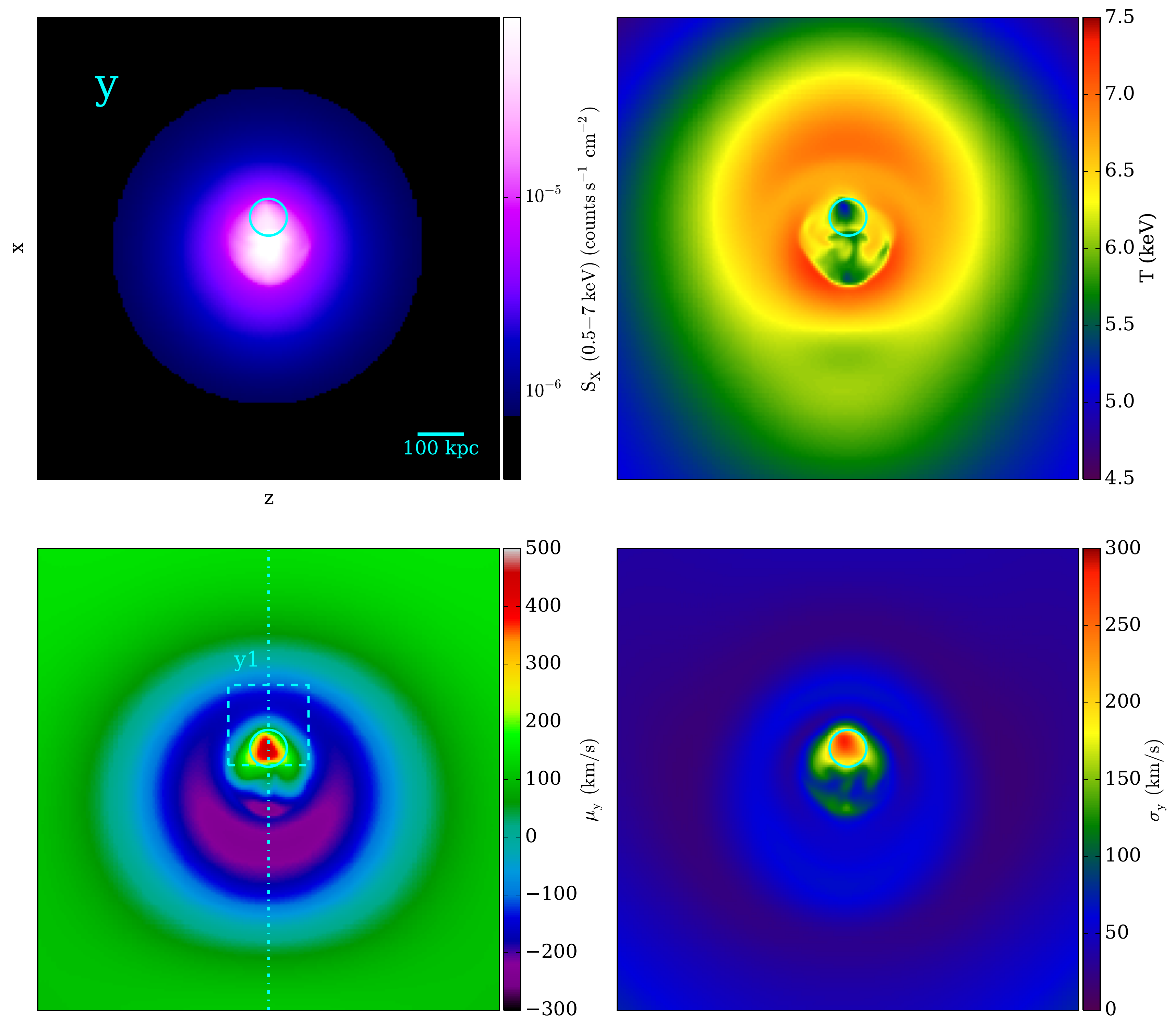}
\caption{Projections of various quantities along the $y$-axis of the viscous simulation, $\sim$1.6~Gyr after core passage. Upper-left: X-ray surface brightness (0.5-7 keV band), upper-right: spectroscopic temperature, lower-left: line-of-sight velocity, lower-right: line-of-sight velocity dispersion. The square dashed region indicates the location of a simulated {\it Astro-H} SXS pointing (FOV of 3’$\times$3’), and the solid elliptical region corresponds to the location of a region of interest from which we will examine the velocity distribution. The cyan dot-dashed line indicates the location of the slices shown in Figure \ref{fig:vy_visc_dist}.\label{fig:map_y_visc}}
\end{center}
\end{figure*}

\subsection{Simulation of X-ray Photons and Synthetic Observations}\label{sec:xray_sims}

To generate X-ray photons for our simulated observations, we employ the \code{photon\_simulator} analysis module \citep{zuh14} from the \code{yt} software package \citep{tur11}. We will outline in brief the procedure for generating these observations here, but the reader should consult \citet{zuh14}, as well as \citet{bif12}, which presents the original implementation of this algorithm, for more details.

A relevant 3D region of cells are selected, within which the X-ray spectrum is computed for each cell using an \code{APEC} model \citep{smi01}, using the cell's density and temperature, with the effect of thermal broadening on the emission lines included. We assume a spatially uniform metallicity of $Z = 0.3~Z_\odot$, using the abundance ratios from \citet{and89}. The model clusters are situated at the redshift $z_{\rm cosmic} = 0.05$, which sets the angular diameter distance of the source. The corresponding angular scale is $\approx 0.965$~kpc/arcsec ($\approx 57.88$~kpc/arcmin). Initially, a large number of photons is generated within the 3D region, by assuming very large, unrealistic values for the exposure time and effective area of the instrument. We choose values for these parameters such that we obtain roughly $\sim$5-10$\times$ more photons than will be eventually used in our synthetic observations, allowing us to use these initial photons as a statistically representative Monte Carlo sample from which to draw subsets for individual exposures.

Next, we project the photons along the desired line of sight, reducing our 3D position space to 2D, and we Doppler-shift the photon energies using the line-of-sight velocity in the cells from which they originated. In this step, we draw a subset of photons from the original sample, corresponding to a more realistic exposure time (the effective area will be determined by the instrumental responses in the next step). The photon energies are cosmologically redshifted. Lastly, we use the Tuebingen-Boulder ISM absorption model \citep[\code{tbabs},][]{wil00} to model Galactic foreground absorption, assuming a Galactic column density of $N_H = 2 \times 10^{20}$~cm$^{-2}$.

To produce synthetic {\it Astro-H} observations from our simulated event lists, we use \code{SIMX}\footnote{\url{http://hea-www.harvard.edu/simx/}}. The \code{SIMX} software package is capable of simulating the instrumental response of photon-counting detectors on a number of current and future X-ray missions, including {\it Astro-H}. \code{SIMX} simulates the predicted effective area, vignetting, PSF, detector response, and pileup fraction\footnote{for SXS only} of the {\it Astro-H} satellite. We use it in conjunction with our simulated events to produce synthetic SXS spectra. For instrumental responses, we use a 5~eV resolution redistribution matrix file (RMF, \code{ah\_sxs\_5ev\_20130806.rmf}) and an auxiliary response file (ARF, \code{sxt-s\_140505\_ts02um\_intallpxl.arf}), distributed with \code{SIMX}.

The outputs of \code{SIMX} are standard OGIP FITS event files which may be viewed and analyzed with standard tools (e.g., \code{ds9}, \code{XSPEC}, \code{FTOOLS}, \code{CIAO}). \code{SIMX} is also capable of simulating the instrumental background of SXS, but we have determined that for our case this background emission is an order of magnitude or more smaller than the continuum emission in the energy range of interest (6.0-7.0~keV, surrounding the Fe-K lines), so for simplicity it is not included in our simulations. We have also verified that a typical astrophysical background from unresolved AGN sources \citep[see][for an example]{bau09} is also significantly smaller than the continuum emission for our source, so we have also not included it in our simulations.

\begin{figure*}
\begin{center}
\begin{minipage}[b]{0.495\linewidth}
\includegraphics[width=\textwidth]{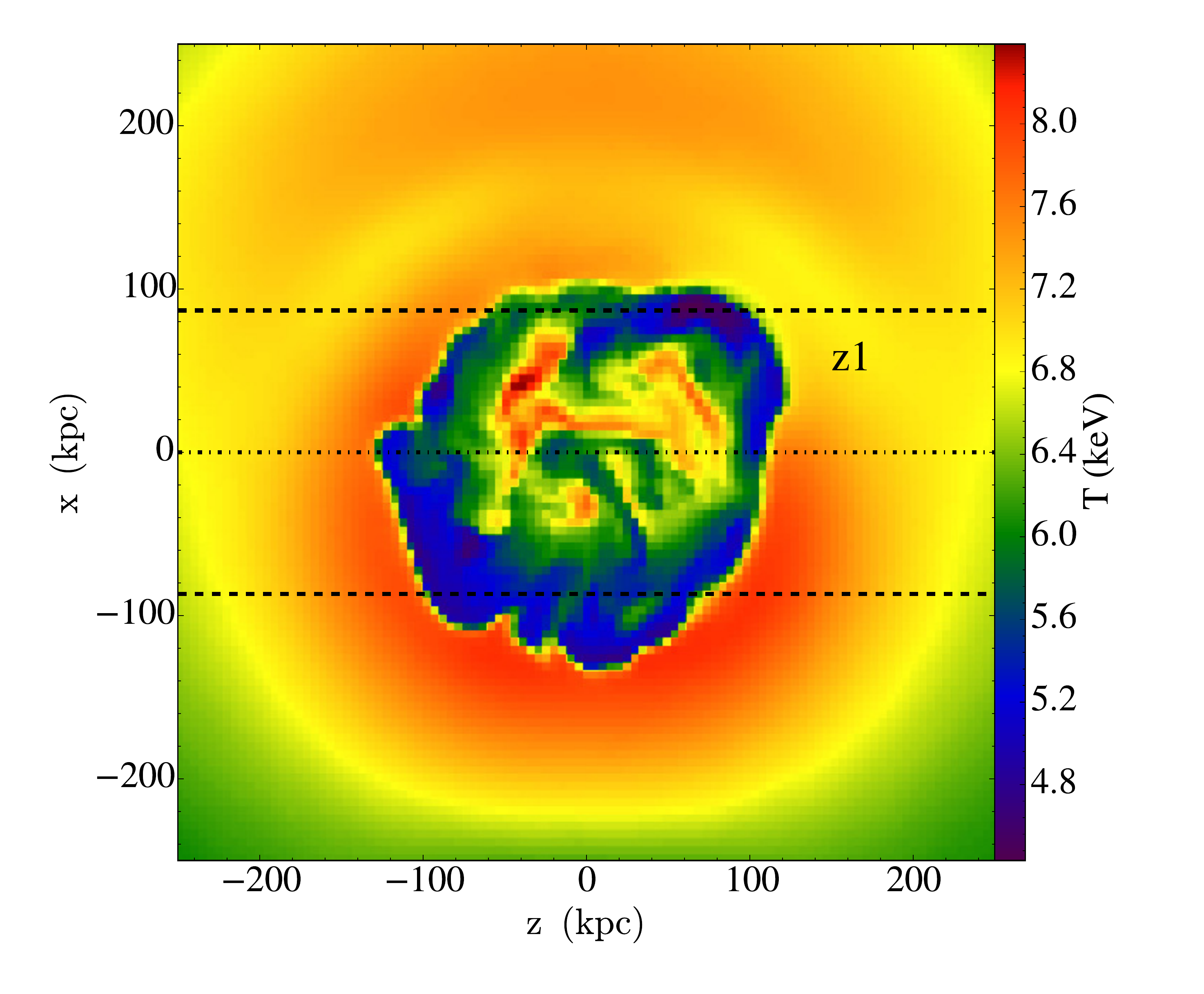}
\end{minipage}
\begin{minipage}[b]{0.495\linewidth}
\includegraphics[width=\textwidth]{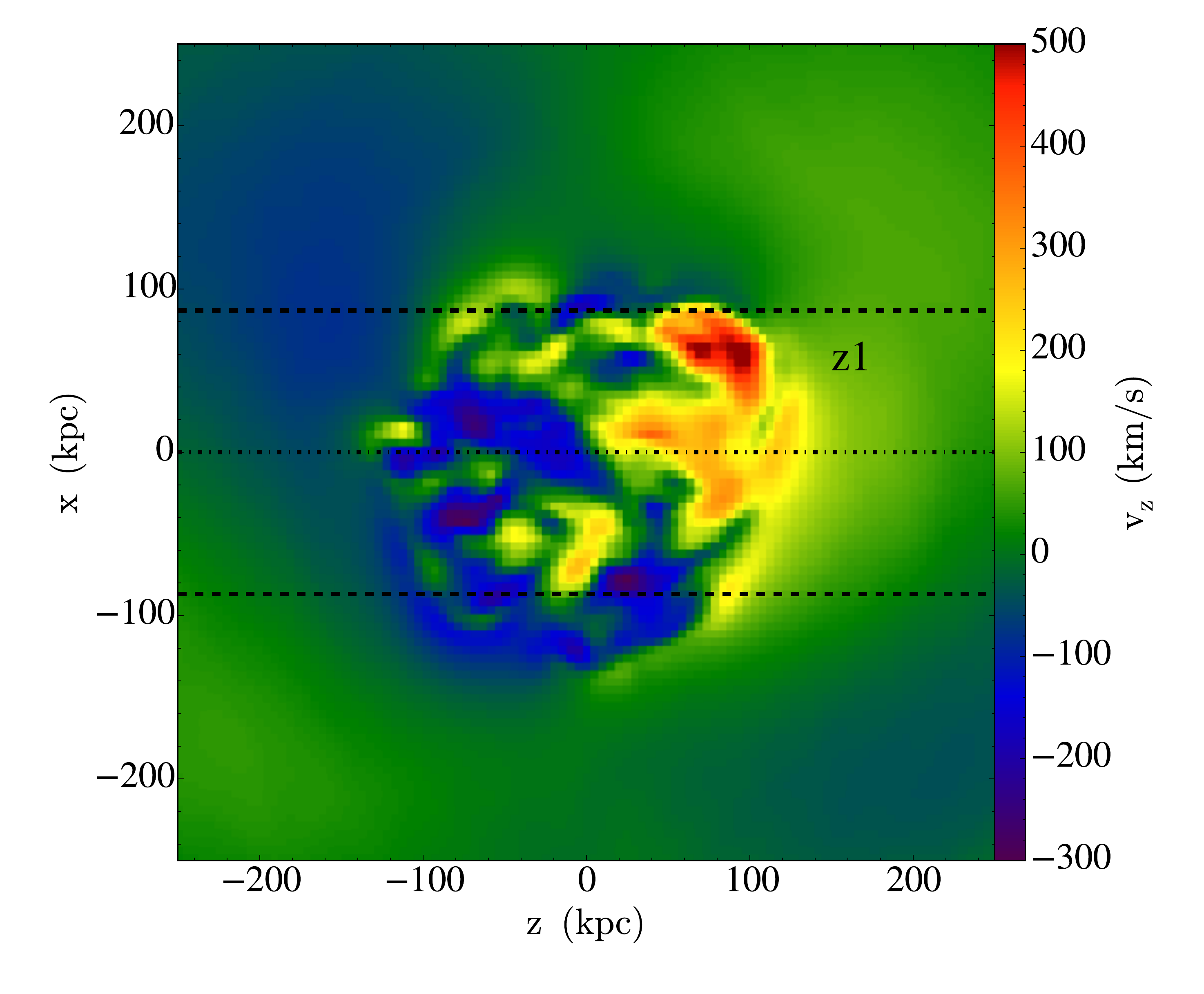}
\end{minipage}
\begin{minipage}[b]{0.51\linewidth}
\includegraphics[width=0.97\textwidth]{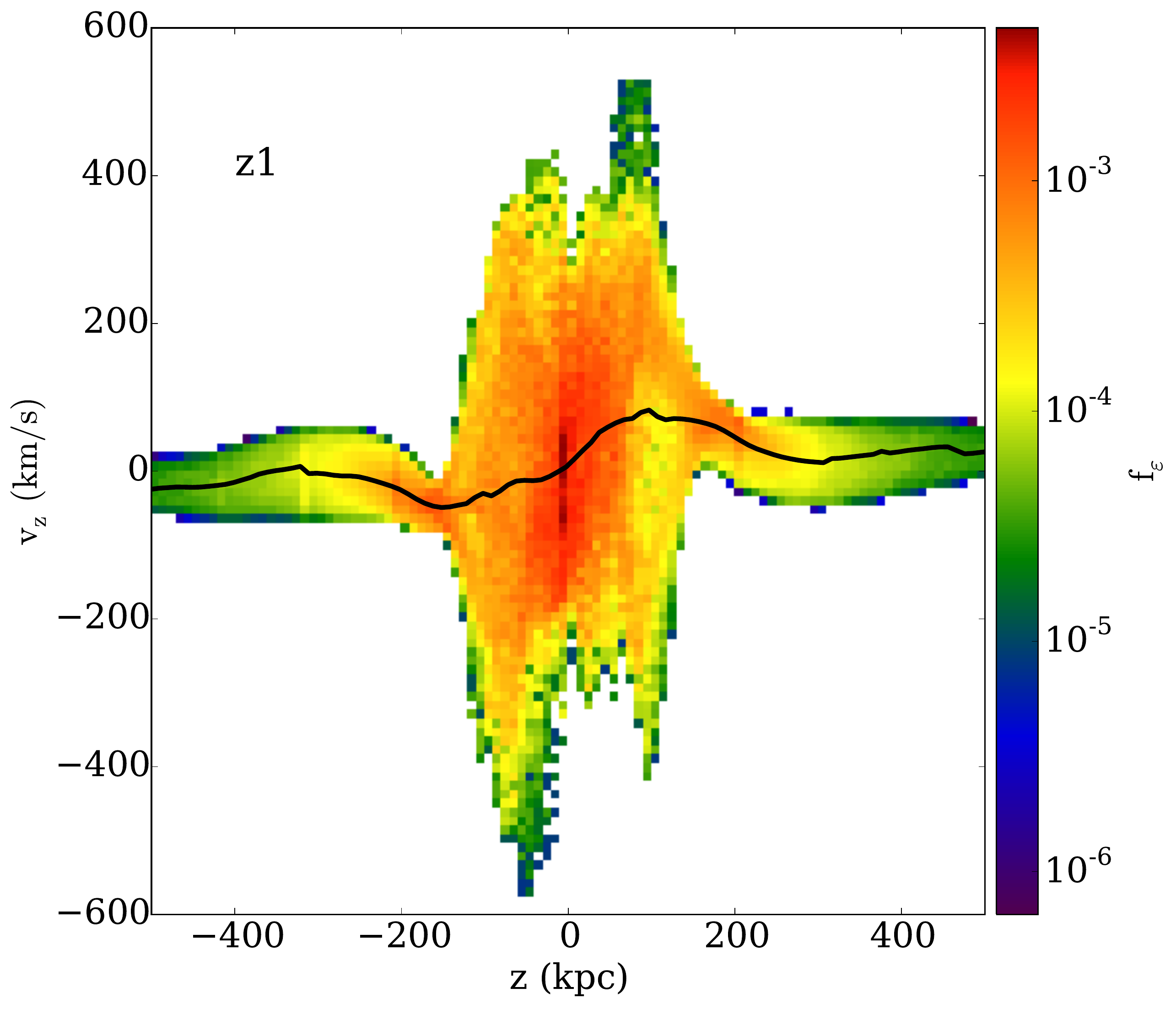}
\end{minipage}
\begin{minipage}[b]{0.47\linewidth}
\includegraphics[width=0.92\textwidth]{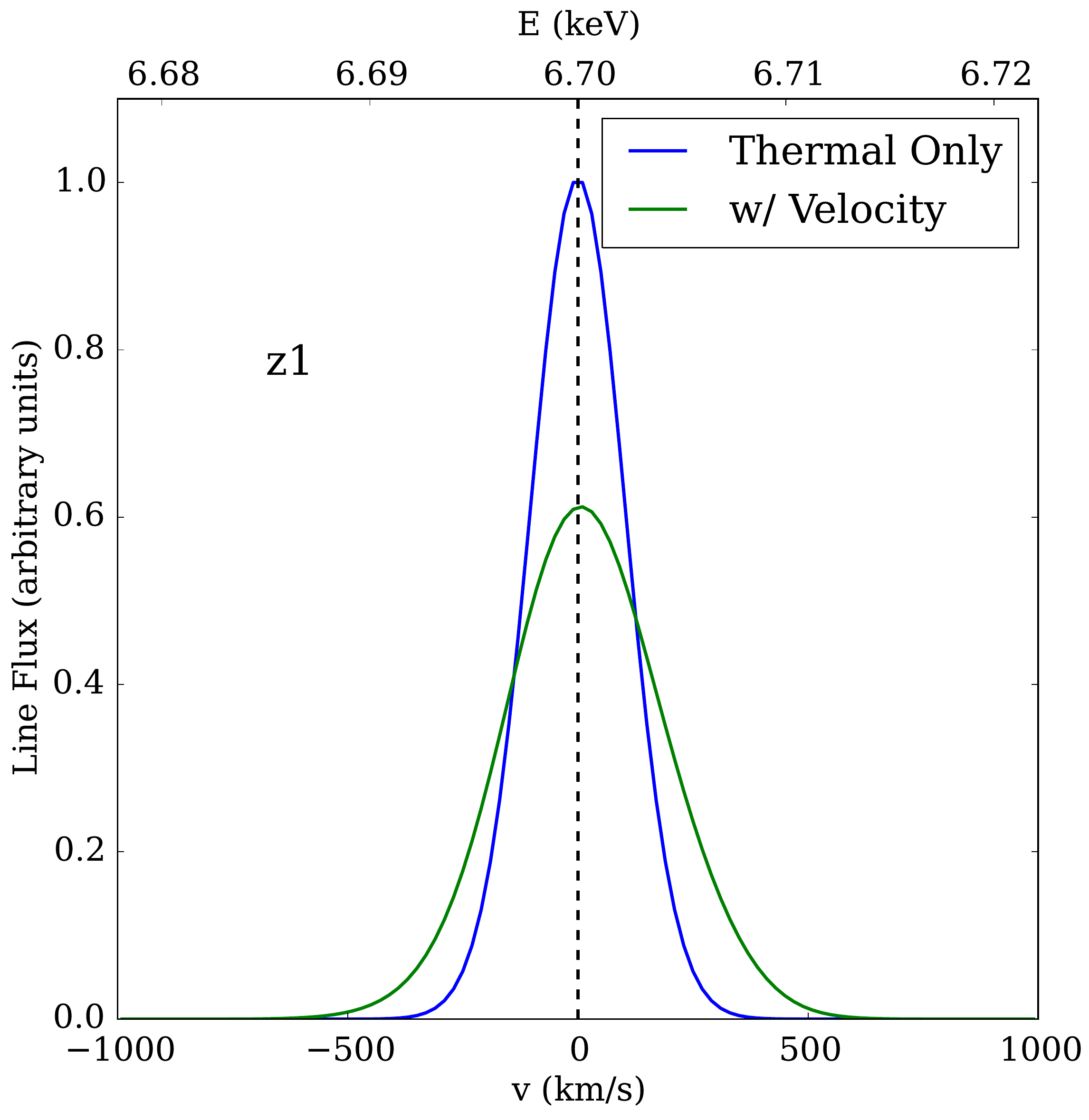}
\end{minipage}
\caption{Characteristics of the velocity field along the $z$-axis of the inviscid simulation, for region ``z1''. Upper panels: slices through the $x-z$-plane at the center of region ``z1'', of temperature (left) and the $z$-component of the velocity (right). Black lines indicate the center and edges of the elliptical cylinder corresponding to the region in Figure \ref{fig:map_z}. Lower-left panel: Phase space plot showing the fraction of emission as a function of position and velocity within the cylinder. The black line indicates the emission-weighted average value. Lower-right panel: Effect of plasma motion on a ``toy'' He-like iron line for the emission with the region.\label{fig:vz_dist1}}
\end{center}
\end{figure*}

\begin{figure*}
\begin{center}
\begin{minipage}[b]{0.495\linewidth}
\includegraphics[width=\textwidth]{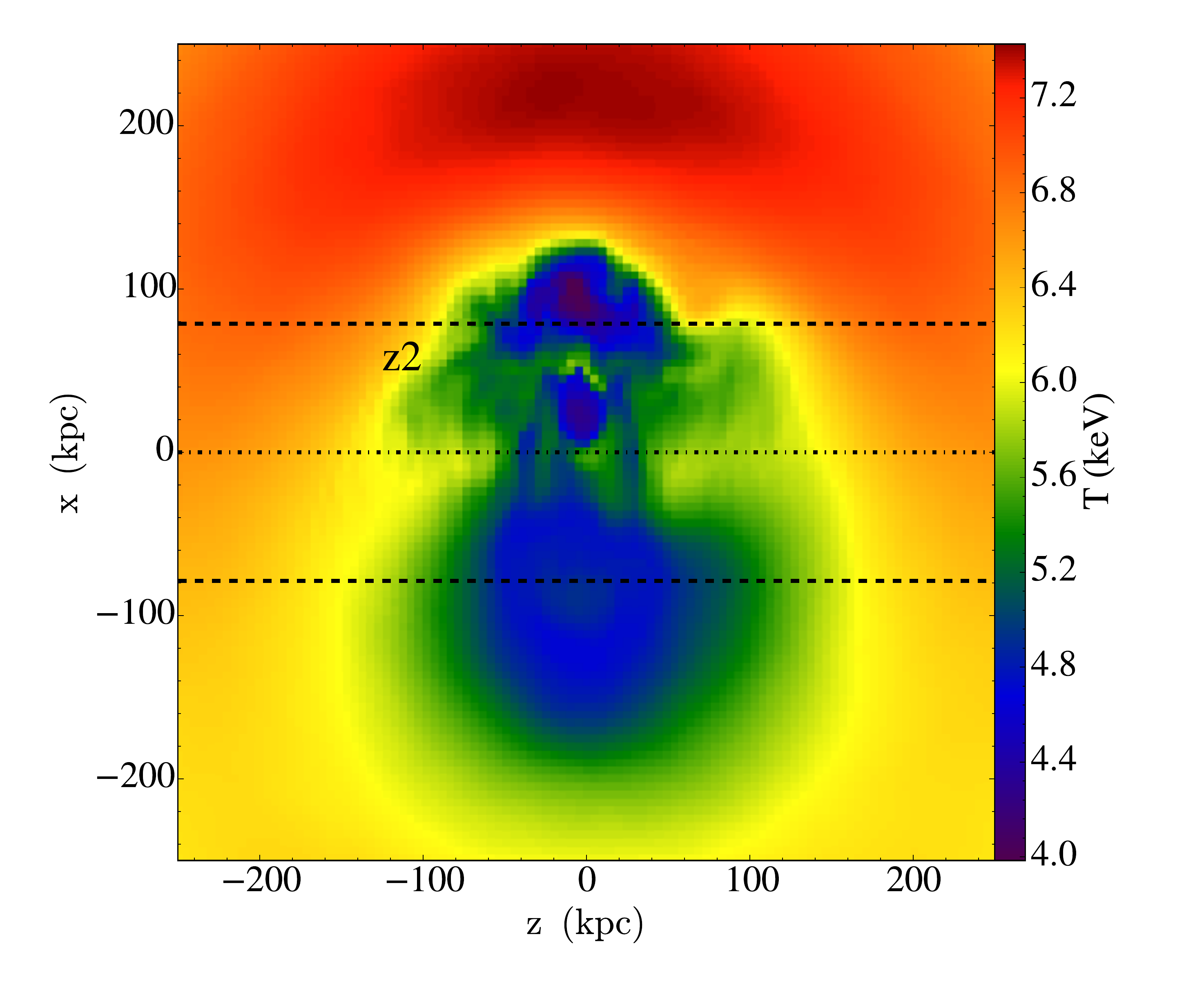}
\end{minipage}
\begin{minipage}[b]{0.495\linewidth}
\includegraphics[width=\textwidth]{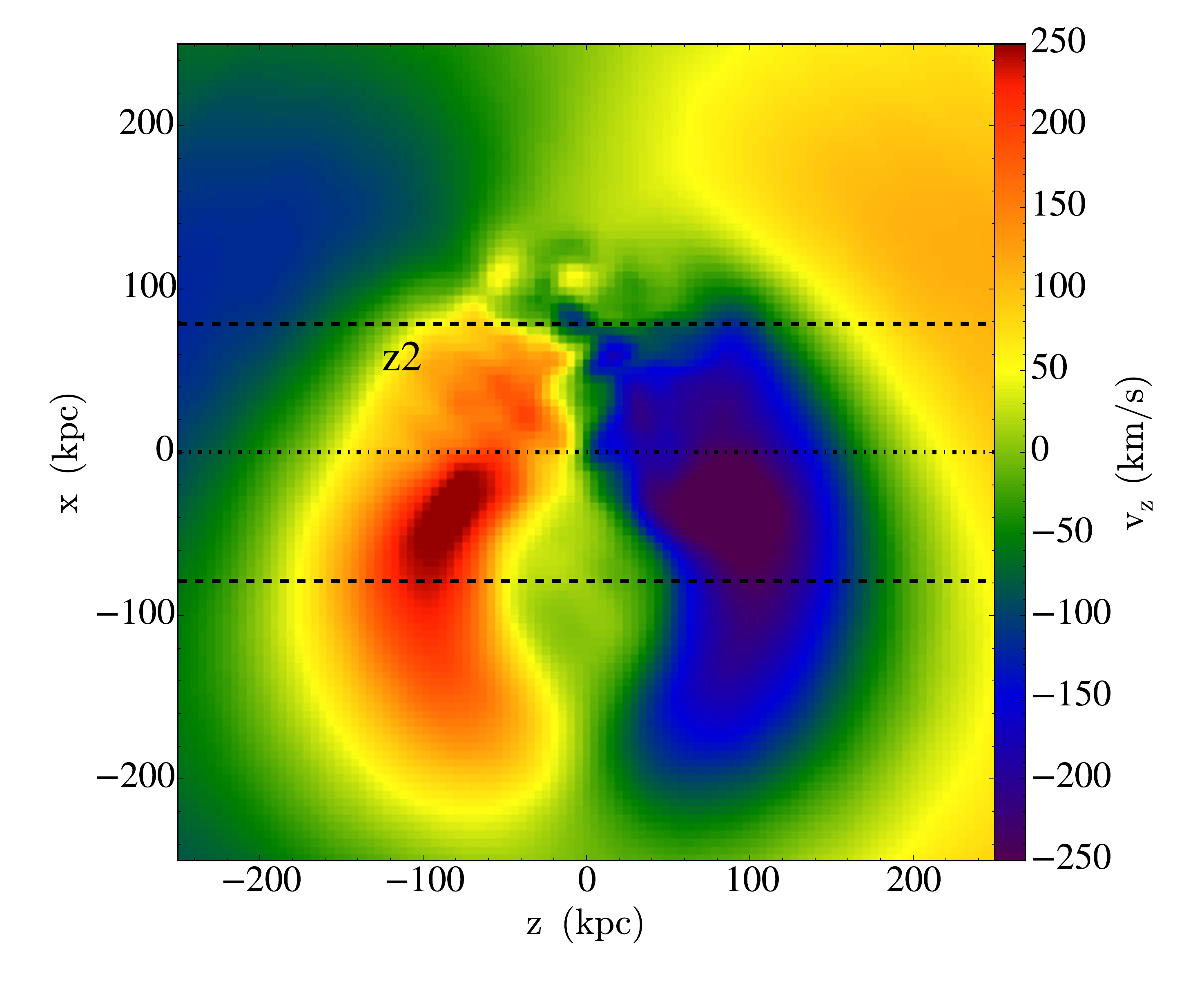}
\end{minipage}
\begin{minipage}[b]{0.51\linewidth}
\includegraphics[width=0.97\textwidth]{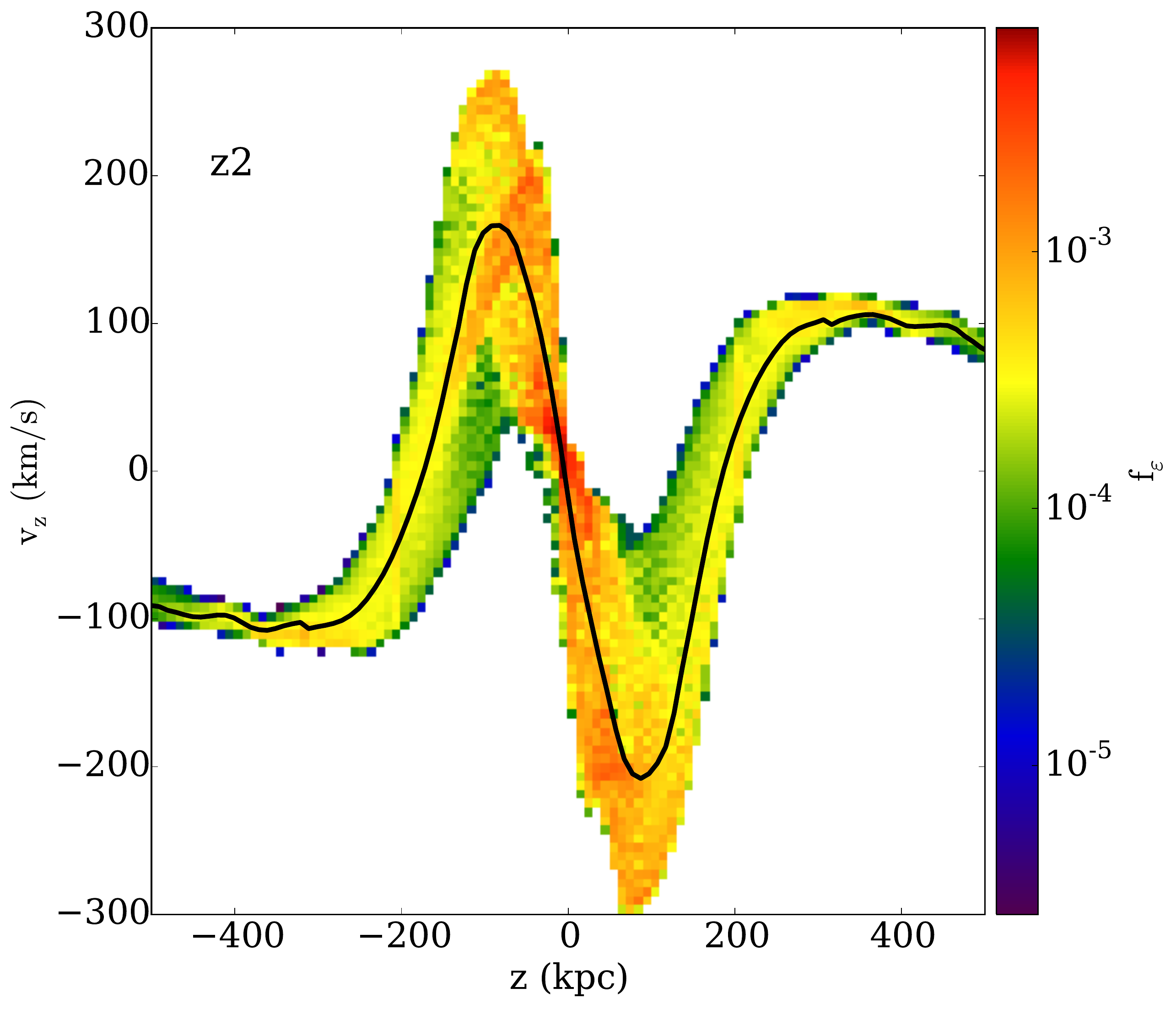}
\end{minipage}
\begin{minipage}[b]{0.47\linewidth}
\includegraphics[width=0.92\textwidth]{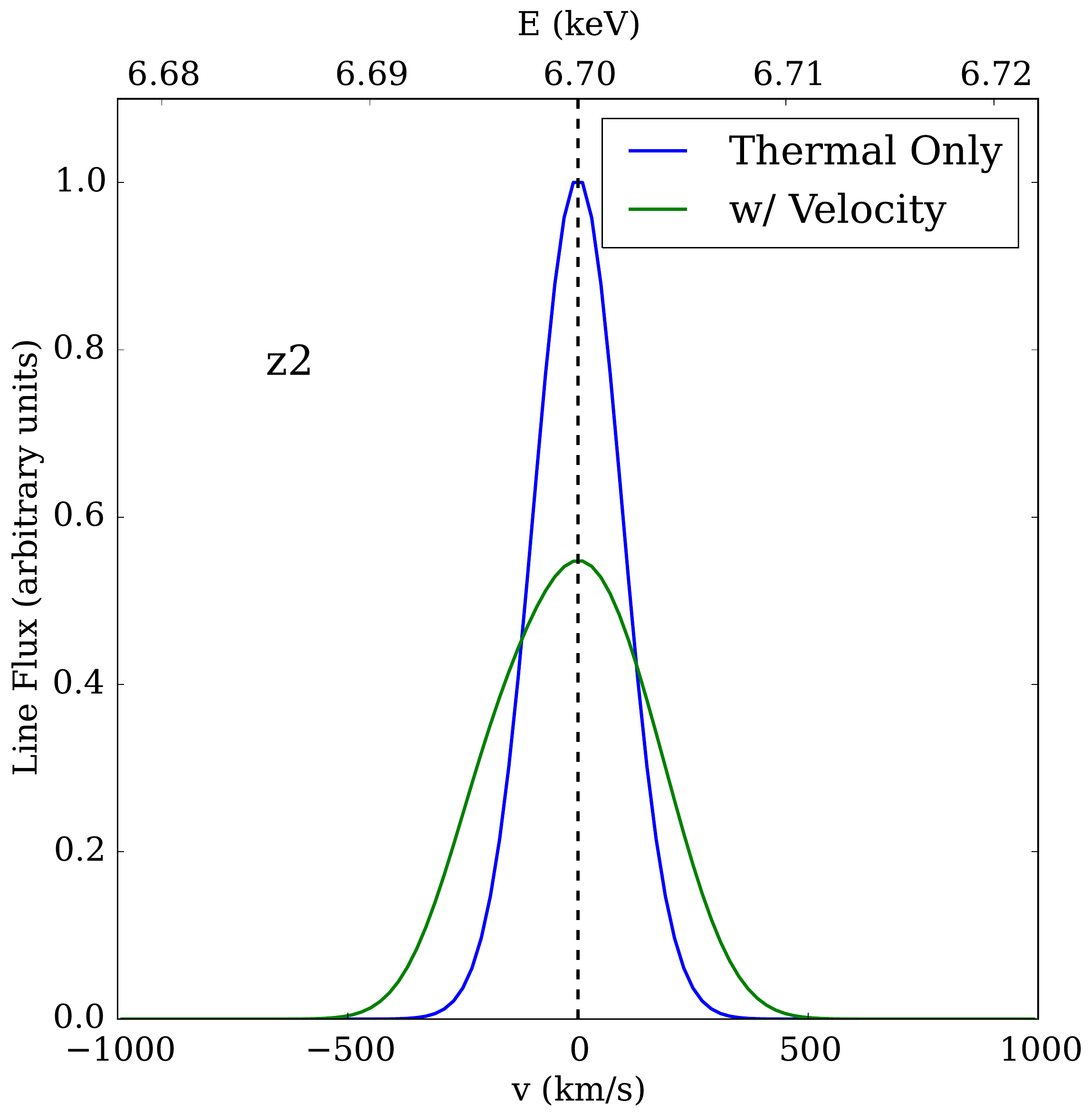}
\end{minipage}
\caption{Characteristics of the velocity field along the $z$-axis of the inviscid simulation, for region ``z2''. Upper panels: slices through the $x-z$-plane at the center of region ``z2'', of temperature (left) and the $z$-component of the velocity (right). Black lines indicate the center and edges of the elliptical cylinder corresponding to the region in Figure \ref{fig:map_z}. Lower-left panel: Phase space plot showing the fraction of emission as a function of position and velocity within the cylinder. The black line indicates the emission-weighted average value. Lower-right panel: Effect of plasma motion on a ``toy'' He-like iron line for the emission with the region.\label{fig:vz_dist2}}
\end{center}
\end{figure*}

\begin{figure*}[h!]
\begin{center}
\begin{minipage}[b]{0.46\linewidth}
\includegraphics[width=\textwidth]{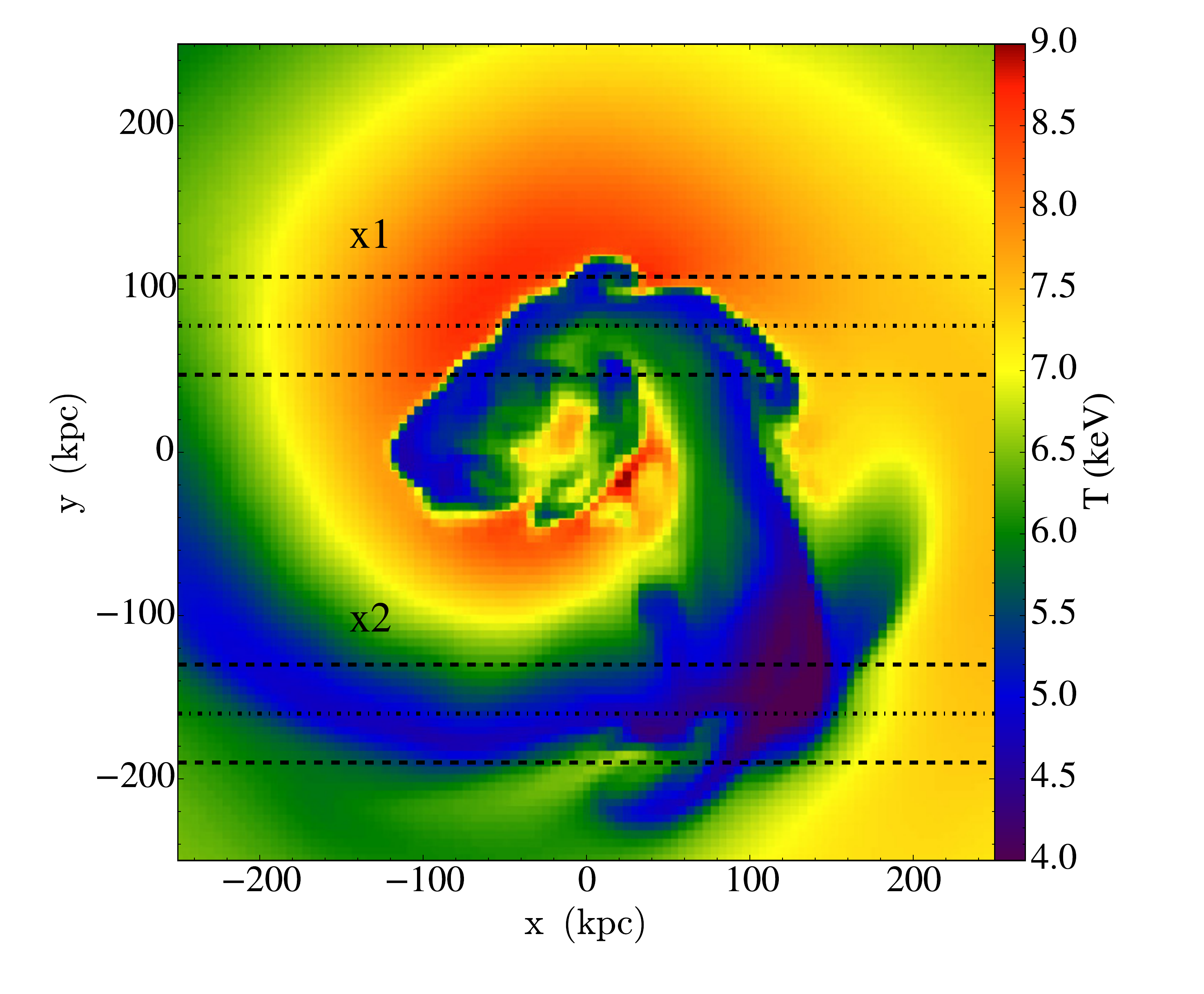}
\end{minipage}
\begin{minipage}[b]{0.46\linewidth}
\includegraphics[width=\textwidth]{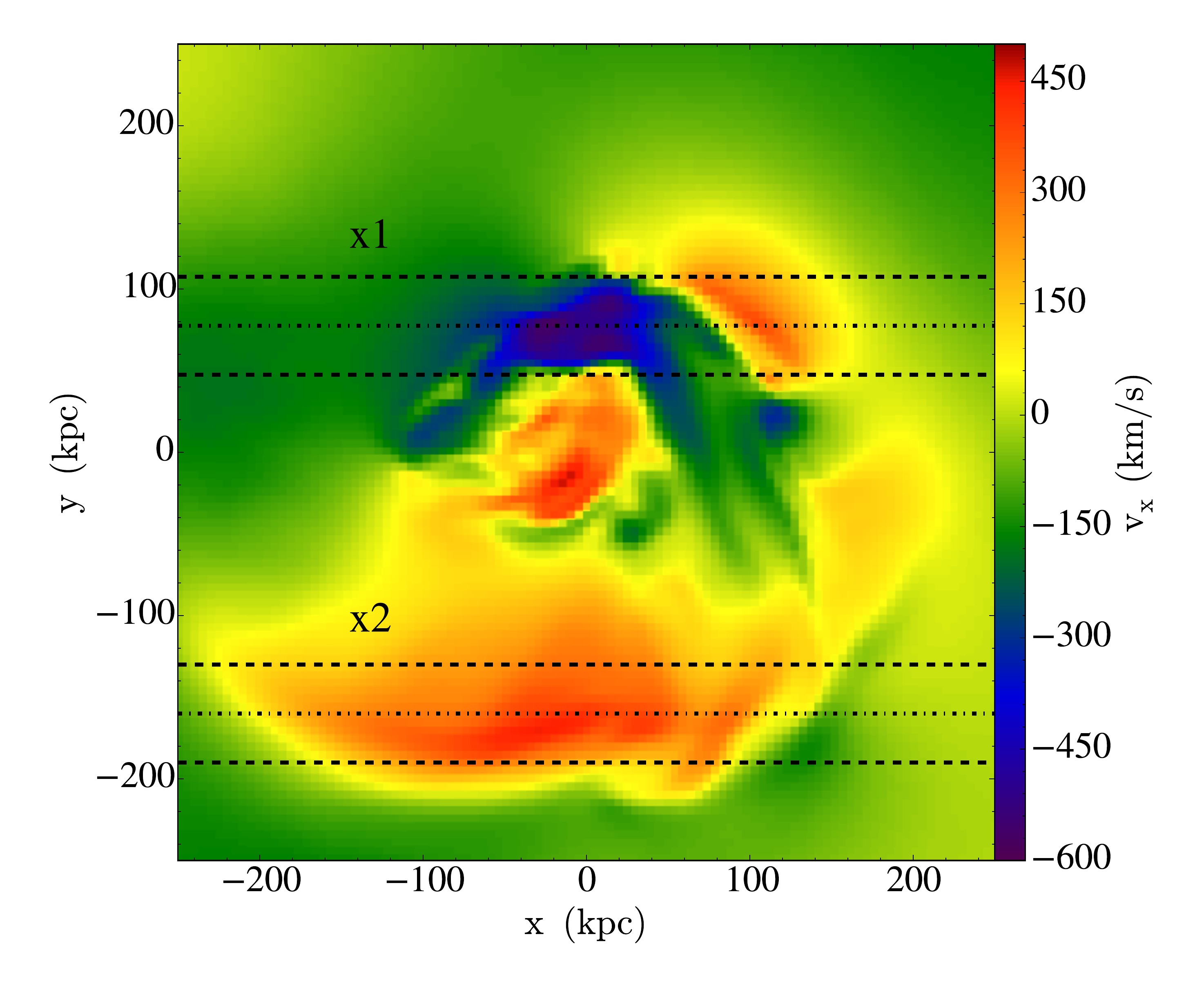}
\end{minipage}
\begin{minipage}[b]{0.46\linewidth}
\includegraphics[width=\textwidth]{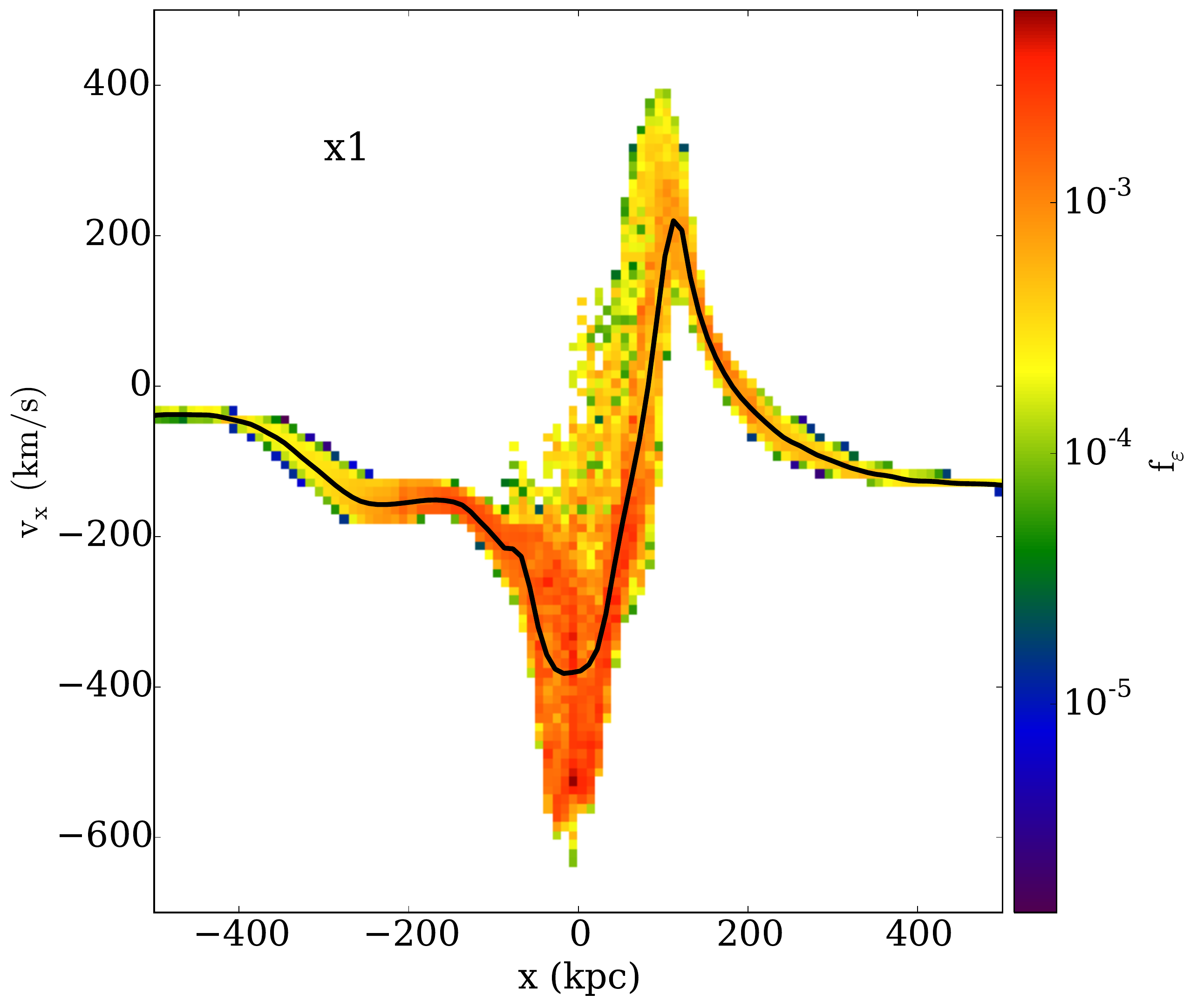}
\end{minipage}
\begin{minipage}[b]{0.46\linewidth}
\includegraphics[width=\textwidth]{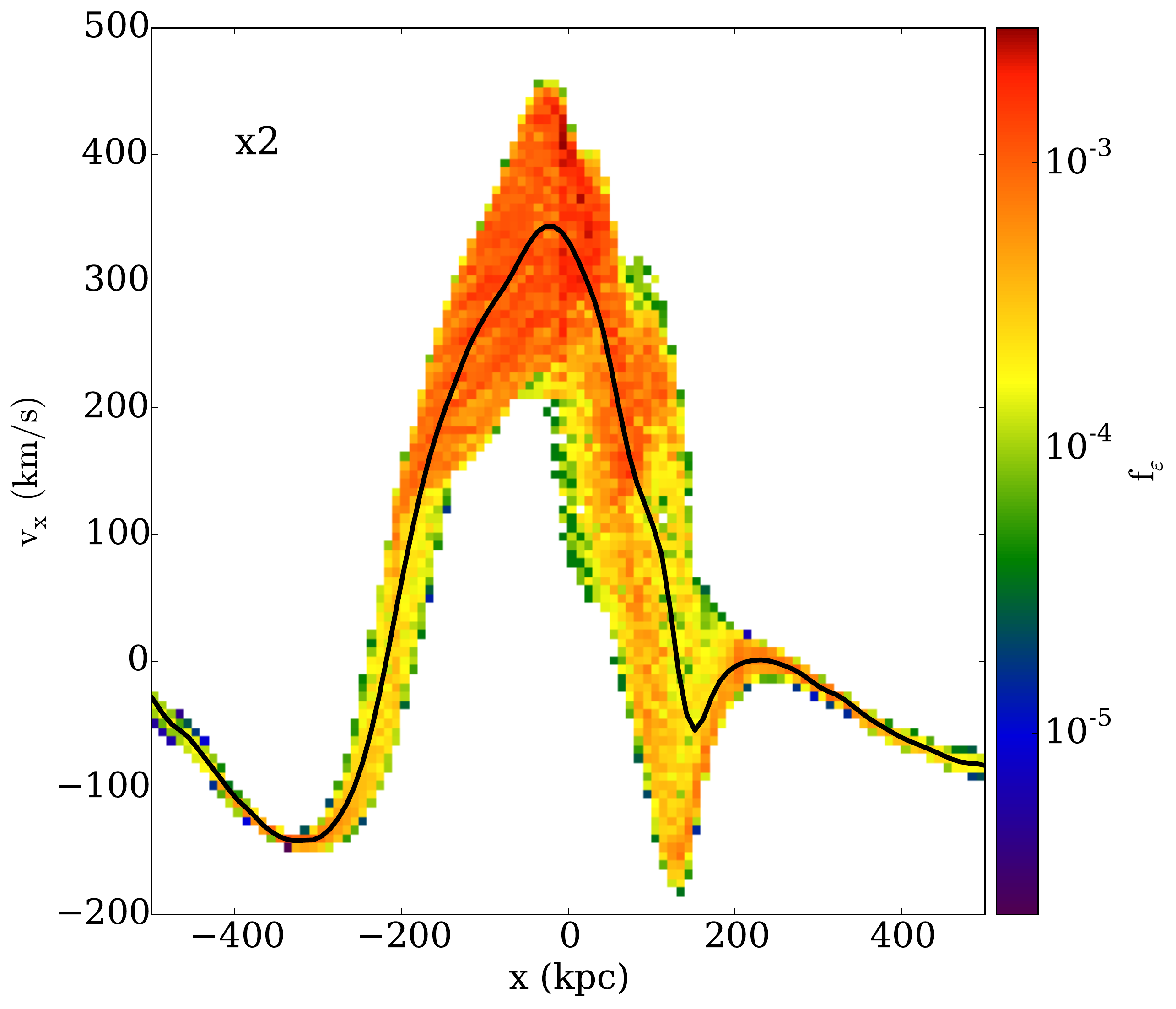}
\end{minipage}
\includegraphics[width=0.9\textwidth]{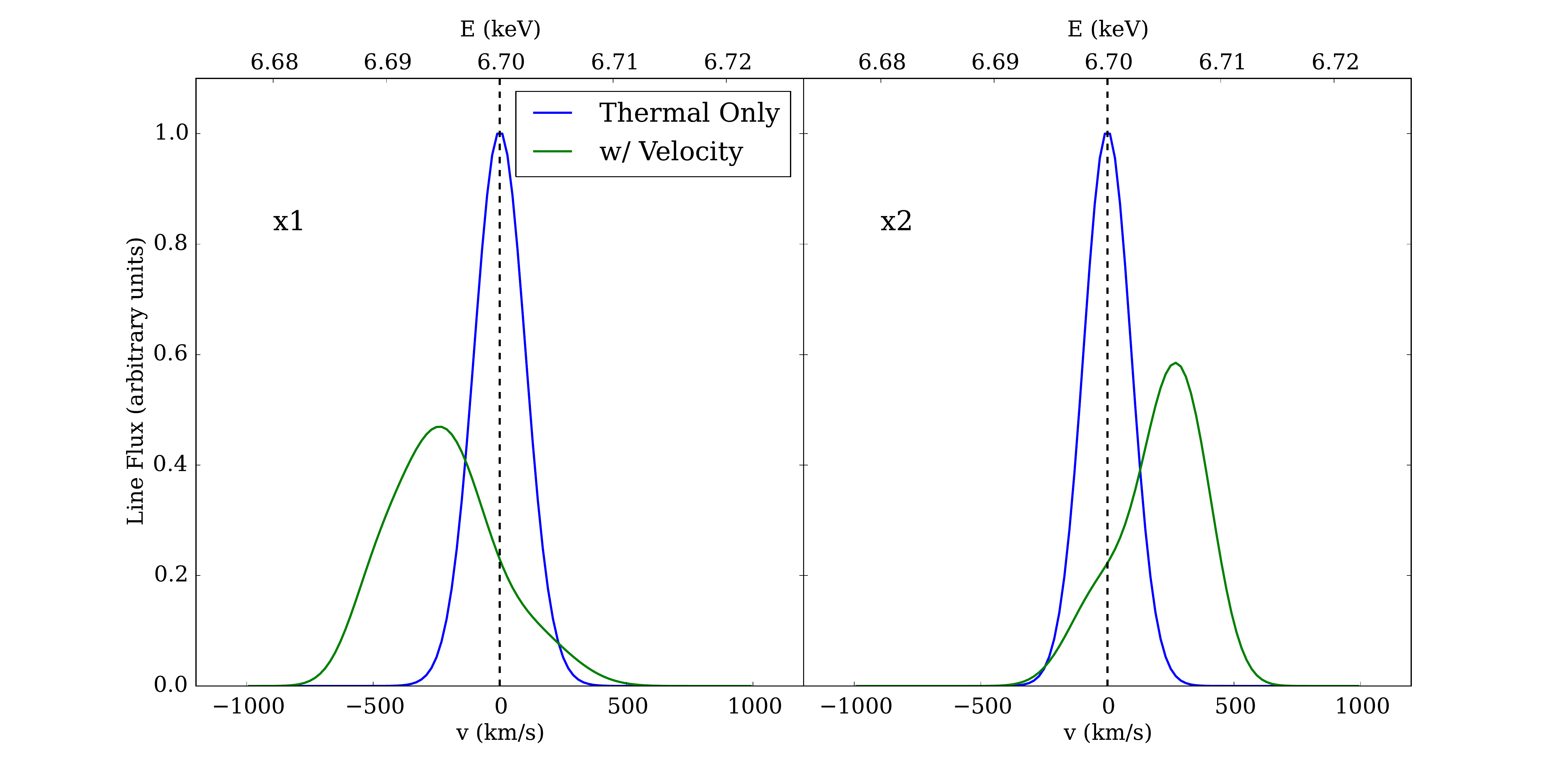}
\caption{Characteristics of the velocity field along the $x$-axis of the inviscid simulation. Upper panels: slices through the $x-y$-plane at $z = 0$, of temperature (left) and the $x$-component of the velocity (right). Black lines indicate the center and edges of the elliptical cylinders corresponding to the regions in Figure \ref{fig:map_x}. Middle panels: Phase space plots showing the fraction of emission as a function of position and velocity within the cylinder. The black line indicates the emission-weighted average value. Lower panels: Effect of plasma motion on a ``toy'' He-like iron line for the emission within the regions.\label{fig:vx_dist}}
\end{center}
\end{figure*}

\section{Results}\label{sec:results}

As a first step, we will examine projections of the velocity field obtained from the raw simulation data, to identify regions which may produce observable consequences in our spectral analysis. Second, we will examine the properties of the velocity field in detail within specific regions. Finally, we will construct and analyze synthetic spectra, making connections where appropriate to our results from the previous steps.

\subsection{Velocity Moment Maps}\label{sec:moment_maps}

Before constructing our synthetic observations, it is instructive to examine the velocity statistics of the raw simulation data along various lines of sight. This will serve as a guide to which regions and projections may yield interesting and observable consequences for {\it Astro-H} spectral analysis of our simulations and future observations of sloshing cold fronts.

Throughout this work, we will refer to the cold gas component of sloshing cold fronts as situated ``below'' the front surface, and the hot component as the gas ``above'' the front surface, where the sense of ``up'' and ``down'' is defined with respect to the direction of the local gravitational acceleration, pointing towards the cluster potential minimum. The sloshing motions have a particular geometry which immediately suggests along which lines of sight the velocity field would be expected to produce the most observable consequences. The mutual orbit of the main cluster and subcluster is situated within the $x-y$ plane, so most of the gas motion is in the $x$ and $y$ directions. Simulations also show that there is a general radial expansion of the cold fronts away from the cluster potential minimum \citep{rod11}. \citet{kes11} demonstrated that such expanding, spiral flows must have a ``bulging'' cylindrical or ``open barrel'' shape as seen from lines of sight parallel to the sloshing plane.

Figures \ref{fig:map_z}, \ref{fig:map_x}, and \ref{fig:map_y} illustrate these characteristics by showing projections along the $z$, $x$, and $y$ axes of the inviscid simulation at an epoch $t \sim 1.6$~Gyr past the core passage. Since the $z$-axis projection shows the characteristic spiral shape of the cold fronts most clearly, we show it first. The top panels of these figures show the X-ray surface brightness (top left) and gas temperature (top right), the latter using a ``spectroscopic-like'' weighting $w_{\rm sl}$ \citep{maz04}:
\begin{eqnarray}
T_{\rm proj}(\boldsymbol{\chi}) &=& \displaystyle\int{T}({\bf r})w_{\rm sl}({\bf r})\hat{\bf n}{\cdot}d{\bf r}\label{eqn:T_sl} \\
w_{\rm sl} &\propto& \rho^2T^{-3/4}
\end{eqnarray}
where $\hat{\bf n}$ is the unit normal vector defining the line of sight direction, and $\boldsymbol{\chi}$ is the 2-D position vector on the sky. The cold fronts appear in all projections, though in the $x$ and $y$-axis projections there is no evidence of a spiral pattern. However, we will see that these projections show the strongest evidence of gas motions from Doppler shifting and broadening of spectral lines.

The bottom left panels of Figures \ref{fig:map_z}, \ref{fig:map_x}, and \ref{fig:map_y} show the line of sight velocity $\mu_n$ (the line shift):
\begin{equation}
\mu_n(\boldsymbol{\chi}) = \displaystyle\int{v_n}({\bf r})w_{\rm Fe}({\bf r})\hat{\bf n}{\cdot}d{\bf r}
\end{equation}
where in this case the weighting function $w_{\rm Fe}$ is proportional to the emission in the He-like Fe line at $\sim$6.7~keV in the cluster rest frame:
\begin{equation}
w_{\rm Fe} \propto \varepsilon_{\rm Fe}(T,Z)
\end{equation}

Throughout this work, the sign convention for line-of-sight velocities is such that gas with a positive (negative) velocity is moving toward (away from) the observer.

As expected, we find fairly large (but still subsonic) line-of-sight velocities, $\mu \sim 300-500$~km/s, for projections parallel to the $x-y$ plane of the simulation. The regions with the highest velocity shifts are located underneath the cold front surfaces, indicating fast motion of the cold gas underneath the fronts. In the $z$-axis projection, perpendicular to the sloshing plane, the velocity shifts are smaller, but $\mu_z$ can vary by as much as $\sim$200~km~s$^{-1}$ on small length scales (around 10~kpc), indicating the presence of turbulence.

The bottom right panels of Figures \ref{fig:map_z}, \ref{fig:map_x}, and \ref{fig:map_y} show the line of sight velocity dispersion $\sigma_n$ (the line width):
\begin{equation}\label{eqn:sigma}
\sigma^2_n(\boldsymbol{\chi}) = \displaystyle\int{v_n^2}({\bf r})w_{\rm Fe}({\bf r})\hat{\bf n}{\cdot}d{\bf r} - \mu_n^2(\boldsymbol{\chi})
\end{equation}

Moderate velocity dispersions ($\sigma \sim 200-300$~km/s) are observed in the $x$ and $y$-axis projections, mostly spatially coincident with the regions of large velocity shifts, again underneath the cold front surfaces. In the $z$-axis projection, similar velocity dispersions are also observed, both in the central core region, and underneath the surface of the largest cold front.

Figures \ref{fig:map_z_visc}-\ref{fig:map_y_visc} show maps of the same quantities at the same epoch as Figures \ref{fig:map_z}-\ref{fig:map_y} for the viscous simulation. The surface brightness and temperature maps reveal cold fronts that are smooth and free of instabilities, as seen in previous works \citep{zuh10,rod13}. The maps of line-of-sight velocity and velocity dispersion exhibit a lack of small-scale structure, indicating that turbulence is indeed strongly suppressed. However, the line shifts of the gas components underneath the cold front surfaces are similar to those in the inviscid simulation, and the velocity dispersion in these regions is still fairly significant.

\begin{figure*}
\begin{center}
\begin{minipage}[b]{0.49\linewidth}
\includegraphics[width=0.96\textwidth]{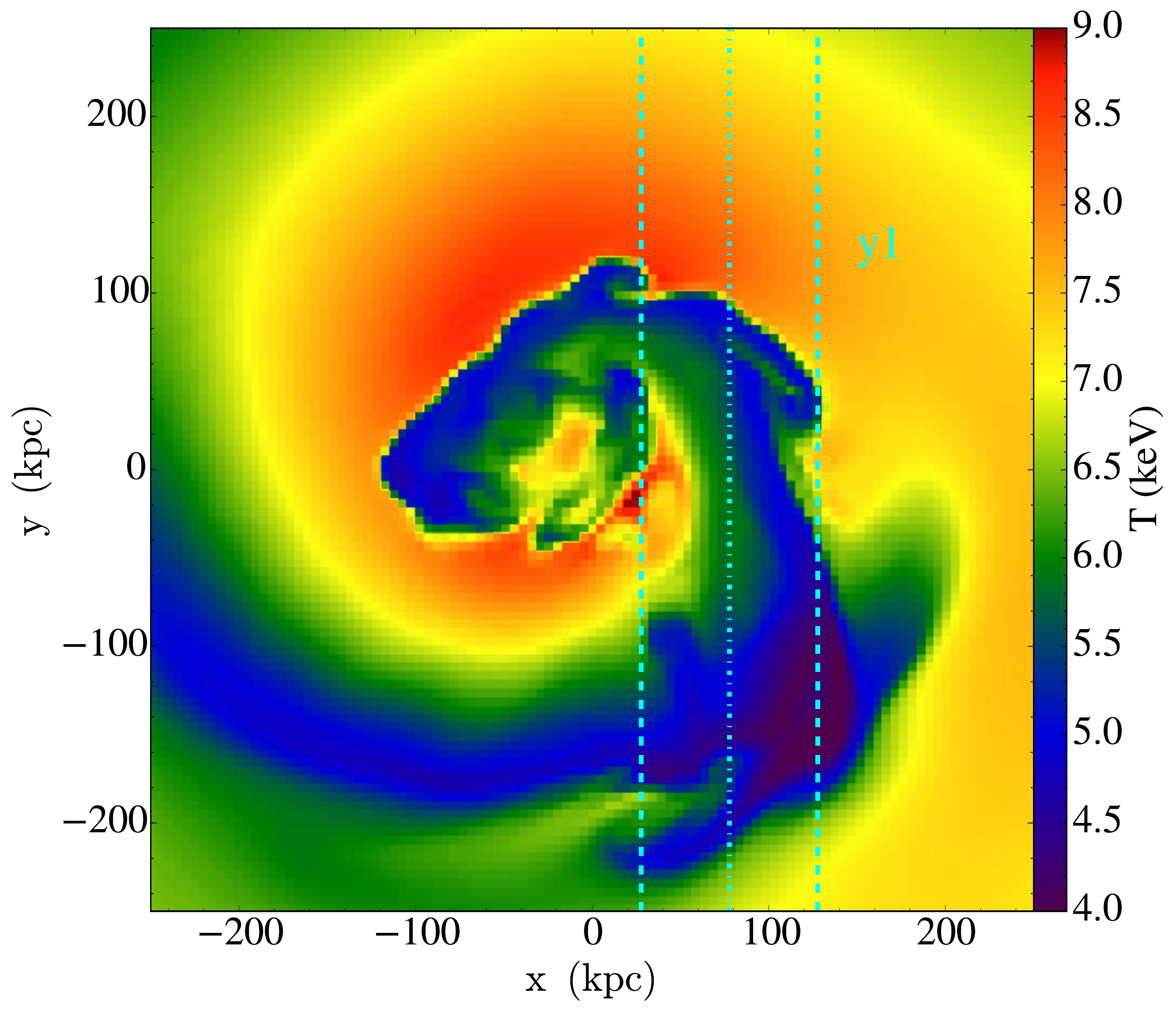}
\end{minipage}
\begin{minipage}[b]{0.49\linewidth}
\includegraphics[width=\textwidth]{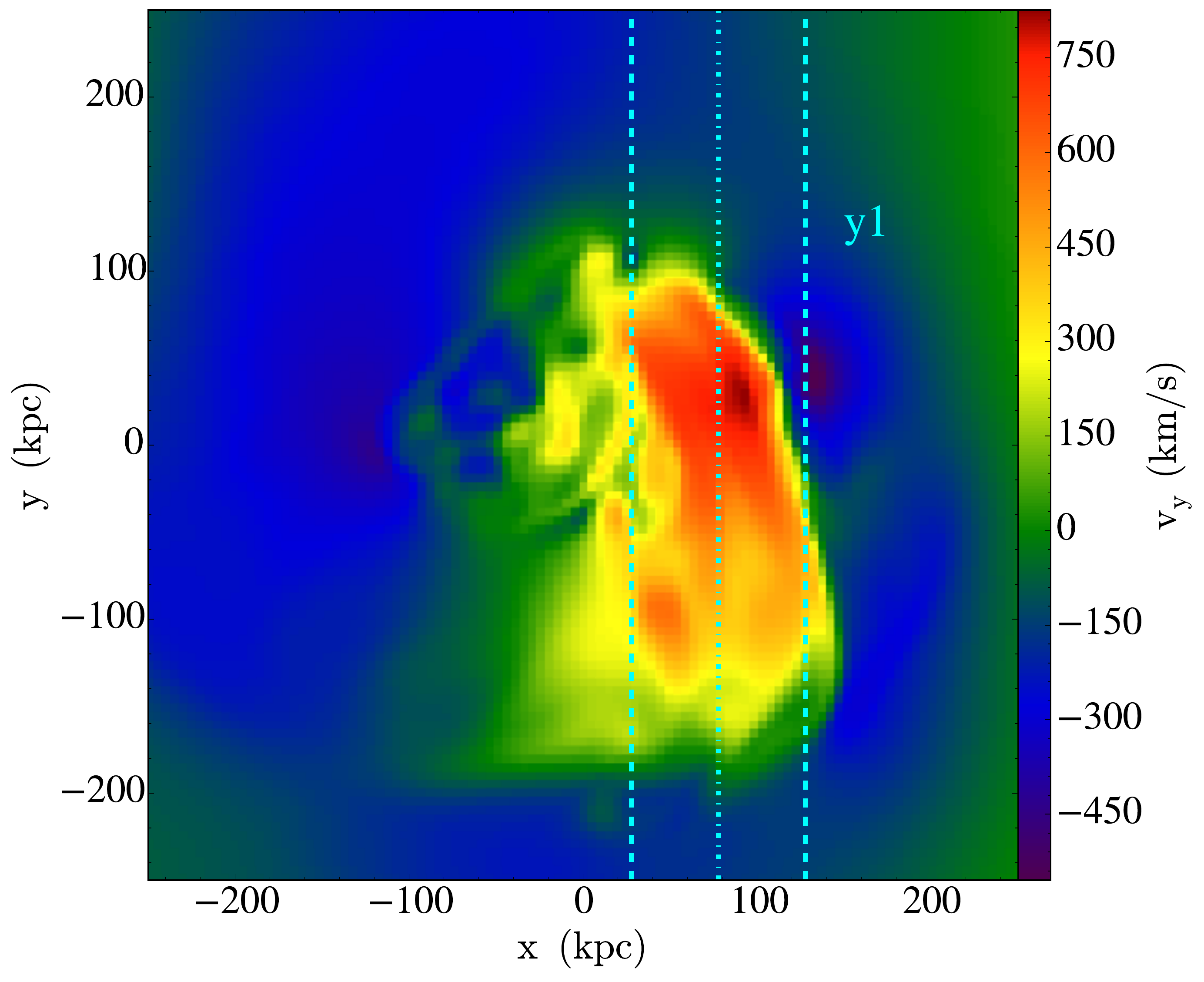}
\end{minipage}
\begin{minipage}[b]{0.51\linewidth}
\includegraphics[width=0.97\textwidth]{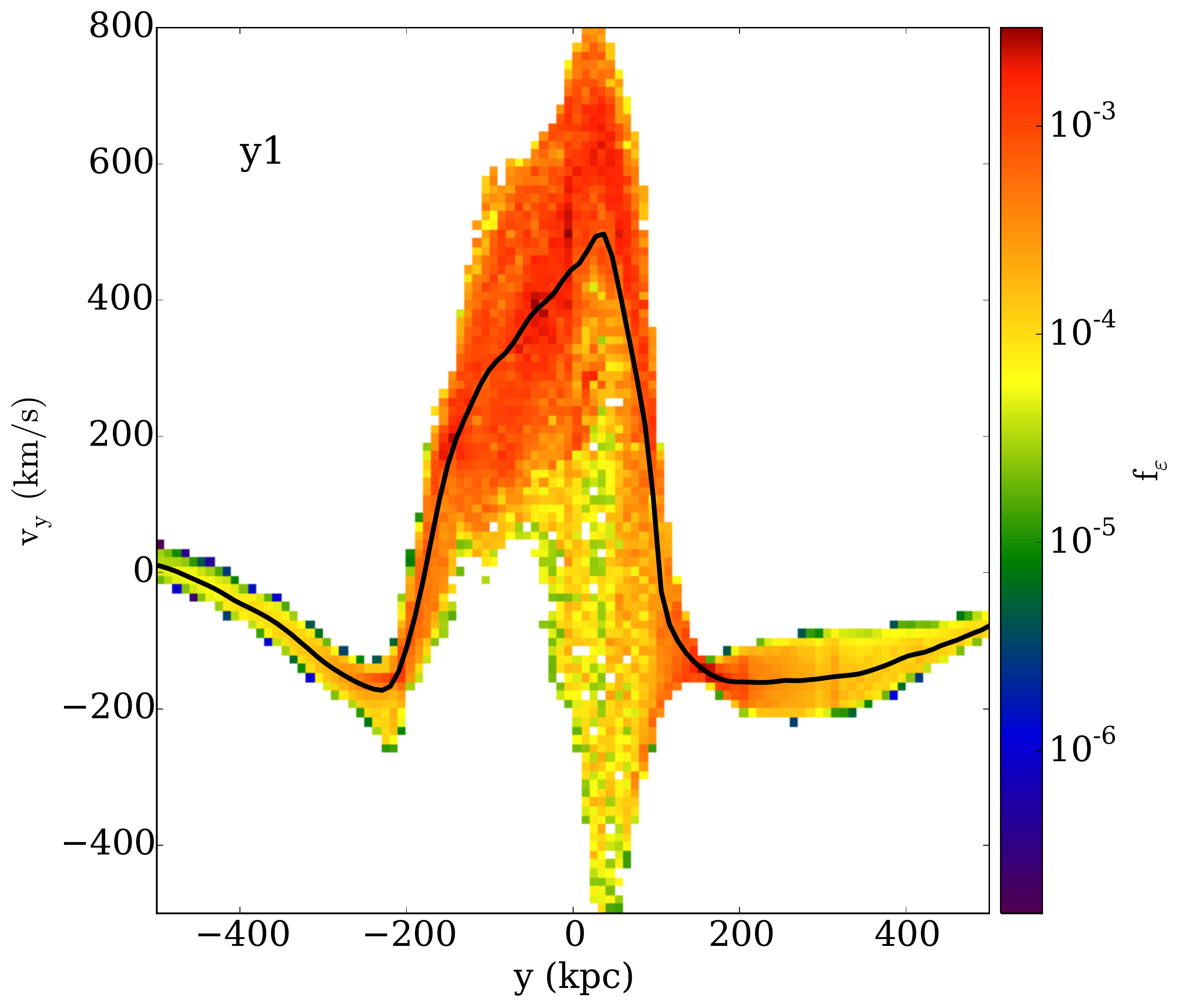}
\end{minipage}
\begin{minipage}[b]{0.47\linewidth}
\includegraphics[width=0.92\textwidth]{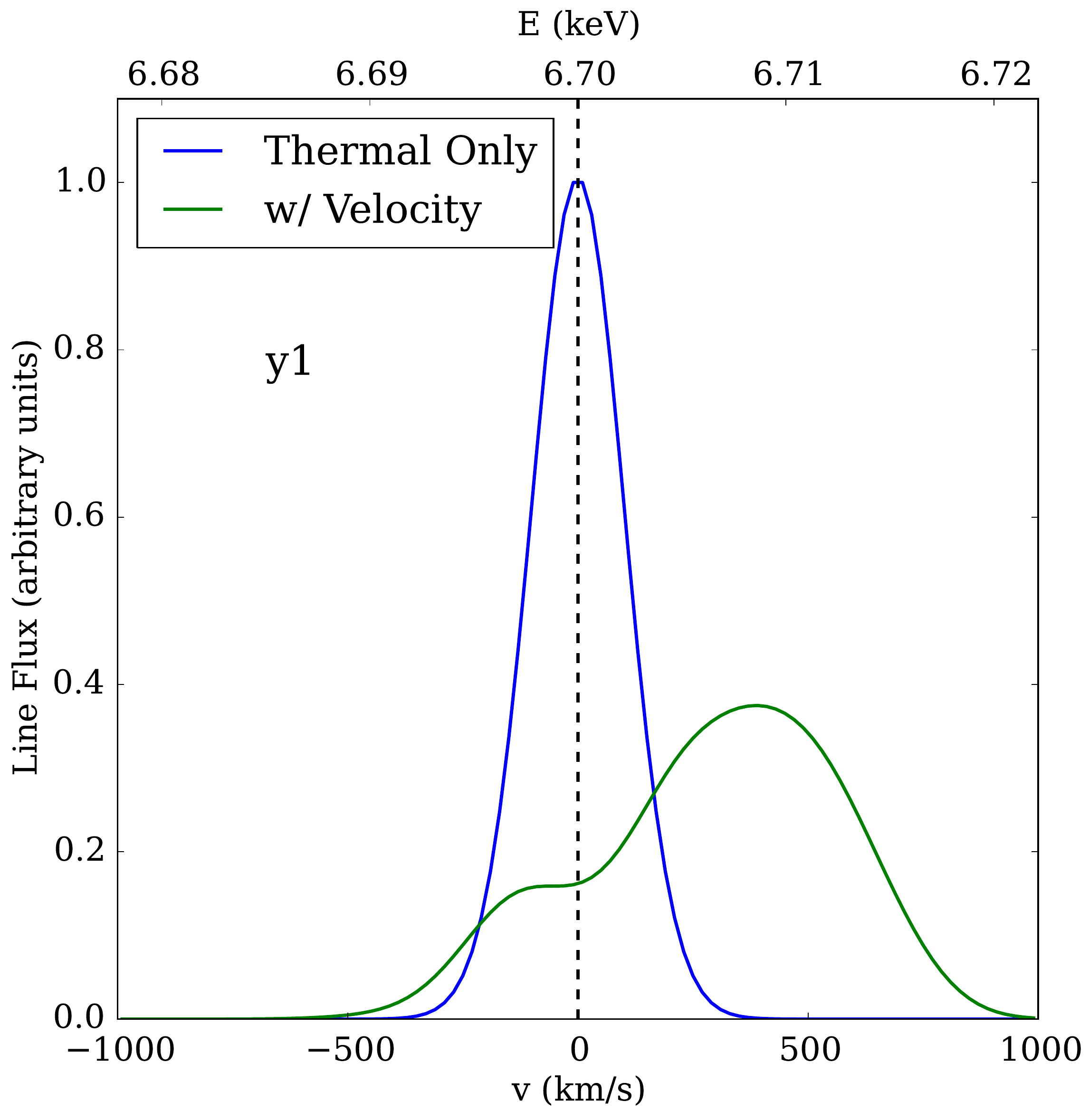}
\end{minipage}
\caption{Characteristics of the velocity field along the $y$-axis of the inviscid simulation. Upper panels: slices through the $x-y$-plane at $z = 0$, of temperature (left) and the $y$-component of the velocity (right). Cyan lines indicate the center and edges of the elliptical cylinder corresponding to the region in Figure \ref{fig:map_y}. Lower-left panel: Phase space plot showing the fraction of emission as a function of position and velocity within the cylinder. The black line indicates the emission-weighted average value. Lower-right panel: Effect of plasma motion on a ``toy'' He-like iron line for the emission with the region.\label{fig:vy_dist}}
\end{center}
\end{figure*}

\begin{figure*}[h!]
\begin{center}
\begin{minipage}[b]{0.46\linewidth}
\includegraphics[width=\textwidth]{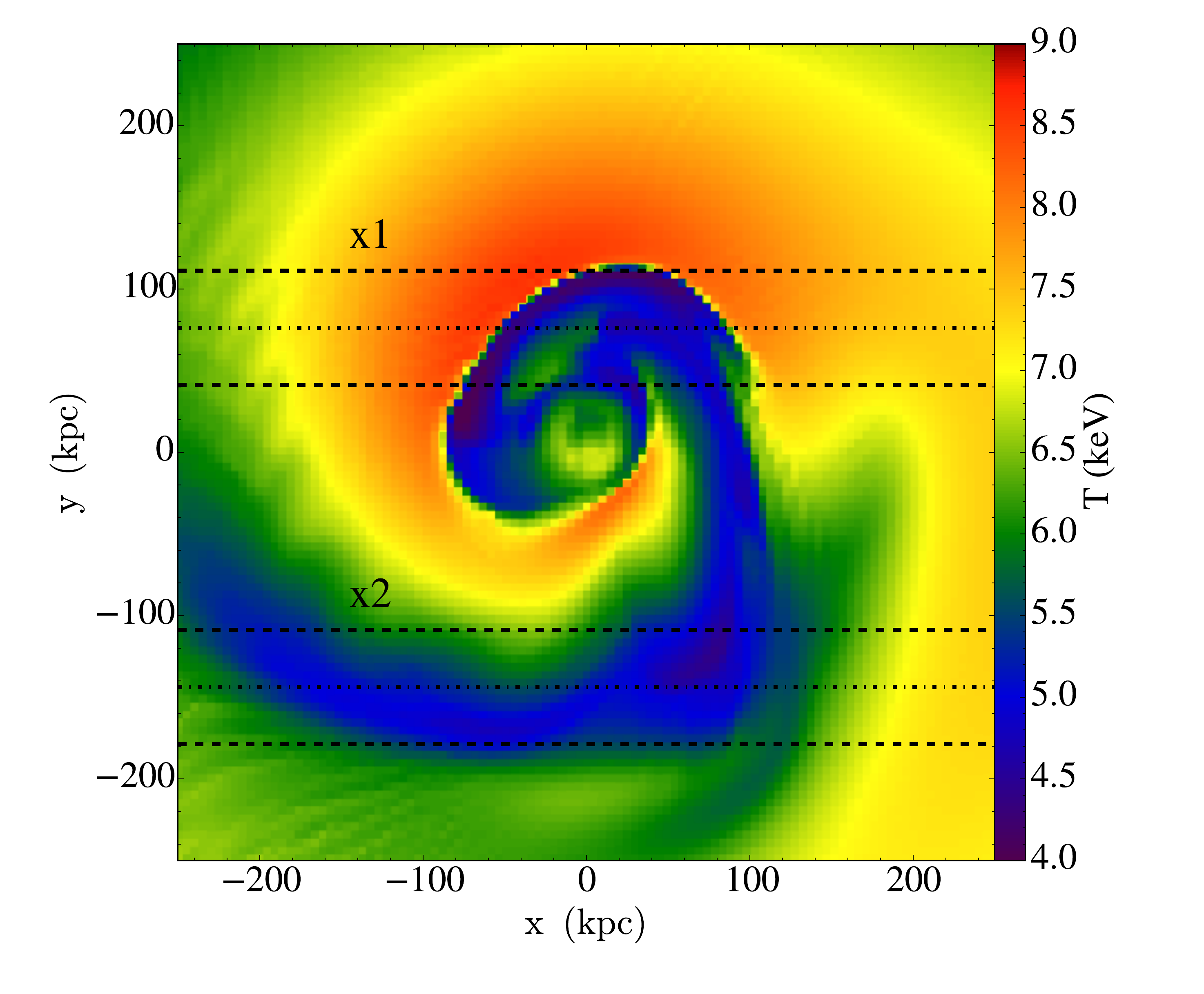}
\end{minipage}
\begin{minipage}[b]{0.46\linewidth}
\includegraphics[width=\textwidth]{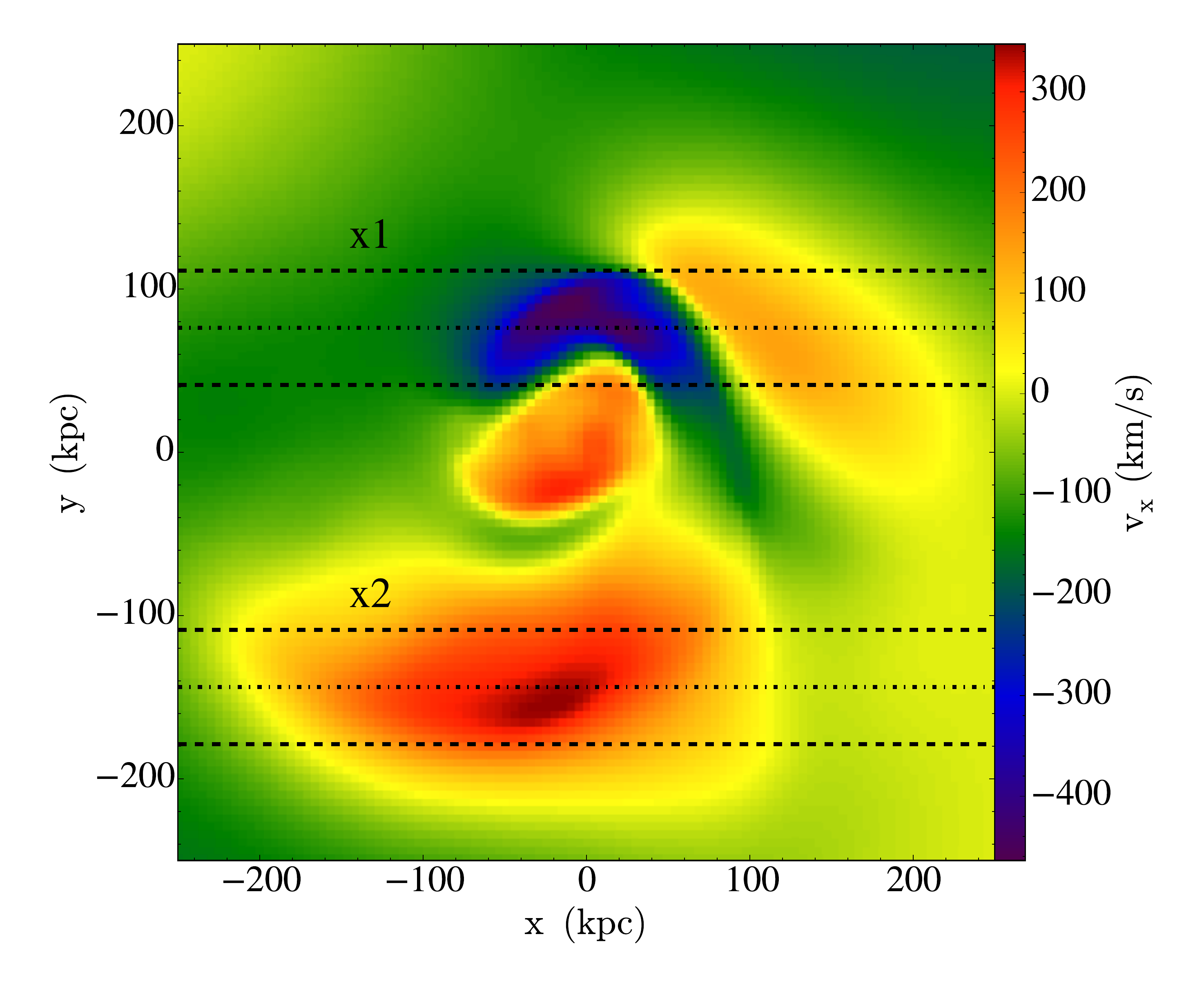}
\end{minipage}
\begin{minipage}[b]{0.46\linewidth}
\includegraphics[width=\textwidth]{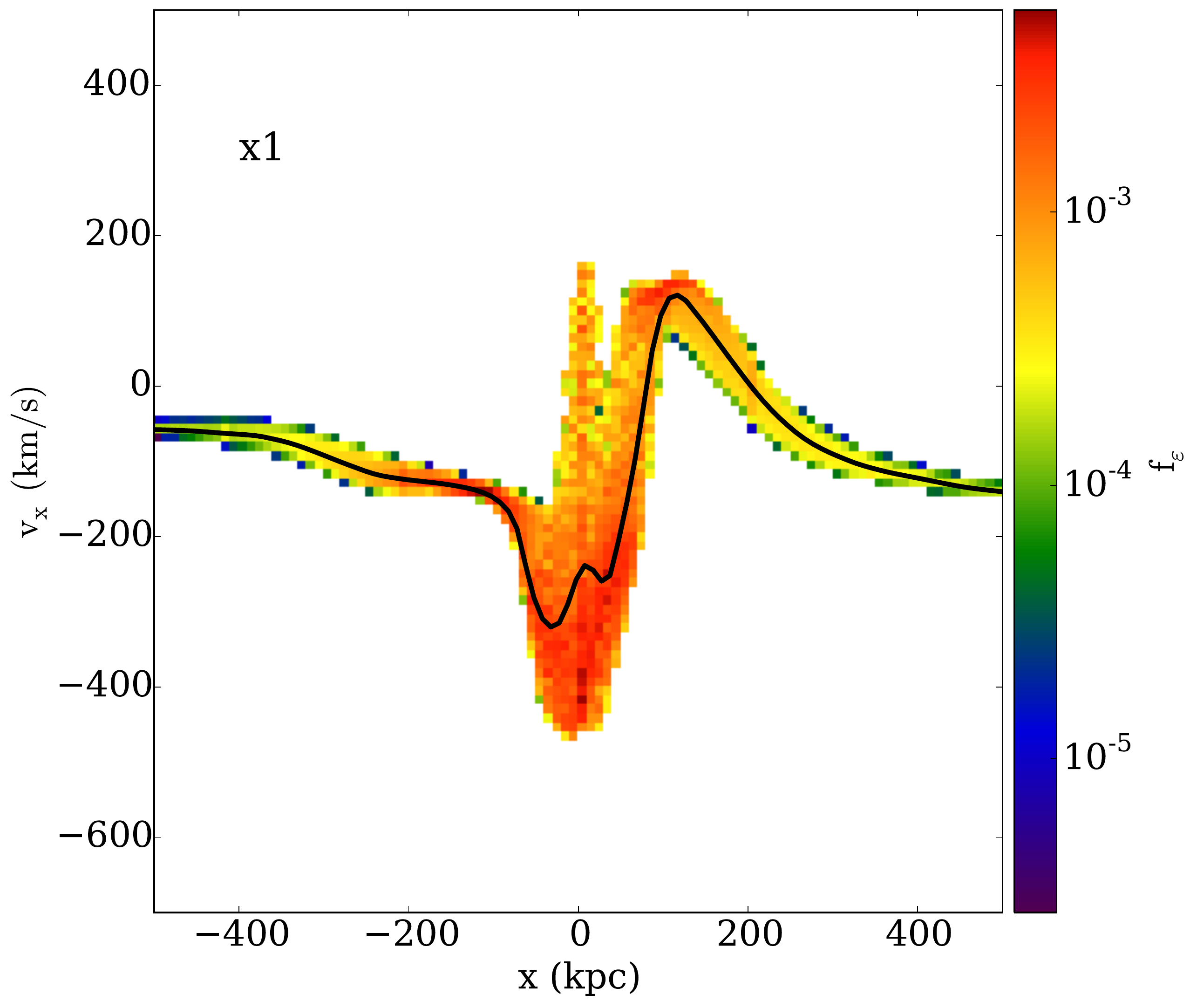}
\end{minipage}
\begin{minipage}[b]{0.46\linewidth}
\includegraphics[width=\textwidth]{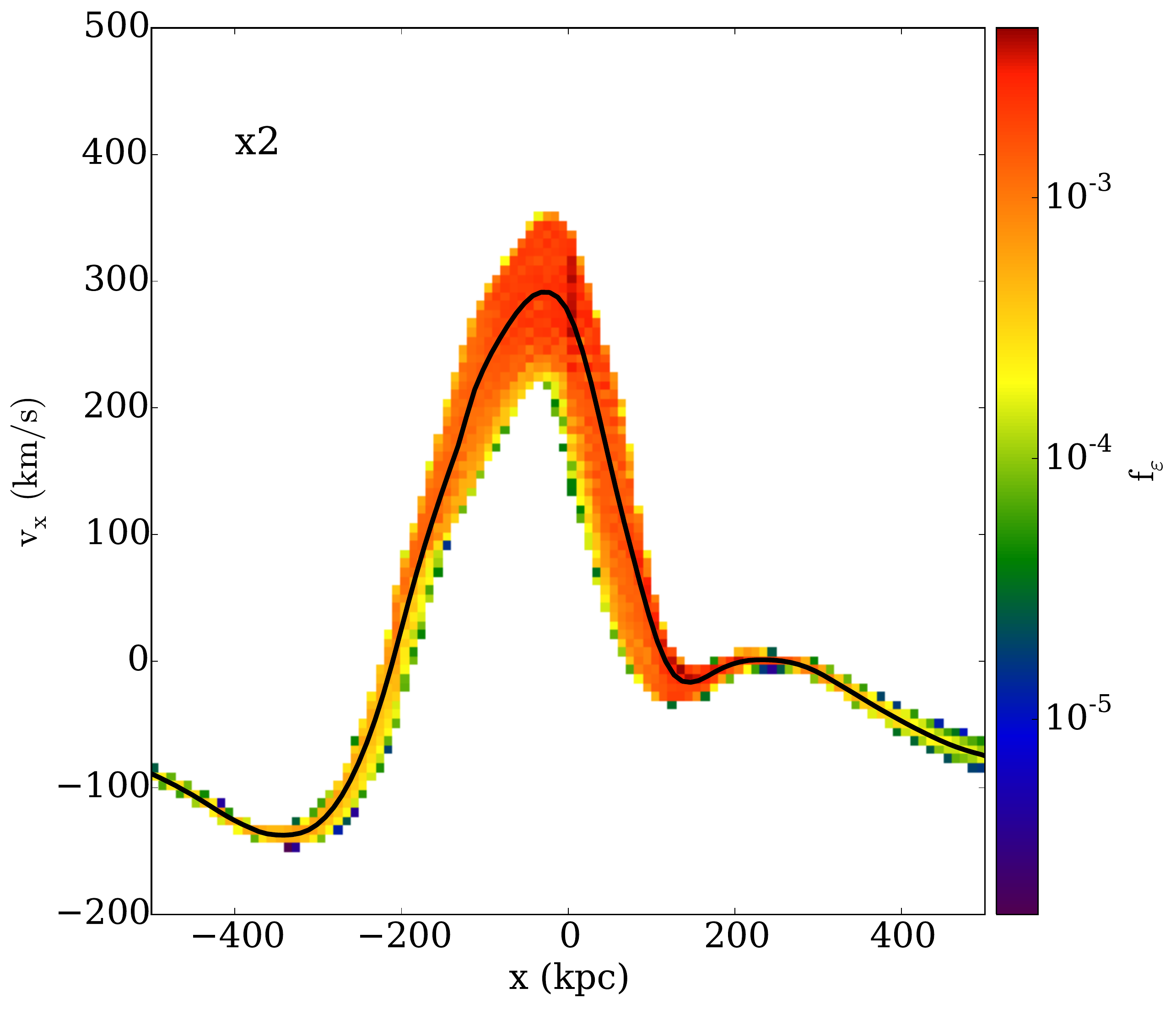}
\end{minipage}
\includegraphics[width=0.9\textwidth]{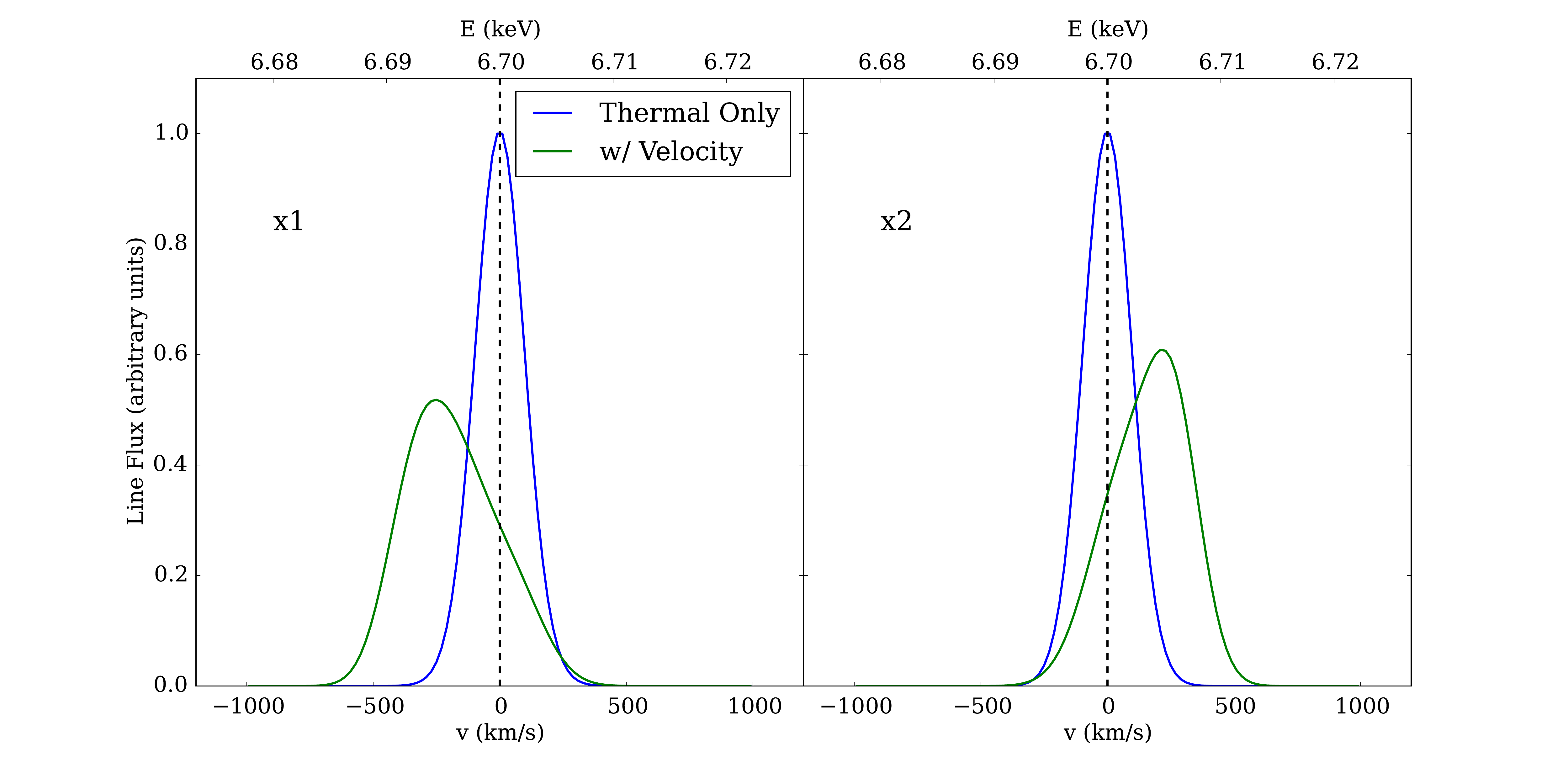}
\caption{Characteristics of the velocity field along the $x$-axis of the viscous simulation. Upper panels: slices through the $x-y$-plane at $z = 0$, of temperature (left) and the $x$-component of the velocity (right). Black lines indicate the center and edges of the elliptical cylinders corresponding to the regions in Figure \ref{fig:map_x_visc}. Middle panels: Phase space plots showing the fraction of emission as a function of position and velocity within the cylinder. The black line indicates the emission-weighted average value. Lower panels: Effect of plasma motion on a ``toy'' He-like iron line for the emission within the regions.\label{fig:vx_visc_dist}}
\end{center}
\end{figure*}

\subsection{Velocity Distribution}\label{sec:velocity_dist}

\subsubsection{Inviscid Simulation}\label{sec:vel_dist_inviscid}

The moment maps in the previous section indicate a number of interesting locations associated with sloshing cold fronts that produce line shifts and broadening that will be observable by {\it Astro-H}. We will use these maps to choose locations for {\it Astro-H} pointings, and elliptical regions within these pointings from which spectra will be extracted. We will take advantage of the velocity structure revealed by Figures \ref{fig:map_z}-\ref{fig:map_y_visc} from the simulation to decide where to locate the pointings. Of course, we will not know {\it a priori} in real observations where to point in this fashion, but the moment maps show that the regions with the most significant line shifting and broadening are located just underneath the cold front surfaces, the locations of which will be available from higher-resolution imaging of the same clusters from {\it Chandra} and {\it XMM-Newton}.

The chosen pointings and regions for the inviscid simulation are shown in Figures \ref{fig:map_z}-\ref{fig:map_y}. Each elliptical region on the maps in 2D defines a cylindrical region in 3D, extended along the line of sight across the cluster. To get a sense of the underlying velocity distribution sampled by these regions, we examine the velocity field within these cylinders. Figures \ref{fig:vz_dist1} through \ref{fig:vy_dist} show the results of this exercise. In the remainder of this work, we label each region by the axis of projection and the number of the pointing, as identified in Figures \ref{fig:map_z}-\ref{fig:map_y}. For example, pointing 1 on the $z$-axis projection is labeled ``z1''.

\begin{figure*}
\begin{center}
\includegraphics[width=\textwidth]{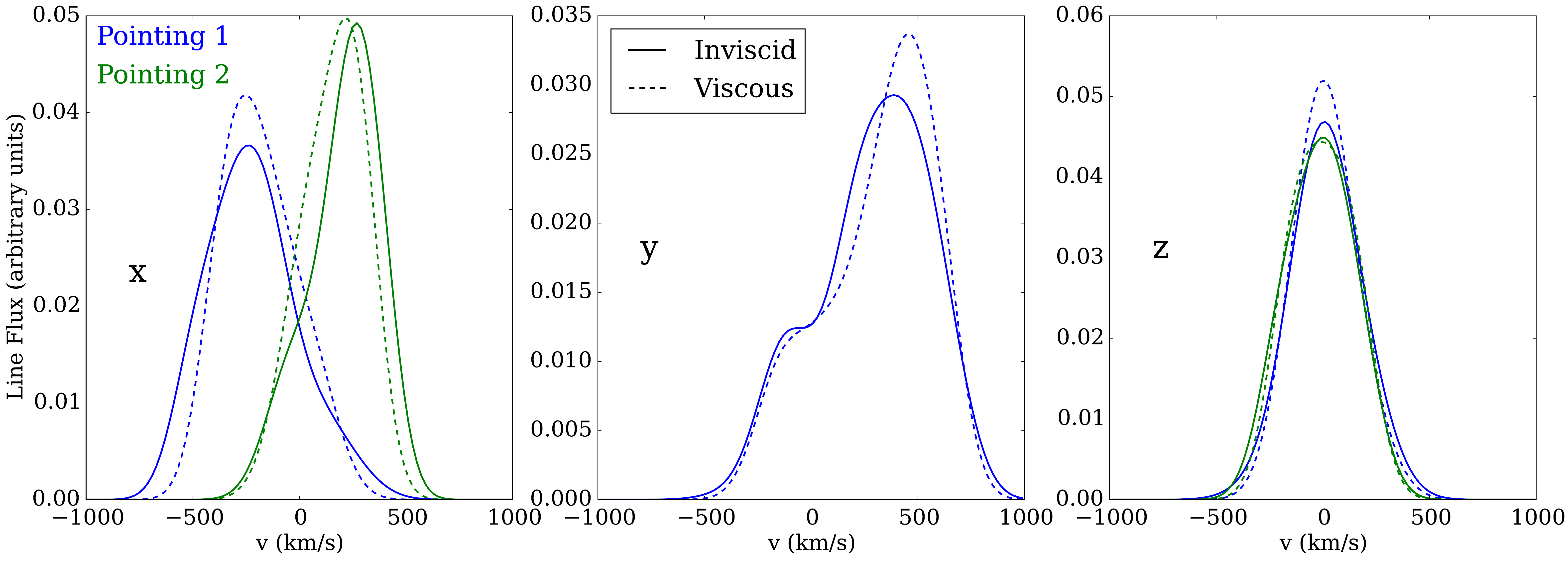}
\caption{Direct comparison of the velocity-broadened lines (see Figures \ref{fig:vz_dist1}-\ref{fig:vx_visc_dist} and \ref{fig:vz_visc_dist1}-\ref{fig:vy_visc_dist}) between the inviscid and viscous simulations. The energy-dependent emission in each line has been renormalized by its total emission.\label{fig:compare_sim_lines}}
\end{center}
\end{figure*}

We take Figure \ref{fig:vz_dist1} as an example, and describe the different panels of this figure in detail; the descrption will apply to all of the related figures. Figures \ref{fig:vz_dist1} and \ref{fig:vz_dist2} examine the $z$-component of the gas velocity. The top panels show slices through the $x-z$ plane of the temperature (left) and $z$-component of the velocity (right). The slice is taken at the $y$ coordinate of the center of the cylindrical region in Figure \ref{fig:map_z} (where lines showing the location of the slice planes for each projection are shown). The center and edges of the cylinder within the slice plane are indicated by black dot-dashed and dashed lines, respectively, in Figure \ref{fig:vz_dist1}. The lower-left panel of Figure \ref{fig:vz_dist1} shows a plot of the velocity distribution phase space in the cylinder, where the color indicates the amount of helium-like iron line emission as a function of both the position along the line of sight and the line of sight velocity. The solid black line indicates the emission-weighted average velocity along the length of the cylinder. Finally, the bottom-right panel of Figure \ref{fig:vz_dist1} shows the shape of a ``toy'' helium-like iron line, in the absence of nearby lines or continuum, computed from the total emission within the cylindrical region. The blue lines show the line shape due to thermal broadening without velocity broadening, whereas the green lines show the combined effects of thermal broadening and plasma motions.

In Figure \ref{fig:vz_dist1}, region ``z1'' is situated within the cold fronts enveloping the cluster center. The $z$-velocity slice plot and phase plot indicate the gas motion is mostly random on scales smaller than the core size within this region, with no significant velocity shift, but a range of gas motion between $v \sim -400$~km/s to $v \sim +400$~km/s. The line emission plot confirms this, showing a symmetric Gaussian-looking line that has been slightly broadened by the turbulent velocity field. Figure \ref{fig:vz_dist2} shows an entirely different velocity field in region ``z2''. In this case, the elliptical region is centered on the outermost part of the spiral to the south, and along the $z$-direction the symmetry of the sloshing motions results in two oppositely directed gas flows of approximately $v_z \sim 200$~km/s on either side of the $x-y$-plane, showed prominently in both the $z$-velocity slice plot and the phase plot. Though the velocity distribution here is very different from that of region ``z1'', the shape of the line emission is nevertheless nearly identical--both are approximately Gaussian in shape.

In Figure \ref{fig:vx_dist}, now examining the $x$-component of the velocity, the slice plots in the top panels are taken through the $x-y$ plane of the two subclusters' mutual orbit, and cuts through the center of both the ``x1'' and ``x2'' cylindrical regions. These regions align with the cold gas component of the sloshing cold fronts, which are regions of significant bulk motion along the $x$-axis. The phase space plots (middle panels) demonstrate that in both regions there is a substantial portion of gas moving along the line of sight in each region, causing the line shifts seen in Figure \ref{fig:map_x}. There is also a considerable amount of variance in the velocity field, both along the line of sight (indicated by the large variation in the black line) and in the plane perpendicular to the line of sight along the length of the cylinder. For example, at the $x = 0$ position in both regions, there is a $\Delta{v} \sim 200$~km/s spread in velocity from the average value. In the bottom panels of Figure \ref{fig:vx_dist}, the large bulk motions along the line of sight in each region create line centroid shifts of $\Delta{v} \sim 300$~km/s ($\Delta{E} \sim 7$~eV) in either direction, with a fairly significant line broadening of FWHM~$ \sim 400-500$~km/s in both cases. Each line also appears slightly skewed in the direction of the velocity shift. These features of the lines may be identified with features of the phase plots, which both show a broad distribution of gas in velocity space centered around a non-zero mean velocity in either direction.

Lastly, Figure \ref{fig:vy_dist} shows the same plots for the single region ``y1'', for the $y$-component of the velocity. The slice is taken through the $x-y$ plane at $z = 0$. Within this region, there is a large bulk motion in the $+y$ direction. The phase plot shows that within the cold front region the bulk of this gas moves with an average $y$-velocity of $v \sim +400$~km/s, with a significant spread of $v \sim +200-+800$~km/s, seen in the central portion of the panel. A smaller fraction of the X-ray emitting gas is moving in the $-y$ direction with a speed of $v \sim -150$~km/s, with much smaller variation, seen at a distance of $\sim$200~kpc on either side of the center. The shape of the emission line in the lower-right panel of Figure \ref{fig:vy_dist} shows evidence of both of these components.

\subsubsection{Viscous Simulation}\label{sec:vel_dist_viscous}

We also perform the same examination of the viscous simulation. Our results are very similar to those in the previous section, so for brevity we only show the results for the $x$-component of the velocity here in Figure \ref{fig:vx_visc_dist}, and refer the reader to Appendix \ref{sec:visc_figures} for the figures showing the other projections.

Figure \ref{fig:vx_visc_dist} shows the slice, phase space, and line shape plots for the viscous simulation for the $x$-component of the velocity. Overall, due to the damping of turbulence and instabilities by viscosity, the velocity distribution within each region is smoother, though the large-scale motions remain essentially the same. The slice plots are almost devoid of small-scale features, and though the average value of the velocity along the length of each cylindrical region is roughly the same as in Figure \ref{fig:vx_dist} (the black lines in the phase plots), the emission is not quite as spread out in phase space as in the inviscid simulation. This manifests itself in a slight narrowing of the velocity distribution at any given position along the axis in the phase space plots. Nevertheless, the line profiles from the viscous simulation are very similar to those from the inviscid case. Figure \ref{fig:compare_sim_lines} demonstrates this by comparing the line shapes from both simulations directly along all three principal axes. This indicates that in both cases the gas motion on large scales is not only responsible for the shift in the spectral line, but is also largely responsible for the line broadening as well, due to its smooth variation along the line of sight. Due to this similarity, in the next section we will only generate and fit synthetic spectra from the inviscid simulation, for brevity.

\subsection{Synthetic Observations}\label{sec:syn_obs}

We generate our synthetic observations according to the procedure in Section \ref{sec:xray_sims}. For each projection, we generate 200 synthetic event lists, assuming an exposure time of 200~ks. We then perform mock observations at the pointing locations shown in Figures \ref{fig:map_z}-\ref{fig:map_y}, extracting spectra from the elliptical regions within those pointings (the same as those used in the previous section). The 200 realizations of each region will be used to determine the model parameters and their confidence limits.

\begin{figure*}
\begin{center}
\includegraphics[width=0.49\textwidth]{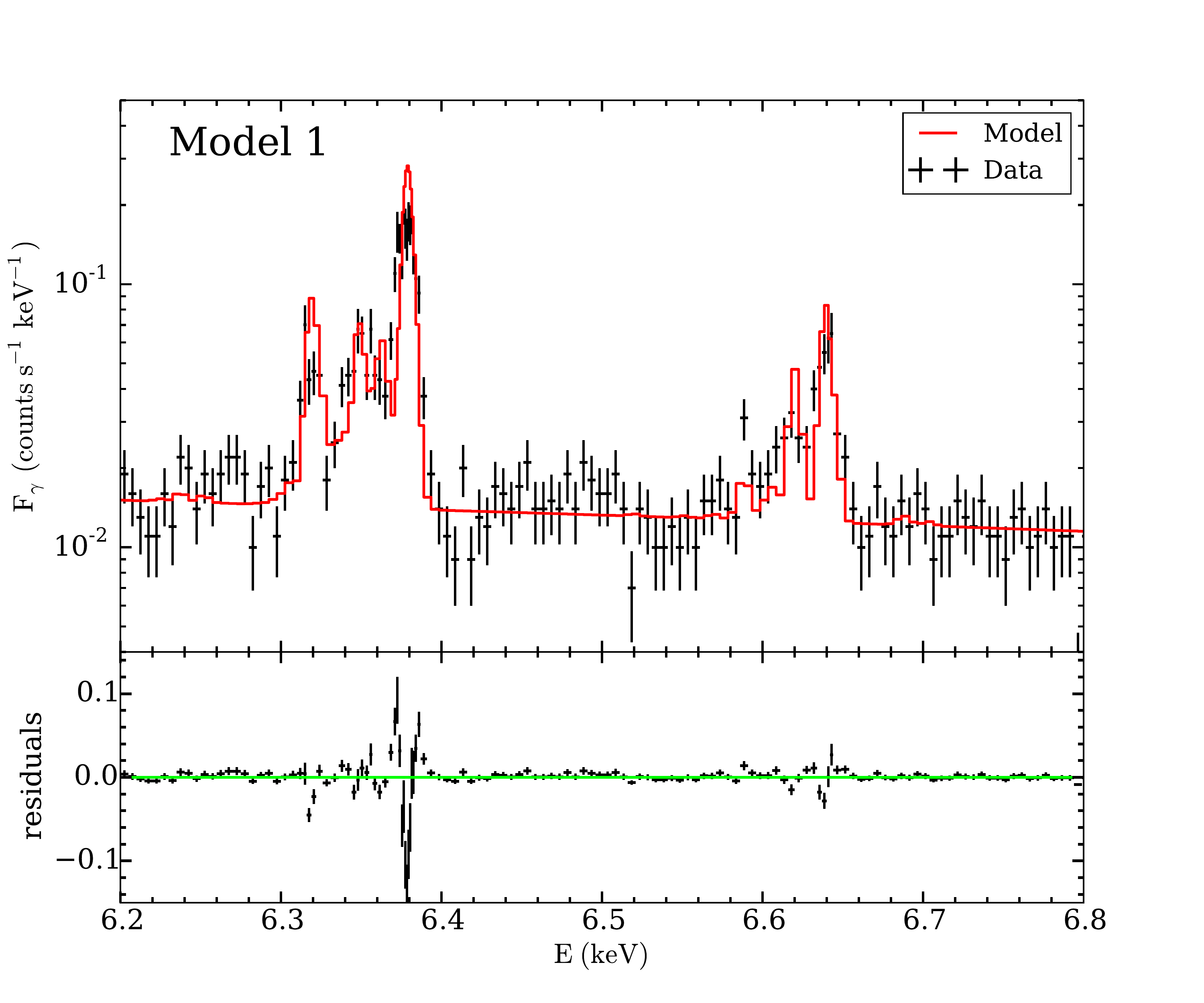}
\includegraphics[width=0.49\textwidth]{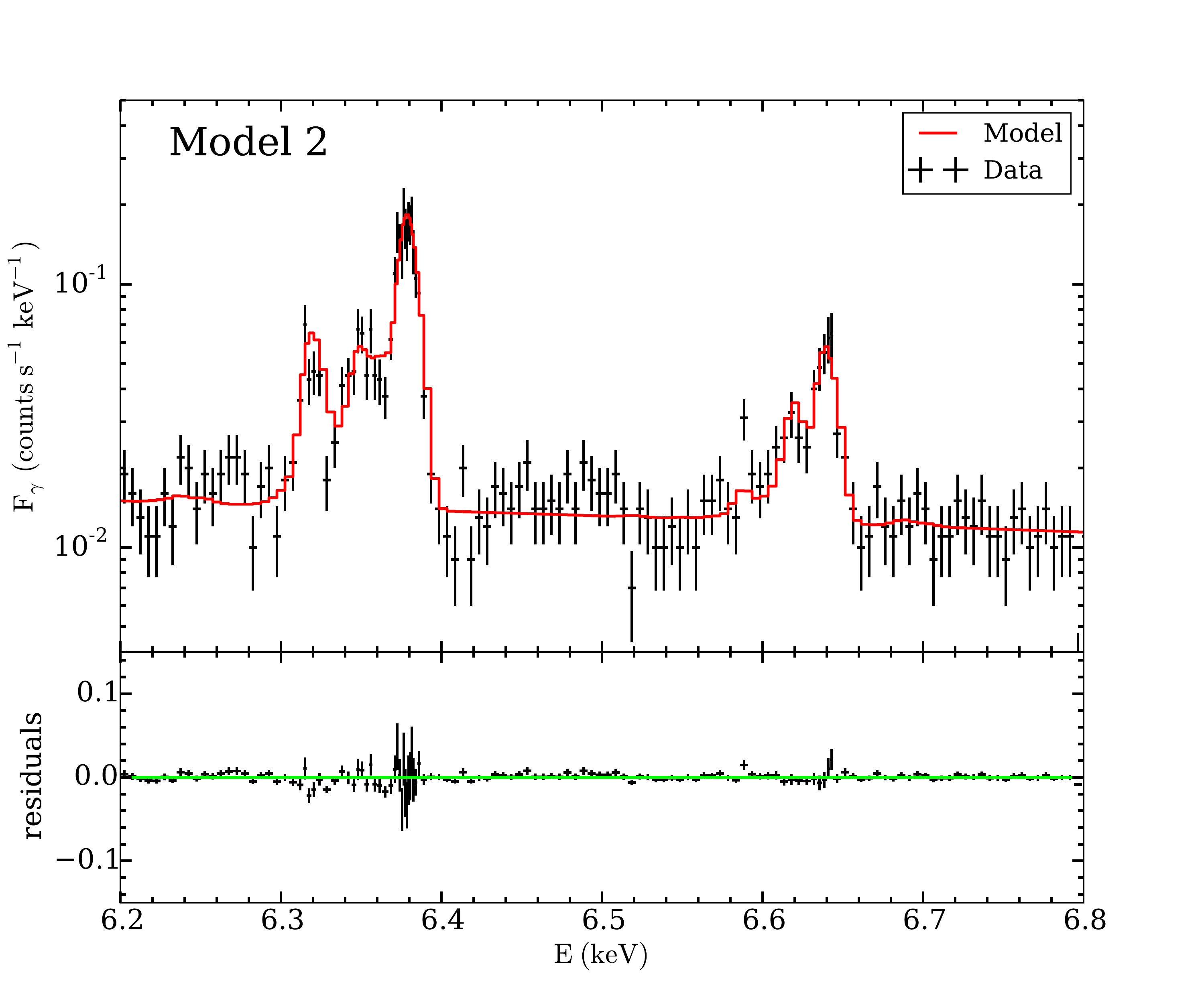}
\includegraphics[width=0.49\textwidth]{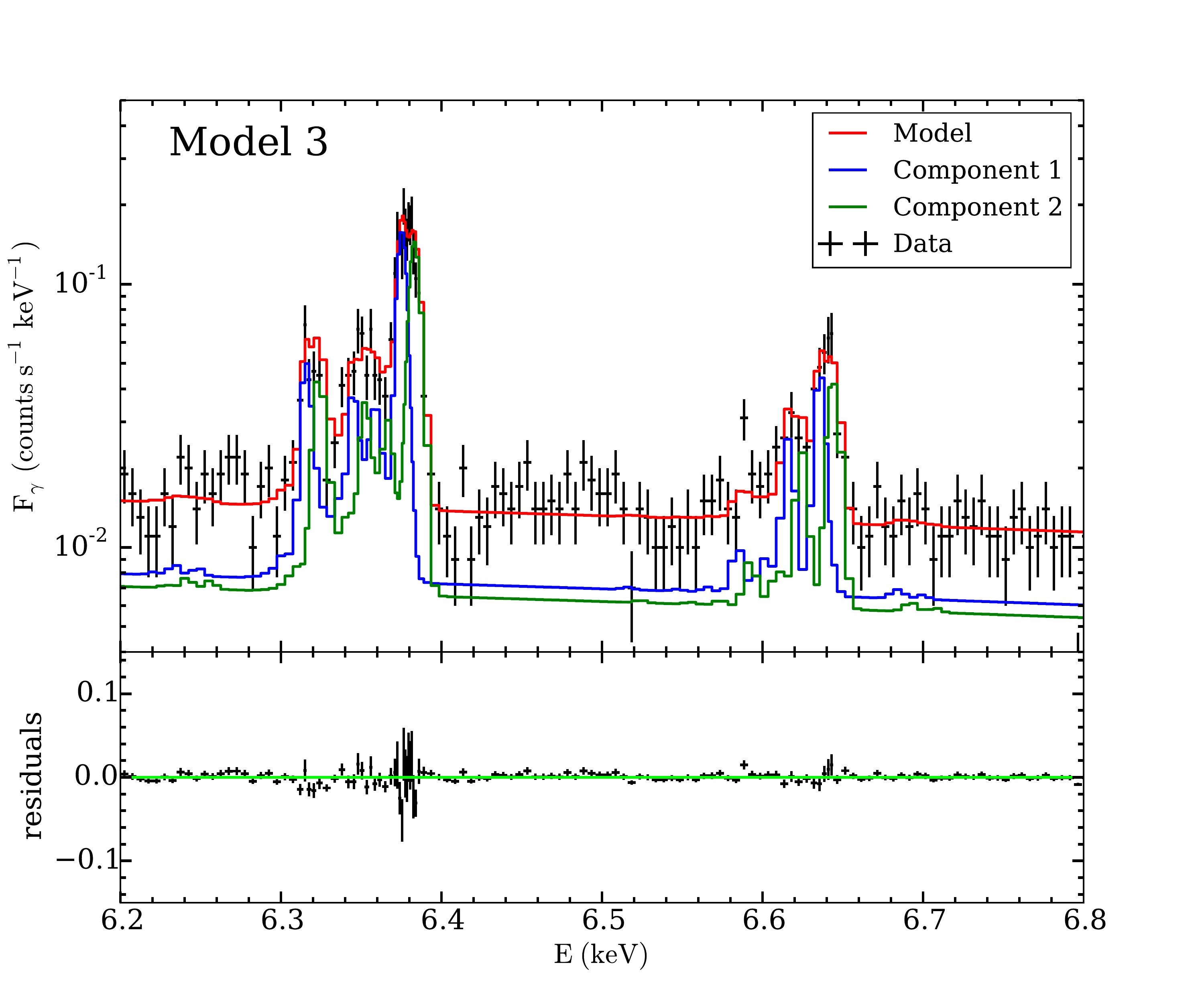}
\includegraphics[width=0.49\textwidth]{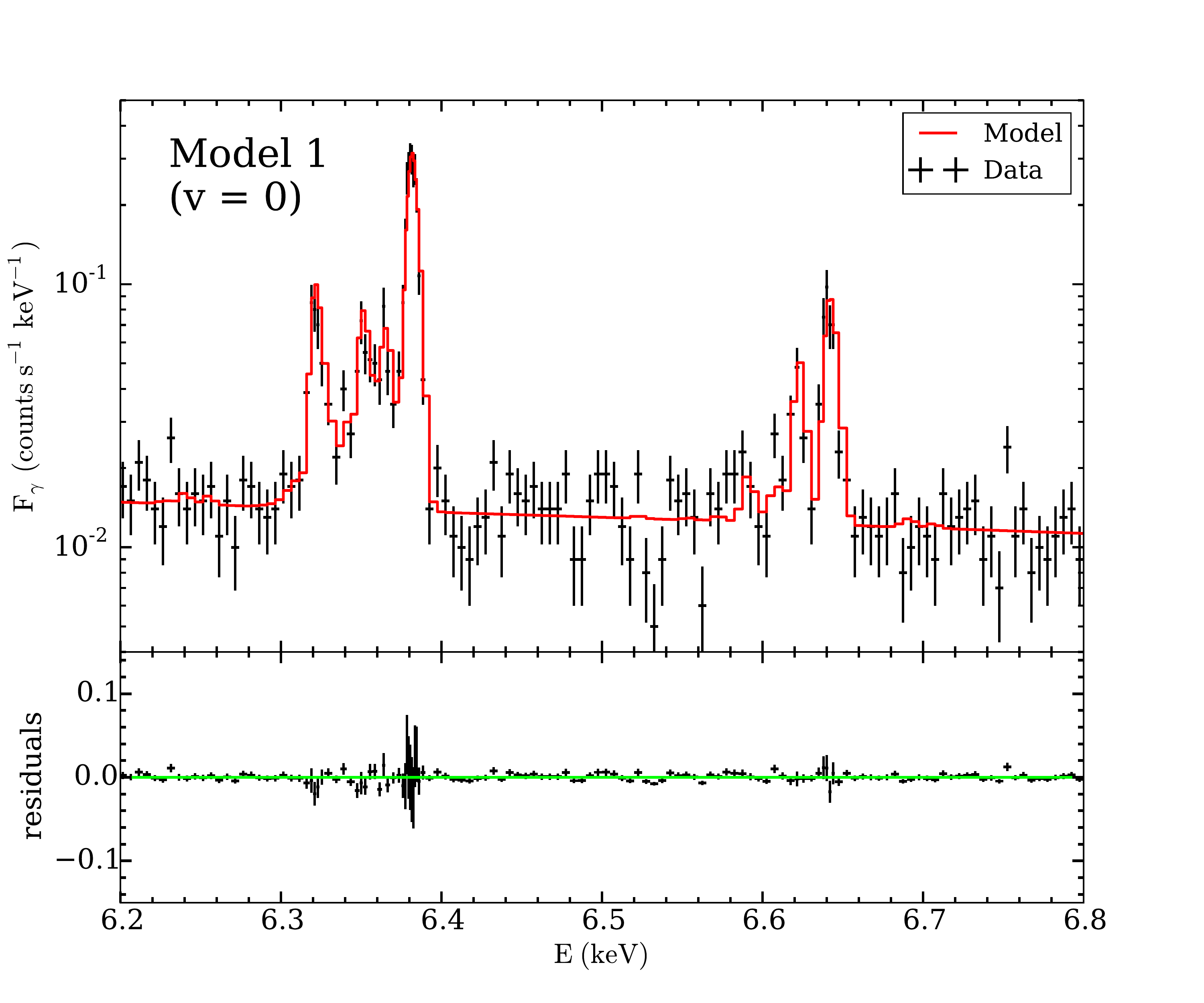}
\caption{Spectrum, fitted models, and residuals around the Fe-K lines of the ``x1'' region. Red curves indicate the total spectral model, and in the lower-left panel, the blue and green curves represent the individual model components for Model 3. In the lower-right panel, the photon energies were generated assuming thermal broadening only. \label{fig:x1_comparison}}
\end{center}
\end{figure*}

\subsubsection{Fitting Spectra}\label{sec:sxs_spectra}

A thermally broadened spectral line with line flux profile $F_0(E)$ in the rest frame of the gas that is Doppler broadened to the new line profile $F(E)$ can be represented by:
\begin{equation}
F(E) = \displaystyle\int_{-\infty}^{\infty}F_0\left[E\left(1-\frac{v}{c}\right)\right]f(v)dv
\end{equation}
which is essentially a convolution with the velocity distribution function $f(v)$. We choose to model the thermal spectrum and the velocity distribution and fit these models to our synthetic spectra in XSPEC, using the \code{bapec}\footnote{\burl{http://heasarc.gsfc.nasa.gov/xanadu/xspec/manual/XSmodelApec.html}} thermal plasma model, which accounts for thermal broadening of spectral lines and allows for velocity broadening to be accounted for with a single parameter $\sigma$ that represents the standard deviation of a Gaussian velocity distribution. We constrain the velocity shift $\mu$ of the spectral lines using the difference of the fitted redshift parameter $z$ of the \code{bapec} model to the cosmological redshift of the cluster model, $\mu = -(z-z_{\rm cosmic})c$, where the sign reflects our convention that positive velocities are toward the observer, and $z_{\rm cosmic} = 0.05$.

We test three different models for the velocity distribution $f(v)$. The first model for $f(v)$, ``Model 1'', is a single-valued velocity distribution at the shift parameter $\mu$:
\begin{equation}\label{eqn:model1}
f(v) = \delta(v-\mu)~{\rm (Model~1)}
\end{equation}
This corresponds to a \code{bapec} model with the velocity broadening parameter set to zero and frozen. This model serves as a test of whether the data is consistent with the assumption of no velocity broadening, even with {\it Astro-H}'s improved spectral resolution. The second model, ``Model 2'', incorporates velocity broadening by thawing the $\sigma$ parameter and fitting it along with the others:
\begin{equation}\label{eqn:model2}
f(v) = G(v;\mu,\sigma^2)~{\rm (Model~2)}
\end{equation}
where $G$ is a normalized Gaussian distribution. The third model, ``Model 3'', is a weighted sum of two ``Model 1'' components:
\begin{equation}\label{eqn:model3}
f(v) = w_1\delta(v-\mu_1) + w_2\delta(v-\mu_2)~{\rm (Model~3)}
\end{equation}
The total Model 3 therefore represents a single-temperature and single-metallicity plasma which is a mixture of two single-valued velocity components (the redshift parameters) with different normalizations. In XSPEC, this is achieved by adding two \code{bapec} models, tying together the temperature and abundance parameters of each component, and allowing the redshift and normalization parameters of each component to vary separately. In this model, the velocity broadening parameter of each component is frozen at zero. Each of our three models also incorporates a \code{tbabs} absorption model, where the $N_H$ parameter is held fixed at the value of $2 \times 10^{20}$~cm$^{-2}$. All other parameters are free to vary, unless noted above.

We minimize the $C$-statistic \citep{cas79} to determine model parameter values. Though the $C$-statistic does not have utility as a goodness-of-fit test as the $\chi^2$-statistic does, is works properly with Poisson statistics, and does not require rebinning the spectrum to approximate Gaussian statistics. This enables us to use the full spectral resolution of {\it Astro-H} in our model fits. The spectra are fit within a broad energy band of 0.3-10.0~keV.

We also must ensure that our observations are long enough so that our estimates of the line shift and width have reasonable statistical accuracy. \citet{ota15} showed that $\simgt$~􏰅200 counts in the He-like iron line complex are required to achieve 20\% accuracy on the measurement of the turbulent velocity for values of $\sigma \sim$~200~km~s$^{-1}$. Table \ref{tab:counts_table} shows the number of counts in this line complex for each of our spectra, all of which easily exceed this requirement. In real clusters, the metallicity in the core region will be nearly solar, instead of the 0.3~$Z_\odot$ assumed in this work, so this requirement will be fulfilled for even shorter exposure times in the centers of clusters.

It should also be noted that our procedure does not take into account other systematic uncertainties that will be important. These include the systematic uncertainty on the line shift due to uncertainty in the SXS gain stability, and the uncertainty on the line width due to uncertainty in the line spread function. The gain uncertainty is of particular concern. Its required accuracy is 2~eV, with a goal of 1~eV \citep{mit14,tak14}. These values would correspond to an accuracy on the line shift of $\Delta{v}_{\rm sys} \sim$90(45)~km~s$^{-1}$ at 6.7~keV. In what follows, we will see that this uncertainty will dominate the statistical uncertainty on the line shift. Since we are focused in this work on identifying the systematic effects arising from projection and non-Gaussian intrinsic line shapes, we refer the reader to \citet{kit14} for a thorough discussion of the instrumental and calibration uncertainties.

\renewcommand{\arraystretch}{1.5}
\begin{table}[thdp]
\tabletypesize{\scriptsize}
\caption{Total Counts in He-like Fe Line Complex\label{tab:counts_table}}
\begin{center}
\begin{tabular}{cc}
\hline
\hline
Spectrum & Line Counts \\
\hline
x1 & 874 \\
x2 & 498 \\
y1 & 1355 \\
z1 & 4315 \\
z2 & 421 \\
\hline
\end{tabular}
\end{center}
\end{table}

Another consideration that must be taken into account is the broadness of the SXS PSF, with a width of roughly $\sim$1'. Most of our cold front pointings are offset from the cluster center, which has a much higher flux, and inspection of Figures \ref{fig:map_z} though \ref{fig:map_y_visc} shows that the velocity structure of the core is generally very different from that of the cold front regions. A number of photons will be scattered by the PSF from the central region into our offset pointings, biasing the line shift and width. To account for this effect, for each region we performed a second set of otherwise identical spectral simulations, except that we have turned off the effect of vignetting and artificially reduced the FWHM of the SXS PSF to 0.01", to compare to our uncorrected spectra. For spectral analysis of real observations, such an artifical reduction of the PSF scattering is obviously unavailable, so the core region and the cold front regions will have to be modeled concurrently. We show an example of such an analysis in Section \ref{sec:joint_modeling}.

The value of the reduced $C$-statistic for each model fit is tabulated in Table \ref{tab:cstat_table}, for spectral simulations with and without PSF scattering.\footnote{The XSPEC implementation of the $C$-statistic is normalized such that it provides a $\chi^2$-like goodness-of-fit value in the limit of Gaussian statistics, hence we provide both the value of the statistic and the number of degrees of freedom. See \url{http://heasarc.gsfc.nasa.gov/xanadu/xspec/manual/XSappendixStatistics.html} for a more detailed discussion.} Figure \ref{fig:x1_comparison} shows example spectra and fitted models for the ``x1'' region, within the $E \sim 6.2-6.8$~keV band, which surrounds the Fe-K lines (at our redshift of $z_{\rm cosmic}$ = 0.05). We will describe each model in turn, referring to the figure as needed.

\renewcommand{\arraystretch}{1.5}
\begin{table}[thdp]
\tabletypesize{\scriptsize}
\caption{$C$-statistic/d.o.f. for Different Model Fits\label{tab:cstat_table}}
\begin{center}
\begin{tabular}{cccc}
\hline
\hline
Spectrum & Model 1 & Model 2 & Model 3 \\
\hline
\multicolumn{4}{c}{With PSF Scattering} \\
\hline
x1 & 10513/9695 & 10291/9694 & 10312/9693 \\
x2 & 10051/9695 & 9982/9694 & 9978/9693 \\
y1 & 10984/9695 & 10359/9694 & 10430/9693 \\
z1 & 10528/9695 & 10162/9694 & 10174/9693 \\
z2 & 9963/9695 & 9941/9694 & 9941/9693 \\
x1$^*$ & 10301/9695 & N/A & N/A \\
\hline
\multicolumn{4}{c}{Without PSF Scattering} \\
\hline
x1 & 10829/9695 & 10570/9694 & 10587/9693 \\
x2 & 10331/9695 & 10258/9694 & 10255/9693 \\
y1 & 11137/9695 & 10567/9694 & 10623/9693 \\
z1 & 10542/9695 & 10189/9694 & 10205/9693 \\
z2 & 9985/9695 & 9958/9694 & 9959/9693 \\
\hline
\end{tabular}
\\[10pt]
$^*$In this case, velocity broadening and shifting has been turned off.
\end{center}
\end{table}

\renewcommand{\arraystretch}{1.5}
\begin{table*}
\tabletypesize{\scriptsize}
\caption{Model 2 Parameters and Simulation Values\label{tab:model2_params}}
\begin{center}
\begin{tabular}{ccccccccc}
\hline
\hline
Spectrum & $T_{\rm fit}$ & $T_{\rm sim}$ & $Z_{\rm fit}$ & $Z_{\rm sim}$ & $\mu_{\rm fit}$ & $\mu_{\rm sim}$ & $\sigma_{\rm fit}$ & $\sigma_{\rm sim}$ \\
& (keV) & (keV) & (Z$_\odot$) & (Z$_\odot$) & (km/s) & (km/s) & (km/s) & (km/s) \\
\hline
\multicolumn{9}{c}{With PSF Scattering} \\
\hline
x1 & 6.20 $\substack{+0.05 \\ -0.05}$ & 6.18 & 0.293 $\substack{+0.009 \\ -0.009}$ & 0.3 & -156 $\substack{+12 \\ -12}$ & -221 & 203 $\substack{+12 \\ -12}$ & 197 \\
x2 & 5.41 $\substack{+0.07 \\ -0.06}$ & 5.21 & 0.292 $\substack{+0.012 \\ -0.012}$ & 0.3 & 156 $\substack{+13 \\ -14}$ & 206 & 157 $\substack{+14 \\ -14}$ & 149 \\
y1 & 6.12 $\substack{+0.04 \\ -0.04}$ & 6.16 & 0.296 $\substack{+0.007 \\ -0.007}$ & 0.3 & 231 $\substack{+13 \\ -14}$ & 290 & 291 $\substack{+12 \\ -10}$ & 254 \\
z1 & 6.35 $\substack{+0.03 \\ -0.02}$ & 6.40 & 0.295 $\substack{+0.004 \\ -0.004}$ & 0.3 & 15 $\substack{+4 \\ -4}$ & 16 & 134 $\substack{+4 \\ -4}$ & 139 \\
z2 & 5.64 $\substack{+0.06 \\ -0.07}$ & 5.73 & 0.292 $\substack{+0.014 \\ -0.008}$ & 0.3 & -9 $\substack{+13 \\ -11}$ & -10 & 118 $\substack{+14 \\ -15}$ & 110 \\
\hline
\multicolumn{9}{c}{Without PSF Scattering} \\
\hline
x1 & 6.12 $\substack{+0.05 \\ -0.05}$ & 6.18 & 0.297 $\substack{+0.009 \\ -0.009}$ & 0.3 & -240 $\substack{+13 \\ -13}$ & -221 & 203 $\substack{+9 \\ -12}$ & 197 \\
x2 & 5.19 $\substack{+0.05 \\ -0.06}$ & 5.21 & 0.299 $\substack{+0.011 \\ -0.013}$ & 0.3 & 224 $\substack{+14 \\ -13}$ & 206 & 147 $\substack{+17 \\ -13}$ & 149 \\
y1 & 6.10 $\substack{+0.05 \\ -0.03}$ & 6.16 & 0.298 $\substack{+0.009 \\ -0.008}$ & 0.3 & 322 $\substack{+14 \\ -15}$ & 290 & 266 $\substack{+12 \\ -9}$ & 254 \\
z1 & 6.35 $\substack{+0.03 \\ -0.03}$ & 6.40 & 0.295 $\substack{+0.004 \\ -0.005}$ & 0.3 & 14 $\substack{+4 \\ -3}$ & 16 & 133 $\substack{+4 \\ -4}$ & 139 \\
z2 & 5.63 $\substack{+0.09 \\ -0.07}$ & 5.73 & 0.292 $\substack{+0.013 \\ -0.011}$ & 0.3 & -6 $\substack{+9 \\ -18}$ & -10 & 122 $\substack{+15 \\ -16}$ & 110 \\
\hline
\end{tabular}
\end{center}
\end{table*}

\renewcommand{\arraystretch}{1.5}
\begin{table*}
\tabletypesize{\scriptsize}
\caption{Model 3 Parameters and Simulation Values\label{tab:model3_params}}
\begin{center}
\begin{tabular}{ccccccccc}
\hline
\hline
Spectrum & $T_{\rm fit}$ & $T_{\rm sim}$ & $Z_{\rm fit}$ & $Z_{\rm sim}$ & $\mu_{\rm 1,fit}$ & $\mu_{\rm 2,fit}$ & $w_{\rm 1,fit}$ & $w_{\rm 2,fit}$ \\
& (keV) & (keV) & (Z$_\odot$) & (Z$_\odot$) & (km/s) & (km/s) & & \\
\hline
x1 & 6.13 $\substack{+0.05 \\ -0.05}$ & 6.18 & 0.292 $\substack{+0.009 \\ -0.009}$ & 0.3 & -372 $\substack{+19 \\ -28}$ & -7 $\substack{+31 \\ -41}$ & 0.60 $\substack{+0.04 \\ -0.05}$ & 0.40 $\substack{+0.05 \\ -0.04}$ \\
x2 & 5.19 $\substack{+0.05 \\ -0.06}$ & 5.21 & 0.298 $\substack{+0.011 \\ -0.013}$ & 0.3 & 318 $\substack{+22 \\ -20}$ & 2 $\substack{+48 \\ -43}$ & 0.68 $\substack{+0.06 \\ -0.09}$ & 0.32 $\substack{+0.09 \\ -0.06}$ \\
y1 & 6.12 $\substack{+0.05 \\ -0.03}$ & 6.16 & 0.290 $\substack{+0.008 \\ -0.007}$ & 0.3 & 492 $\substack{+22 \\ -15}$ & 37 $\substack{+37 \\ -31}$ & 0.61 $\substack{+0.03 \\ -0.03}$ & 0.39 $\substack{+0.03 \\ -0.03}$ \\
z1 & 6.35 $\substack{+0.03 \\ -0.03}$ & 6.40 & 0.293 $\substack{+0.004 \\ -0.005}$ & 0.3 & 153 $\substack{+18 \\ -12}$ & -101 $\substack{+13 \\ -12}$ & 0.47 $\substack{+0.03 \\ -0.05}$ & 0.53 $\substack{+0.05 \\ -0.03}$ \\
z2 & 5.63 $\substack{+0.09 \\ -0.07}$ & 5.73 & 0.291 $\substack{+0.013 \\ -0.010}$ & 0.3 & -123 $\substack{+253 \\ -51}$ & 87 $\substack{+40 \\ -180}$ & 0.5 $\substack{+0.10 \\ -0.16}$ & 0.50 $\substack{+0.16 \\ -0.10}$ \\
\hline
\end{tabular}
\end{center}
\end{table*}

From Table \ref{tab:cstat_table}, we see that the fit to Model 1 is generally poor (except in the ``z2'' case). The upper-left panel of Figure \ref{fig:x1_comparison} shows the spectrum for region ``x1'' with its fitted Model 1, showing significant fit residuals around the Fe emission lines. For most of the fits to Model 1 for the various regions, we are able to recover the correct temperature, since it is strongly constrained by the continuum emission, but the metallicity is always underestimated by approximately 10-15\%. Taking the ``x1'' region as an example, we find that we can recover the expected temperature from the simulation of $kT_{\rm sim} = 6.18$~keV, with $kT_{\rm fit}$ = 6.23$\substack{+0.05 \\ -0.06}$~keV, but the metallicity is underestimated, with $Z_{\rm sim}$ = $0.3~Z_\odot$ and $Z_{\rm fit}$ = 0.266$\substack{+0.008 \\ -0.009}$~Z$_\odot$. We find similar results if we ignore PSF scattering. The fitted value of the velocity shift is strongly biased if PSF scattering is included, with $\mu_{\rm fit}$ = -145$\substack{+14 \\ -12}$~km~s$^{-1}$, far away from the expected value of $\mu_{\rm sim}$ = -221~km~s$^{-1}$. Without PSF scattering, we find $\mu_{\rm fit}$ = -238$\substack{+13 \\ -14}$~km~s$^{-1}$, a much more accurate value. This indicates that modeling of the bright core component will be necessary to accurately measure line shifts of cold fronts.

As a sanity check, if we turn off the effects of Doppler shifting and broadening from gas velocity when creating the synthetic spectrum, Model 1 produces a much better fit to the data than in the case where these effects were included (see the last row of Table \ref{tab:cstat_table}, and the lower-right panel of Figure \ref{fig:x1_comparison}) despite the fact that we are fitting a single-temperature model to a spectrum from gas with a range of temperatures. In this case, the metallicity is correctly recovered with $Z_{\rm fit} = 0.292 \substack{+0.009 \\ -0.008}$~Z$_{\odot}$. Since Model 1 provides a good fit to this spectrum, but not the broadened one, this confirms that {\it Astro-H} is sensitive enough to detect the effects of Doppler shifting and broadening on the emission lines from the subsonic sloshing motions.

\begin{figure*}
\begin{center}
\includegraphics[width=0.33\textwidth]{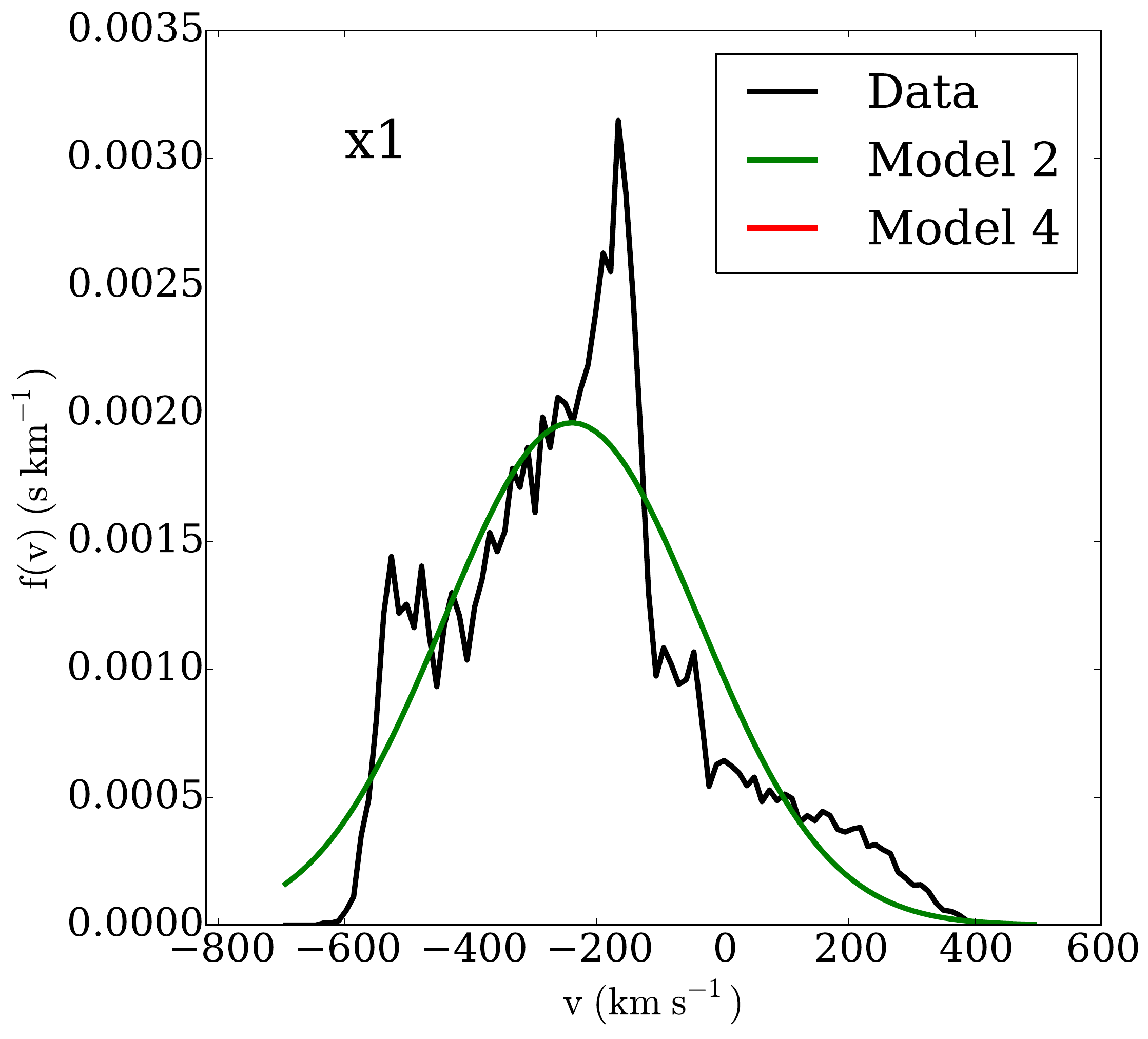}
\includegraphics[width=0.33\textwidth]{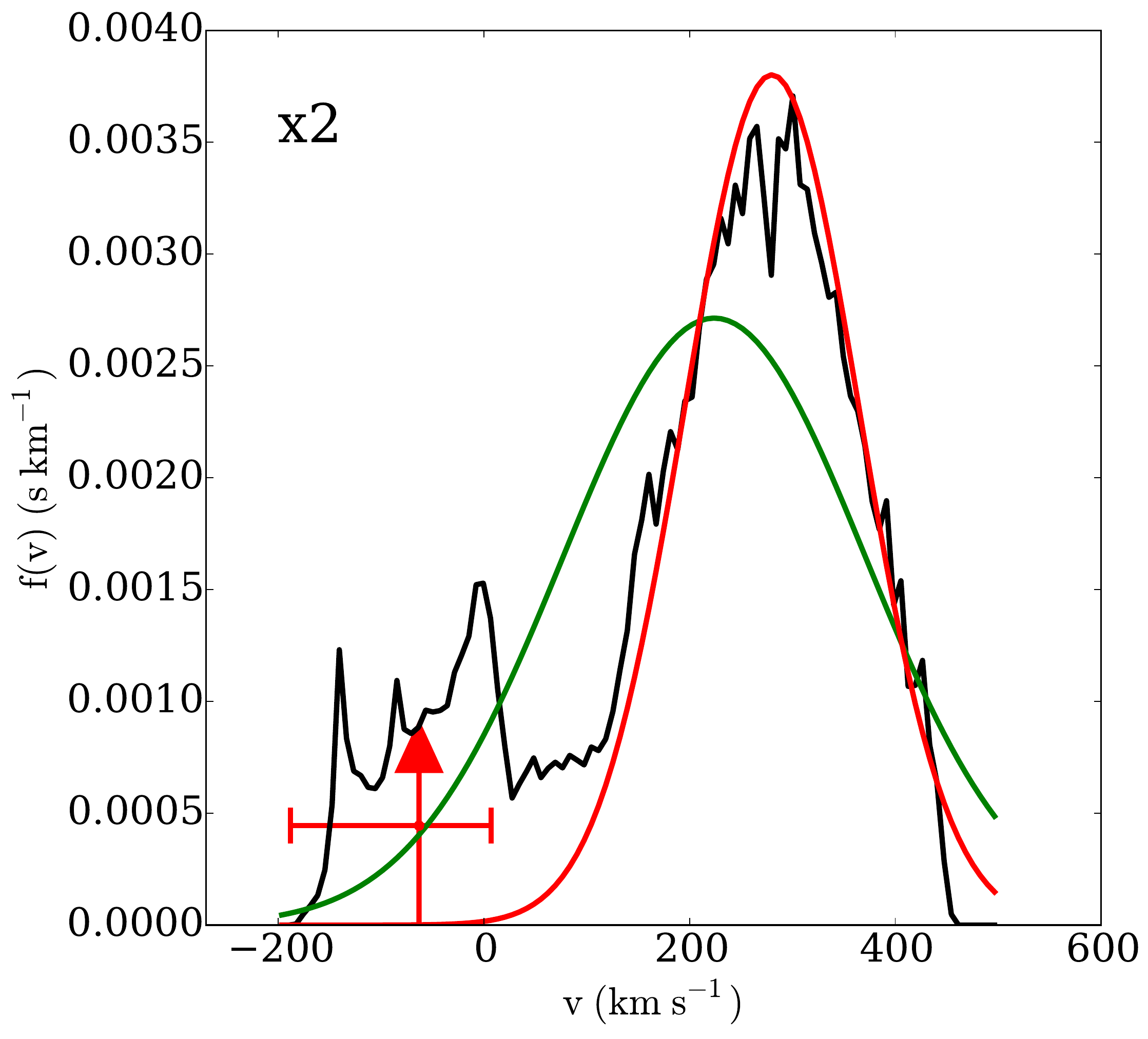}
\includegraphics[width=0.32\textwidth]{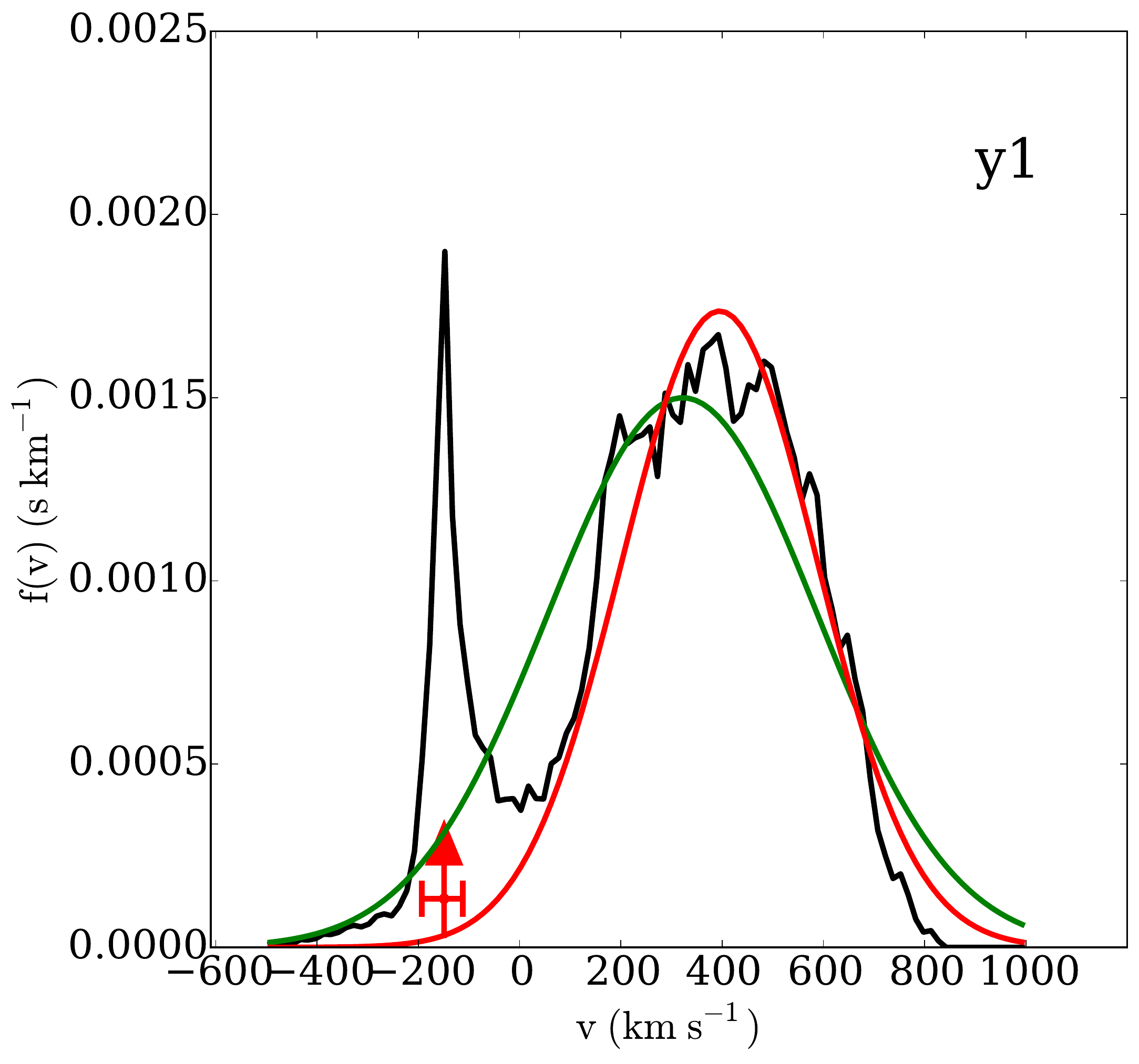}
\includegraphics[width=0.325\textwidth]{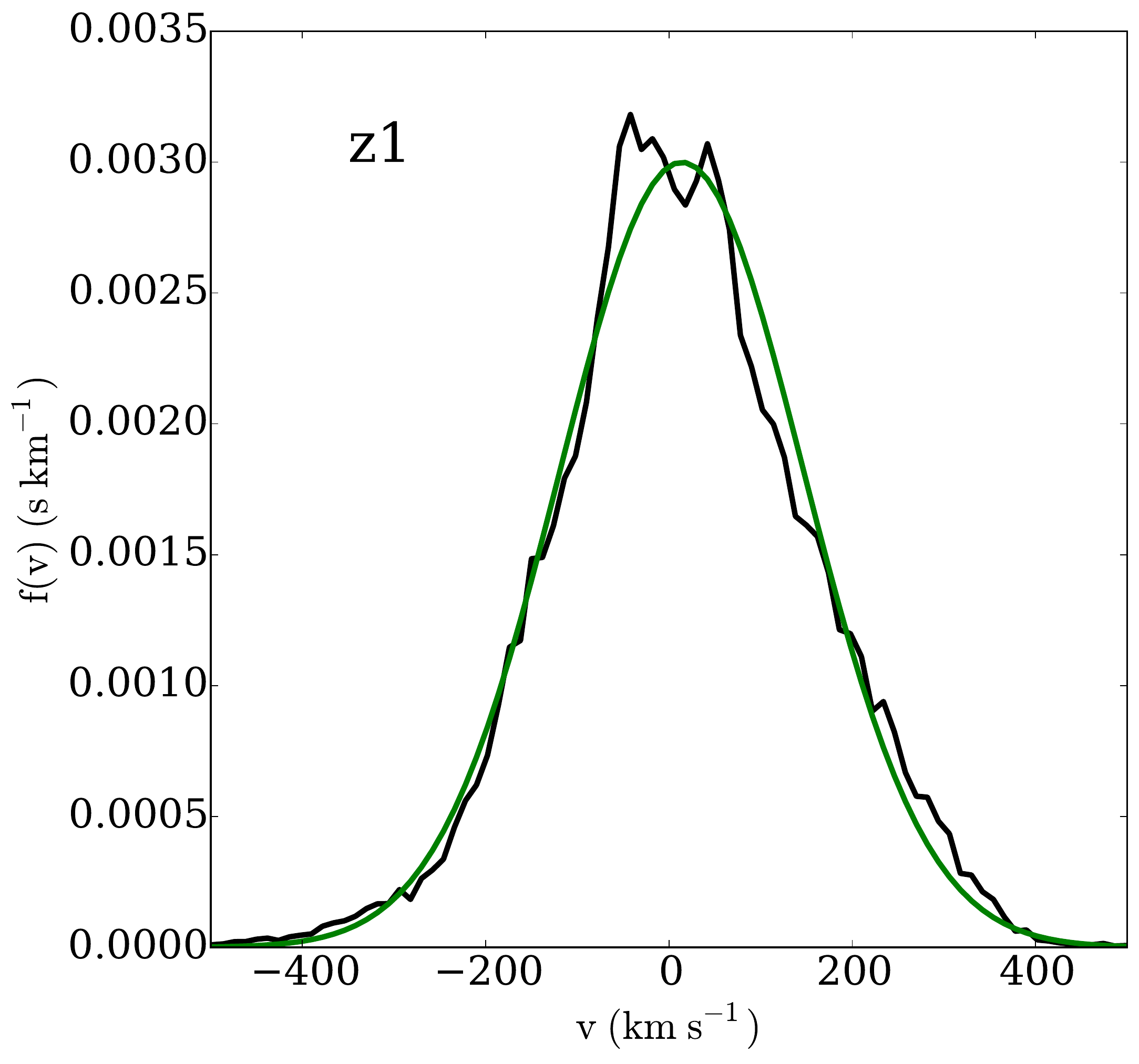}
\includegraphics[width=0.33\textwidth]{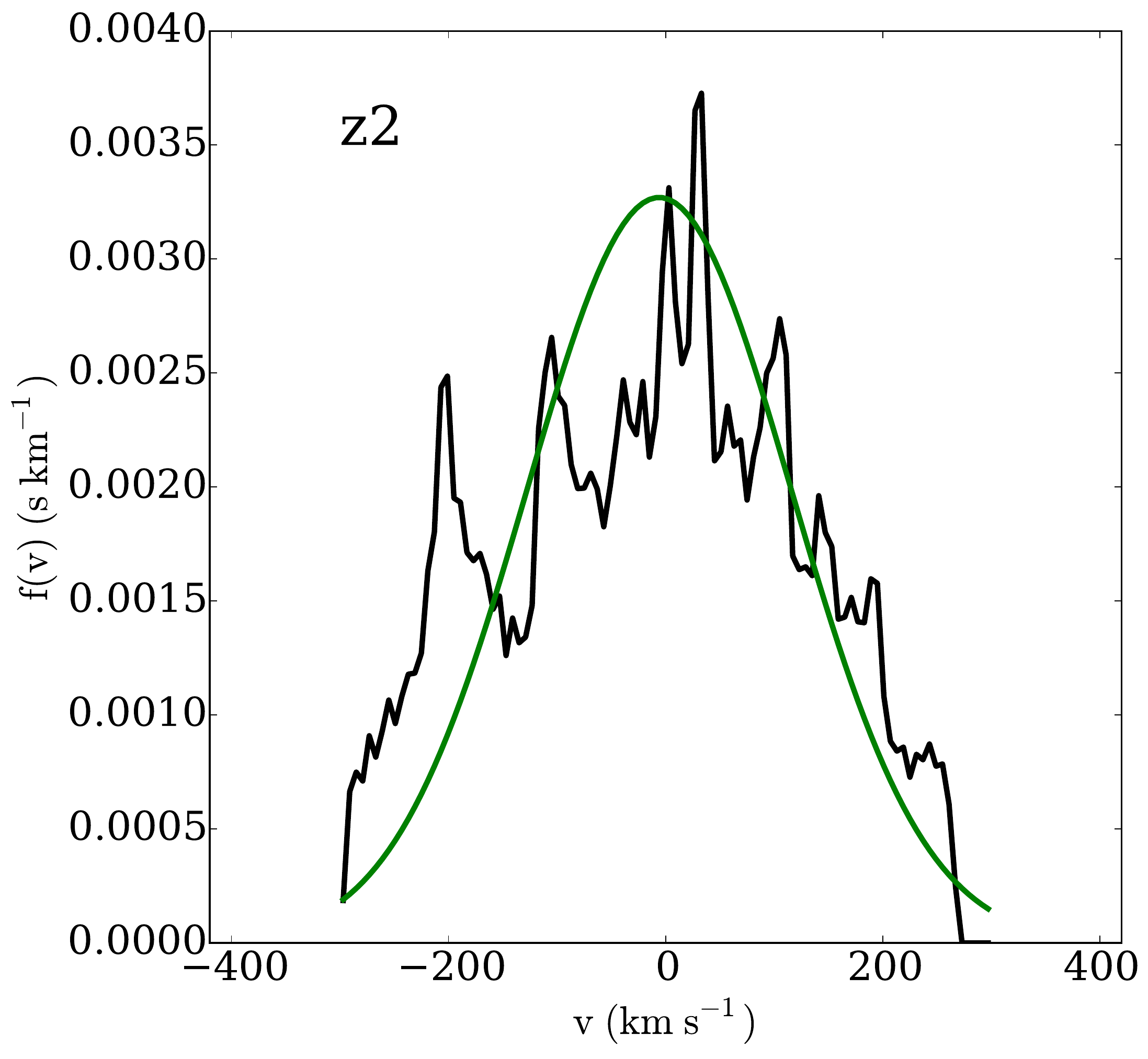}
\caption{Velocity distributions for the five regions from the inviscid simulation with the predicted velocity distributions from Model 2 and, in two cases, Model 4. Vertical arrows indicate the position of the line shift associated with the $\delta$-function component in Model 4, and the error bars on the arrows give the 1-$\sigma$ error on the line shift.\label{fig:fv}}
\end{center}
\end{figure*}

\begin{figure*}
\begin{center}
\includegraphics[width=0.33\textwidth]{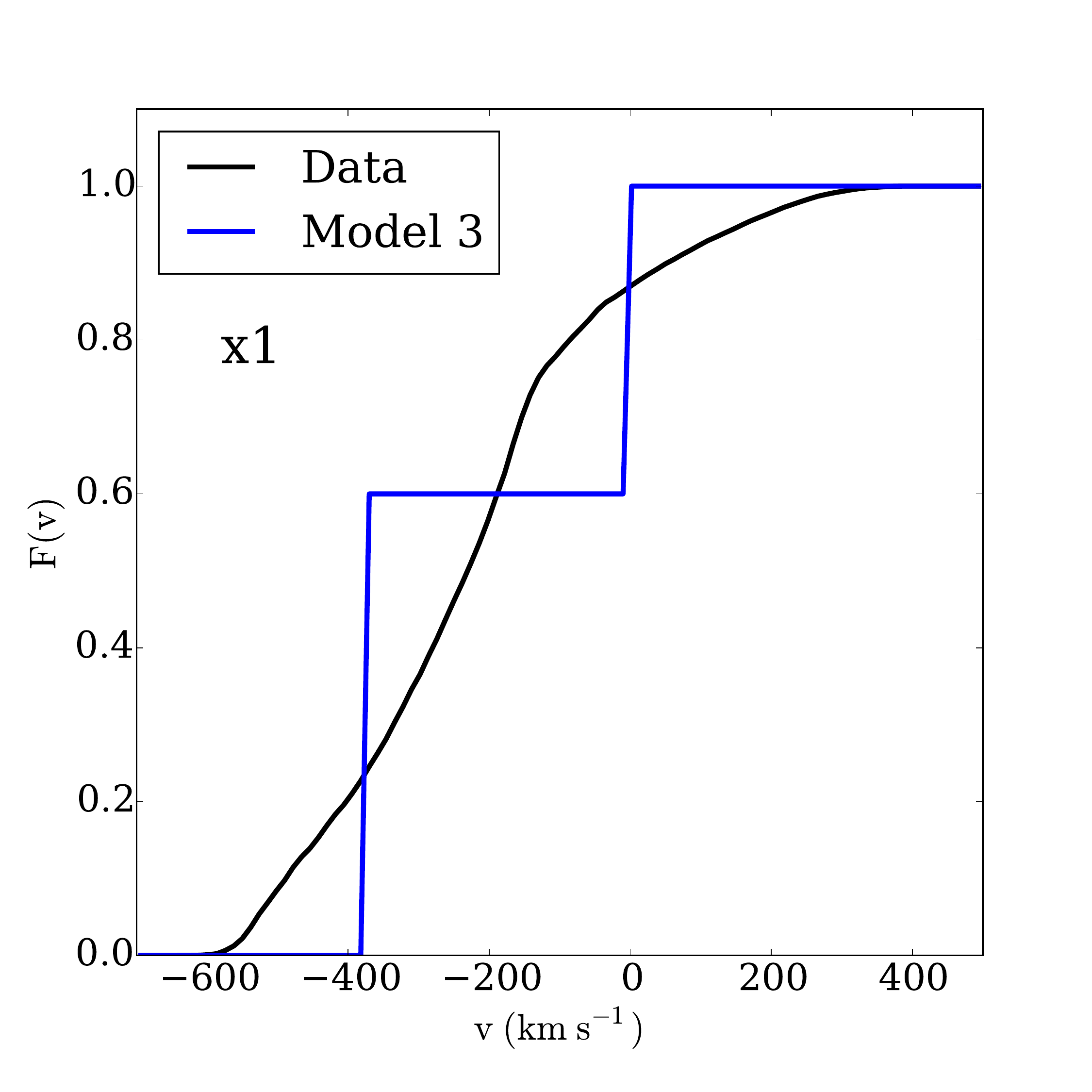}
\includegraphics[width=0.33\textwidth]{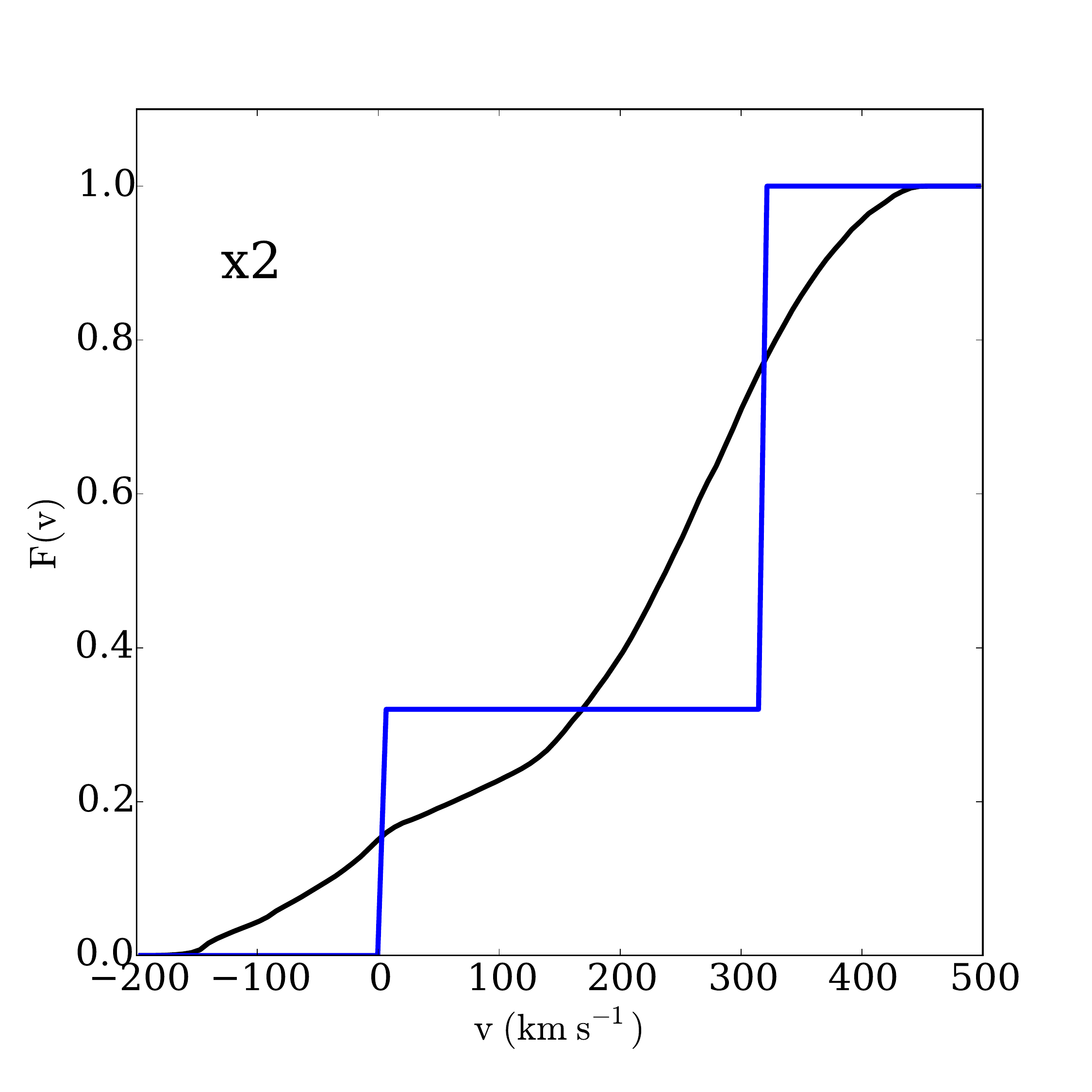}
\includegraphics[width=0.33\textwidth]{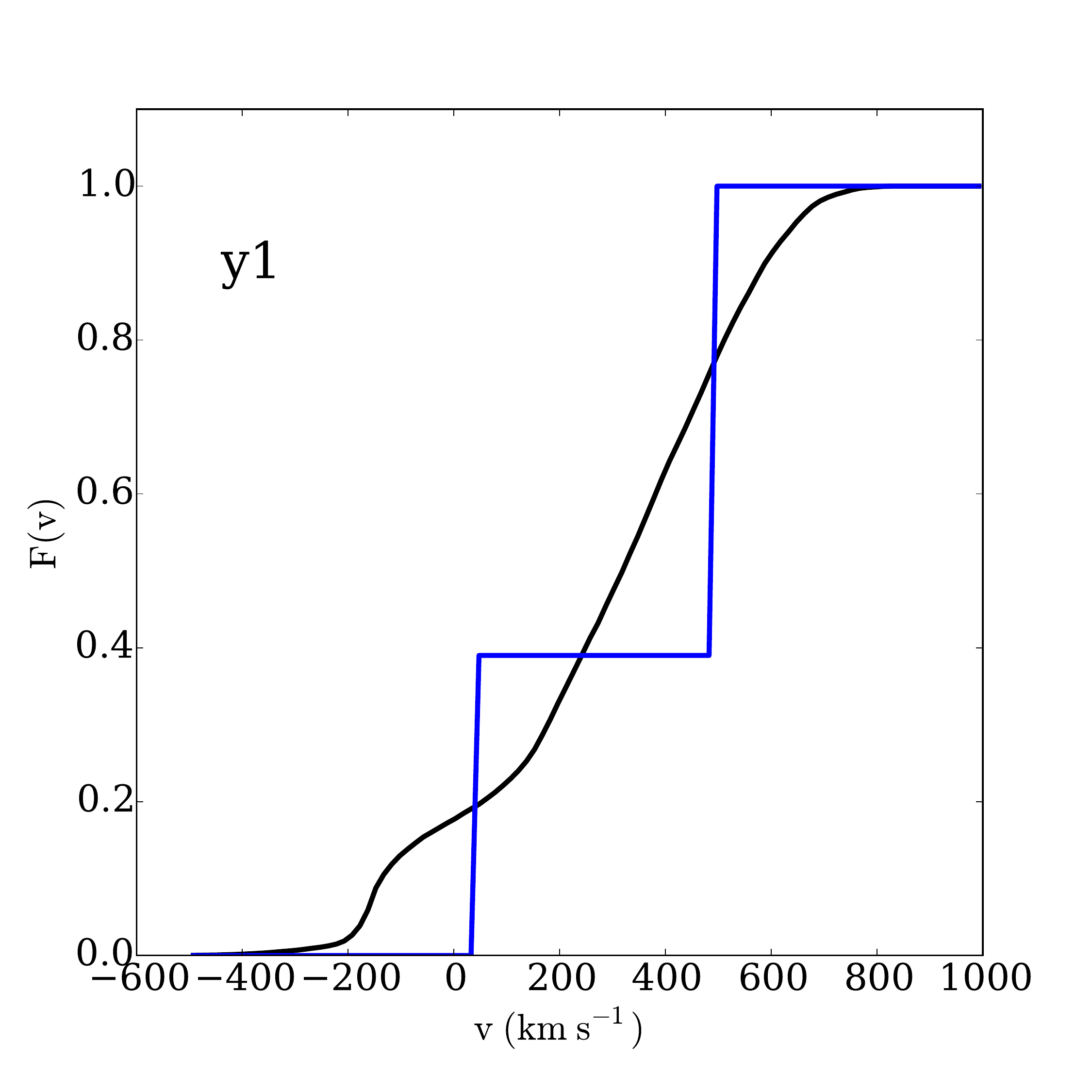}
\includegraphics[width=0.33\textwidth]{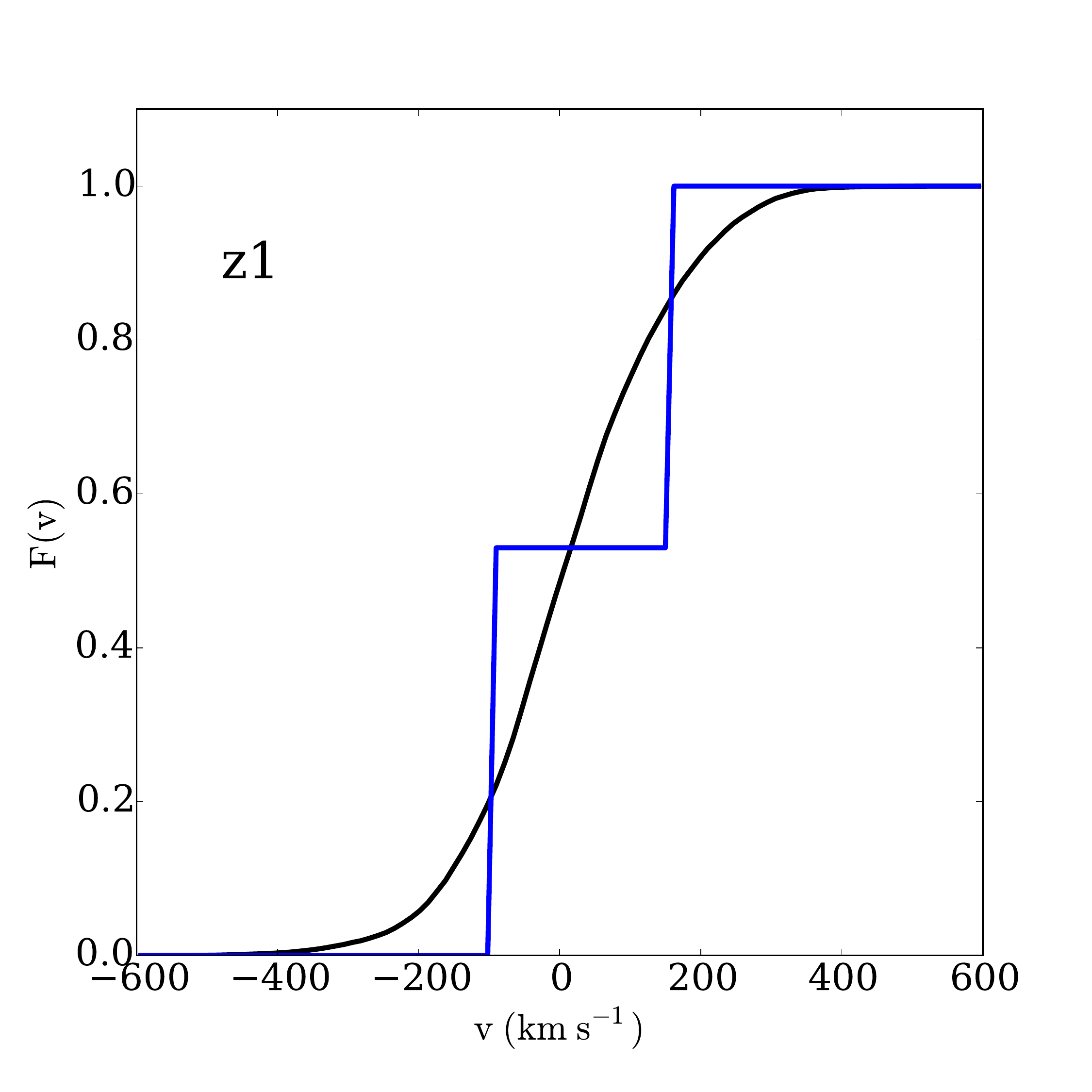}
\includegraphics[width=0.33\textwidth]{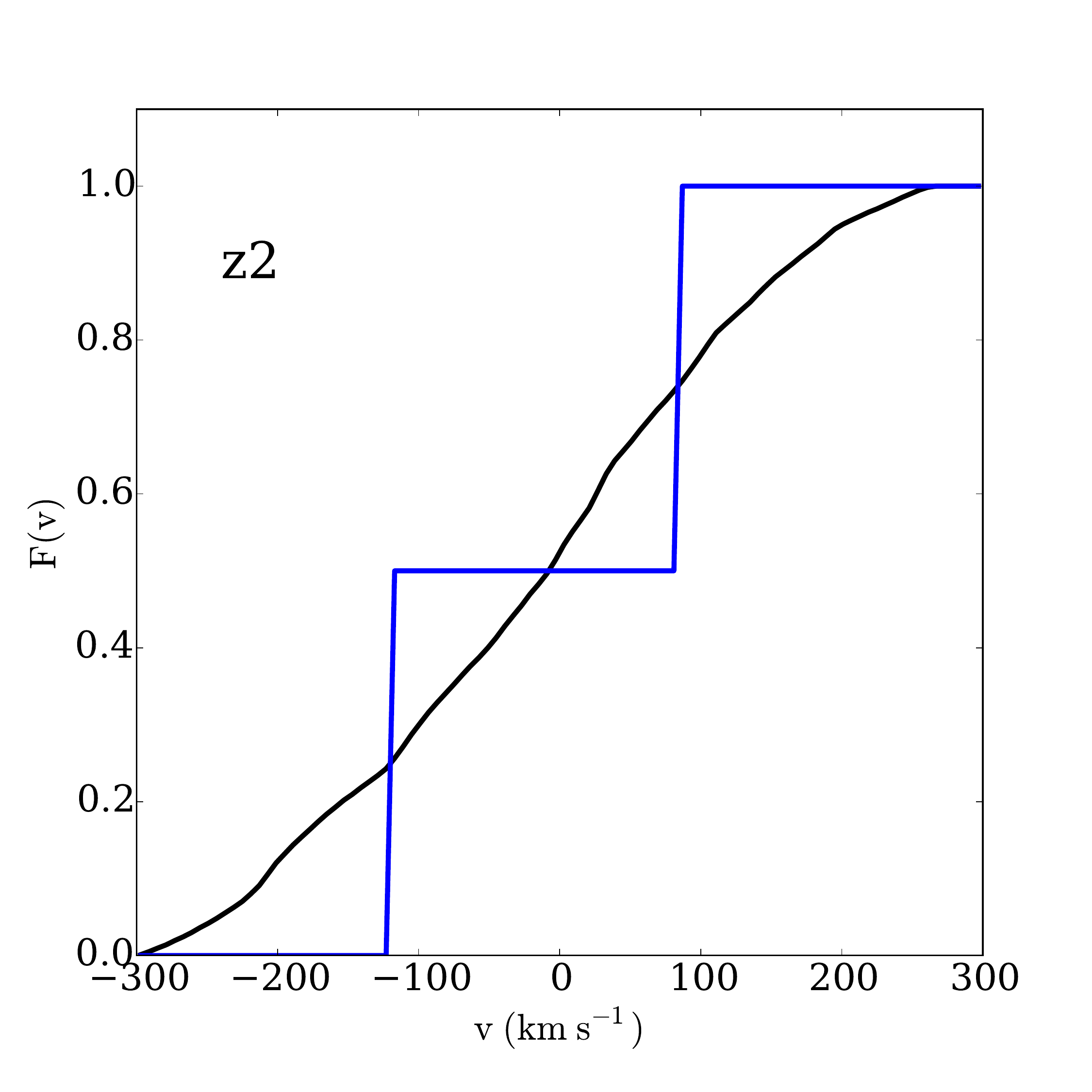}
\caption{Cumulative velocity distributions for the five regions from the inviscid simulation with the predicted cumulative velocity distributions from Model 3.\label{fig:cum_fv}}
\end{center}
\end{figure*}

Fitting the synthetic spectra with Model 2 results in a significant reduction in the fit statistic (except in the ``z2'' case, and in the case of ``x2'' only a modest improvement is achieved), as seen in the top-right panel of Figure \ref{fig:x1_comparison}. Table \ref{tab:model2_params} shows the fitted parameters from Model 2, for spectra with PSF scattering (top rows) and those without (bottom rows). For nearly all the spectra, the fitted temperature $T_{\rm fit}$ agrees with the ``real'' temperature $T_{\rm sim}$, with the exception of the ``x2'' region, where photons scattered from the core by the PSF bias the temperature upward. With PSF scattering turned off, we recover the correct temperature for ``x2''. The metallicity parameters are much closer to the correct value of $Z_{\rm sim} = 0.3$~Z$_\odot$, though they are all still biased lower.

With PSF scattering included in the simulated observation, the velocity shifts of the ``x1'', ``x2'', and ``y1'' regions, all within the sloshing plane, are biased in the direction of zero velocity from their expected values by $\sim$60-70~km~s$^{-1}$, due to the mostly unshifted photons from the bright core scattering into the field of view. The velocity shifts for the spectra without PSF scattering are much closer to the true values, but they still exhibit a small bias of $\sim$20-30~km~s$^{-1}$ from the true line shift, in the direction toward zero velocity. In Section \ref{sec:predict_fv}, we will see that this is due to the fact that the true velocity distributions are not strictly Gaussian. The effect of PSF scattering on the velocity broadening parameter $\sigma$ is not quite as severe. The fitted line shift parameters for the ``z1'' and ``z2'' regions exhibit no bias due to PSF scattering, but this is expected, since the ``z1'' region is centered on the core itself, and the ``z2'' region has nearly the same line shift, due to the symmetry of the sloshing motions. In all cases, however, the velocity bias from PSF scattering is comparable to or even less than that expected from instrumental uncertainties, if $\Delta{v}_{\rm sys} \sim 45(90)$~km~s$^{-1}$ as noted above.

Fitting the spectra with Model 3 also provides a good fit in most cases (see the bottom-left panel of Figure \ref{fig:x1_comparison} for the fit for region ``x1''). The reduction in the $C$-statistic is approximately the same as in Model 2. Though it might not be expected that such a crude velocity distribution function (Equation \ref{eqn:model3}) would provide a good fit to complex velocity distributions, this is made possible because of the blending together of the individual line components of complexes (such as Fe near 7~keV in the rest frame) due to the finite spectral resolution of SXS ($\sim$5~eV) and line broadening. Table \ref{tab:model3_params} shows the fitted parameters from Model 3, with no PSF scattering. In general, the temperature parameters in each spectrum are correctly recovered, and the two velocity components $\mu_1$ and $\mu_2$ have comparable normalization, with a typical difference of $\sim$250-400~km~s$^{-1}$. Though the model provides a good fit, its utility as a physical description of the velocity distribution is limited.

\subsubsection{Predicted Velocity Distributions}\label{sec:predict_fv}

How well do our models reproduce the actual velocity distributions within the regions from which the spectra are extracted? Figure \ref{fig:fv} shows the predicted velocity distribution function $f(v)$ from Model 2 (green curves) for our five regions, compared to the actual $f(v)$ for the same regions (black curves). For the ``x1'', ``x2'', and ``y1'' regions, the velocity distributions are complex, and exhibit deviations from Gaussianity, though the Model 2 fits do approximately capture the width of the distribution. This explains why the Model 2 line shifts from Table \ref{tab:model2_params} (without PSF scattering) do not always agree precisely with the line shift measured from the simulation (though this disagreement is small compared to the expected systematic error due to uncertainty in the line spread function and gain); the mean of the best-fit Gaussian component and the mean of the true distribution are not the same if the underlying distribution is non-Gaussian. The velocity distributions for the ``z1'' and ``z2'' regions are well-fit by Gaussian distributions, and as a result both the line shift and width are correctly recovered by Model 2.

Figure \ref{fig:fv} indicates that the velocity distribution of cold fronts may be better modeled by multiple components. \citet{sha12} used a mixing-model approach to fit velocity distribution models to Doppler-broadened spectral lines. They found that they were able to fit velocity distributions from a variety of simulated velocity fields (including those from core gas sloshing) using a sum of Gaussian models. They also found that for the {\it Astro-H} spatial and spectral resolution, typically only two Gaussian components can be constrained. Compared to \citet{sha12}, we are limited by statistics. For each of their spectra, they assumed 10$^4$ counts in the single He-like iron line at 6.7~keV, corresponding to roughly a megasecond exposure of the entire SXS field of view pointed at the cluster core for a number of nearby clusters (see their Table 2). As noted above, our observations of 200~ks exposure typically have $\sim$10$^3$ counts in the entire He-like iron line complex (see Table \ref{tab:counts_table}). Also, we extract spectra from smaller regions underneath cold front surfaces, which are fainter than the cluster core.

Though we do not employ a mixing-model approach, we can fit a sum of two Gaussian models by extending our Model 3 (Equation \ref{eqn:model3}); we thaw the velocity broadening parameters of the separate \code{bapec} components, allowing them to vary in the fit. Along with the temperature and metallicity parameters, we have six additional parameters: a normalization, line centroid, and line width for each Gaussian velocity component. In all cases, we are unable to constrain all of these parameters uniquely with our simulated spectra. However, we do find that we can constrain {\it one} velocity broadening parameter. This defines our ``Model 4'', a simple generalization of Model 3:
\begin{equation}\label{eqn:model4}
f(v) = w_1\delta(v-\mu_1) + w_2G(v;\mu_2,\sigma_2^2)~{\rm (Model~4)}
\end{equation}
Predicted velocity distributions from Model 4 fits to the ``x2'' and ``y1'' spectra are shown in Figure \ref{fig:fv}, as red curves with arrows showing the positions of the line shift parameter along with its 1-$\sigma$ error bar. We are only able to constrain parameters for these particular regions, as these cases have the clearest evidence for two well-separated velocity components. In these two cases, Model 4 resolves both components well.

To compare Model 3 to the actual velocity distribution, it is more instructive to examine the {\it cumulative} distribution function (CDF) of the velocity, $F(v)$. For Model 3, the CDF of the velocity distribution is
\begin{equation}\label{eqn:model3_cum}
F(v) = w_1H(v-\mu_1) + w_2H(v-\mu_2)~{\rm (Model~3)}
\end{equation}
where $H$ is the Heaviside step function. Figure \ref{fig:cum_fv} shows $F(v)$ for the five different regions, with the Model 3 prediction overlaid. From this figure, we see that Model 3 represents a crude, piecewise-constant approximation to the cumulative distribution of the velocity, which may provide a rough sense of the location and relative importance of the two most dominant velocity components, but in general a Gaussian velocity model or (if possible) a sum of two models is to be preferred, as they will provide a more accurate physical representation of the velocity field for a comparable number of parameters.

\begin{figure*}
\begin{center}
\begin{minipage}{0.49\linewidth}
\includegraphics[width=\textwidth]{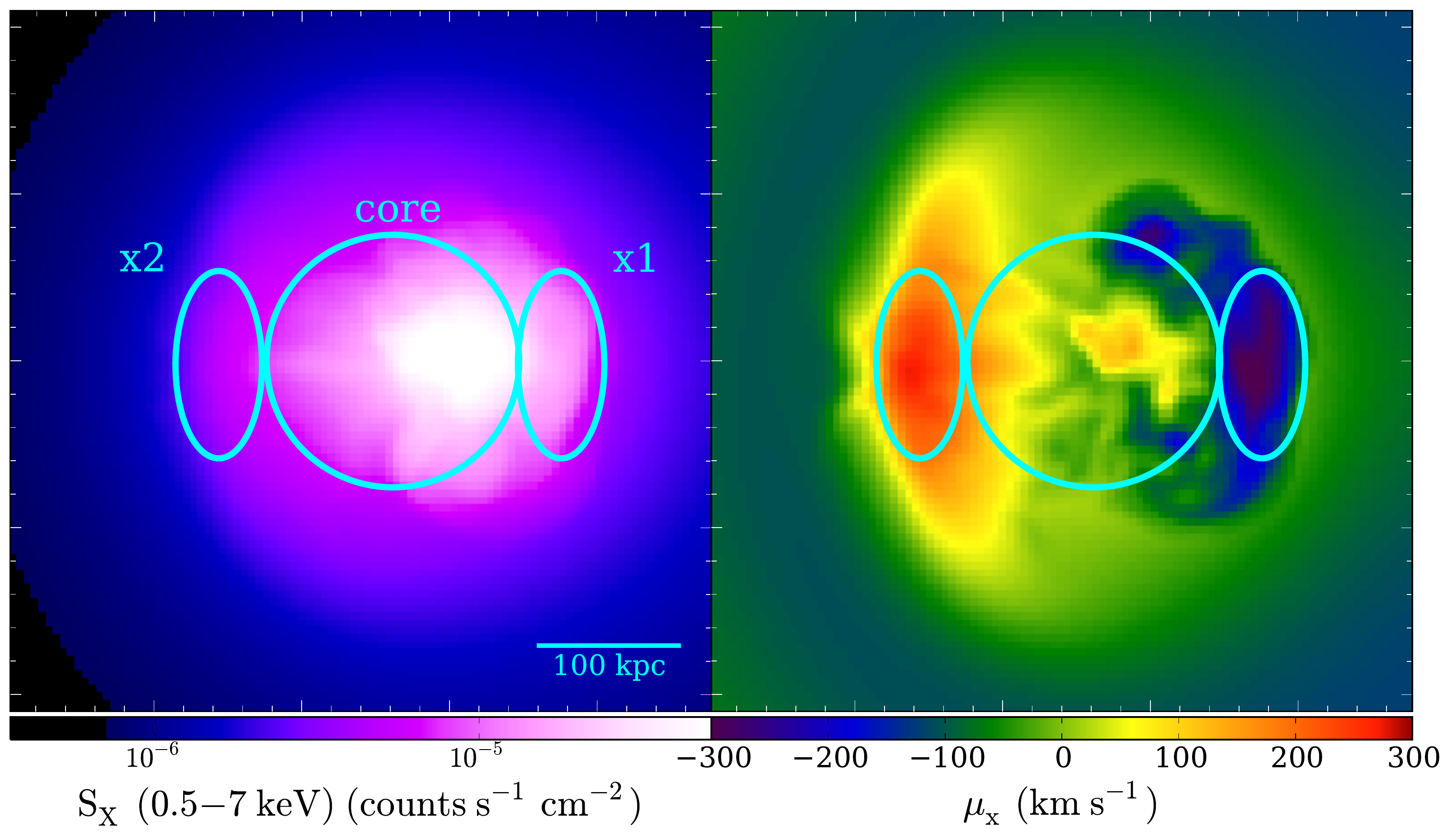}
\end{minipage}
\begin{minipage}{0.49\linewidth}
\includegraphics[width=\textwidth]{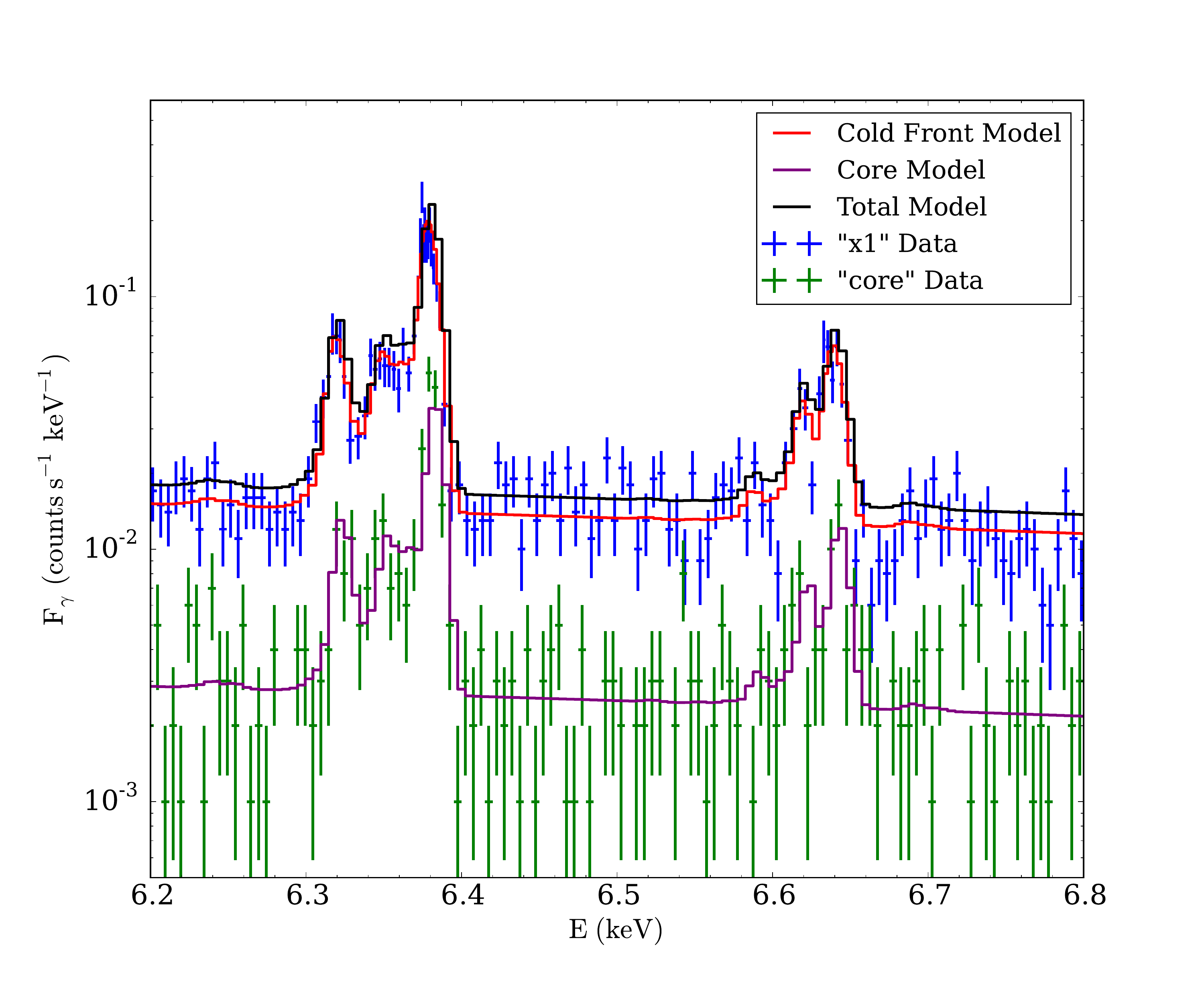}
\end{minipage}
\caption{Joint fits to cold front and core region spectra. Left panel: Projections of X-ray surface brightness and line-of-sight velocity along the $x$-axis, with the regions for spectral extraction overlaid. The ``core'' region has been added. Right panel: Example spectra from the ``x1'' and ``core'' regions with fitted models to both the ``core'' and ``x1'' components.\label{fig:joint_fit}}
\end{center}
\end{figure*}

\renewcommand{\arraystretch}{1.5}
\begin{table*}[t]
\tabletypesize{\scriptsize}
\caption{Accounting for PSF Scattering: Joint Fit Parameters and Simulation Values\label{tab:joint_fit_params}}
\begin{center}
\begin{tabular}{ccccccccc}
\hline
\hline
Spectrum & $T_{\rm fit}$ & $T_{\rm sim}$ & $Z_{\rm fit}$ & $Z_{\rm sim}$ & $\mu_{\rm fit}$ & $\mu_{\rm sim}$ & $\sigma_{\rm fit}$ & $\sigma_{\rm sim}$ \\
& (keV) & (keV) & (Z$_\odot$) & (Z$_\odot$) & (km/s) & (km/s) & (km/s) & (km/s) \\
\hline
x1 & 6.22 $\substack{+0.07 \\ -0.07}$ & 6.18 & 0.293 $\substack{+0.013 \\ -0.013}$ & 0.3 & -192 $\substack{+17 \\ -14}$ & -221 & 196 $\substack{+15 \\ -16}$ & 197 \\
x2 & 5.30 $\substack{+0.06 \\ -0.09}$ & 5.21 & 0.294 $\substack{+0.017 \\ -0.017}$ & 0.3 & 176 $\substack{+20 \\ -20}$ & 206 & 161 $\substack{+20 \\ -22}$ & 149 \\
\hline
\end{tabular}
\end{center}
\end{table*}

\subsubsection{Accounting for PSF Scattering Via Joint Modeling}\label{sec:joint_modeling}

As mentioned above, in SXS observations of cold fronts a non-negligible number of photons may be scattered into the field of view from the nearby core region. Therefore, to achieve accurate measurements of model parameters, it will be necessary to model the spectrum of a cold front region simultaneously with the spectrum of the cluster core. Examples of such analyses can be found in the {\it Astro-H} cluster white paper \citep{kit14}, where {\it Chandra} images were used in conjunction with simulated spectra to estimate the effects of PSF scattering. In this section, we perform such a joint analysis using the ``x1'' and ``x2'' regions as examples.

The left panel of Figure \ref{fig:joint_fit} shows maps of the X-ray surface brightness and the line-of-sight velocity, with the ``x1'' and ``x2'' regions chosen for spectral extraction (which are the same as before), with an additional ``core'' region that contains the bright cluster core. We then perform separate simulations including only the photons originating from this core region that are scattered into regions ``x1'' and ``x2''. We create spectra from each region which contain only photons originating in the core. We find that the photon flux from the core scattered into the ``x1'' and ``x2'' regions is about $\sim$20\% of the total flux of each region. We performed a joint fit in XSPEC of the two spectra using a sum of two \code{bapec} models, where one component models the photons from the core region spectrum and the other models those from the cold front spectrum. The right panel of Figure \ref{fig:joint_fit} shows example spectra for the ``x1'' region and the fitted models. Table \ref{tab:joint_fit_params} shows the fitted parameters for the ``x1'' region resulting from the joint fit, where we only show the parameters for the cold front model component. The joint fit results in an improvement of the estimation of the line shift of the cold front component in both regions.

This effect may be slightly larger than our simulations indicate. In reality, clusters have central metallicities of $\sim{Z_\odot}$ or higher, and the metallicity decreases with radius. In our simulations we have assumed a spatially uniform $Z = 0.3~Z_\odot$. Depending on the difference in metallicity between the core and the cold front for a given cluster, the PSF-scattered flux from the core into the cold front region may be somewhat larger.

\begin{figure*}
\centering
\includegraphics[width=0.424\textwidth]{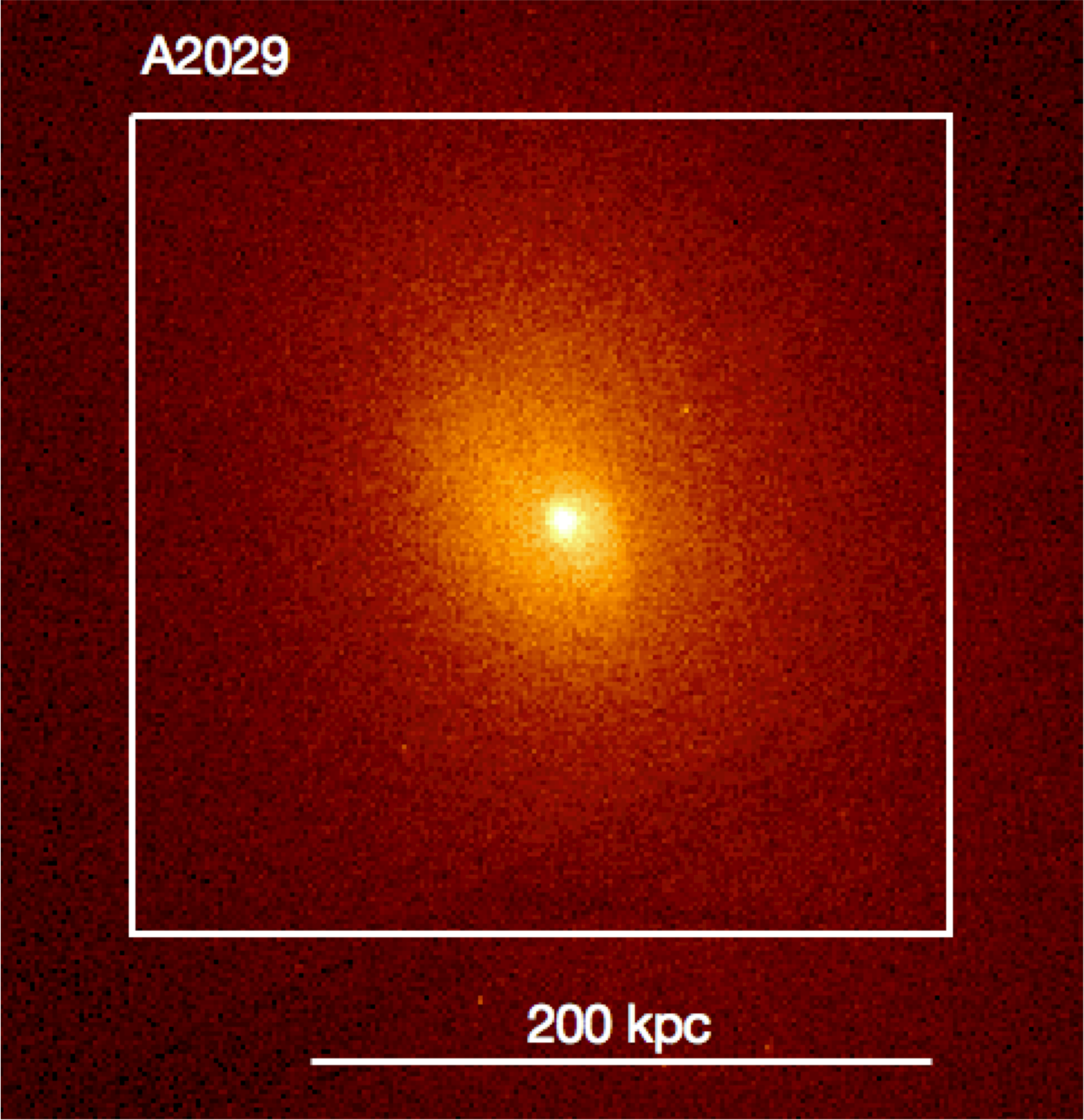}
\includegraphics[width=0.43\textwidth]{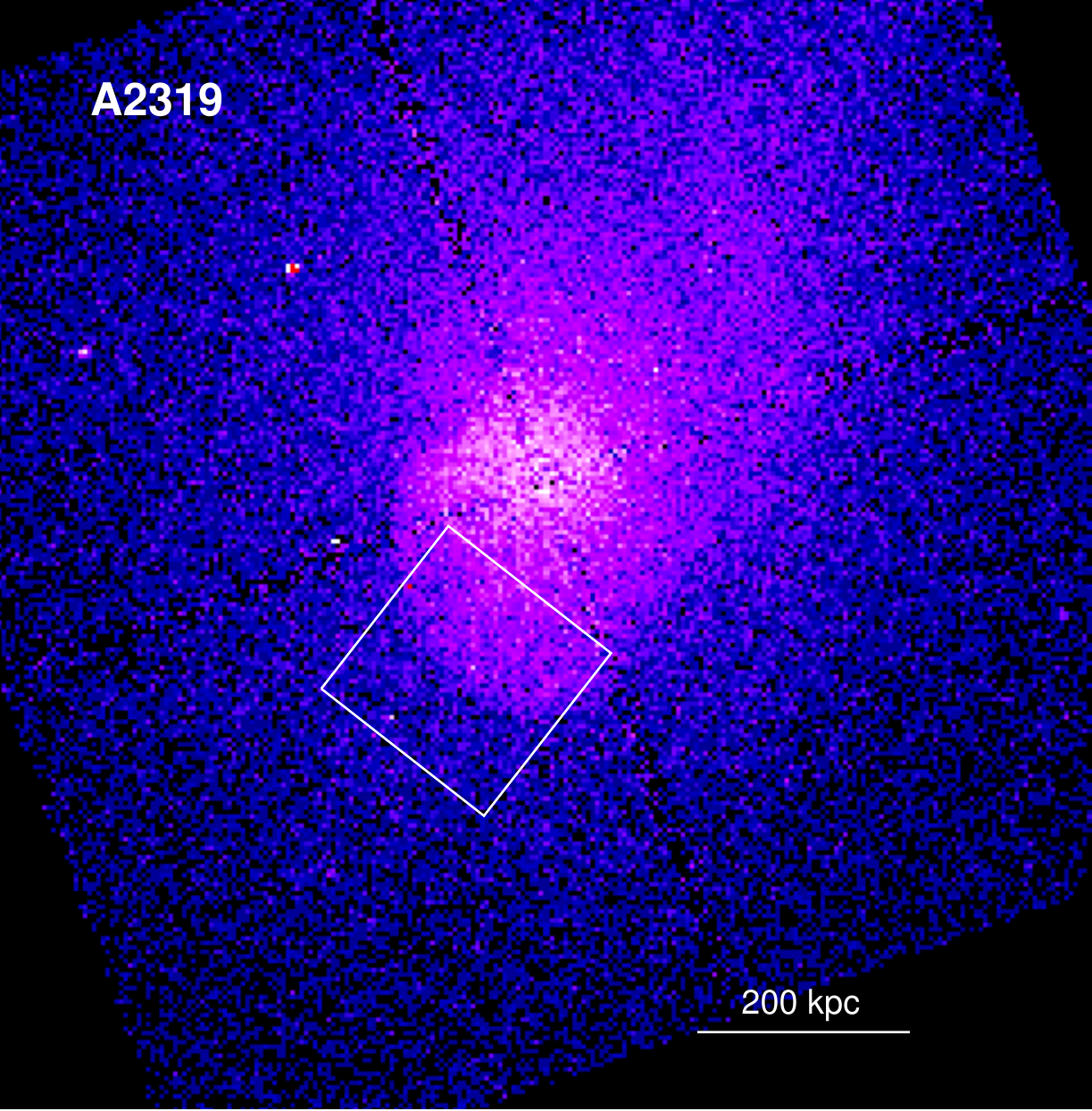}
\includegraphics[width=0.3845\textwidth]{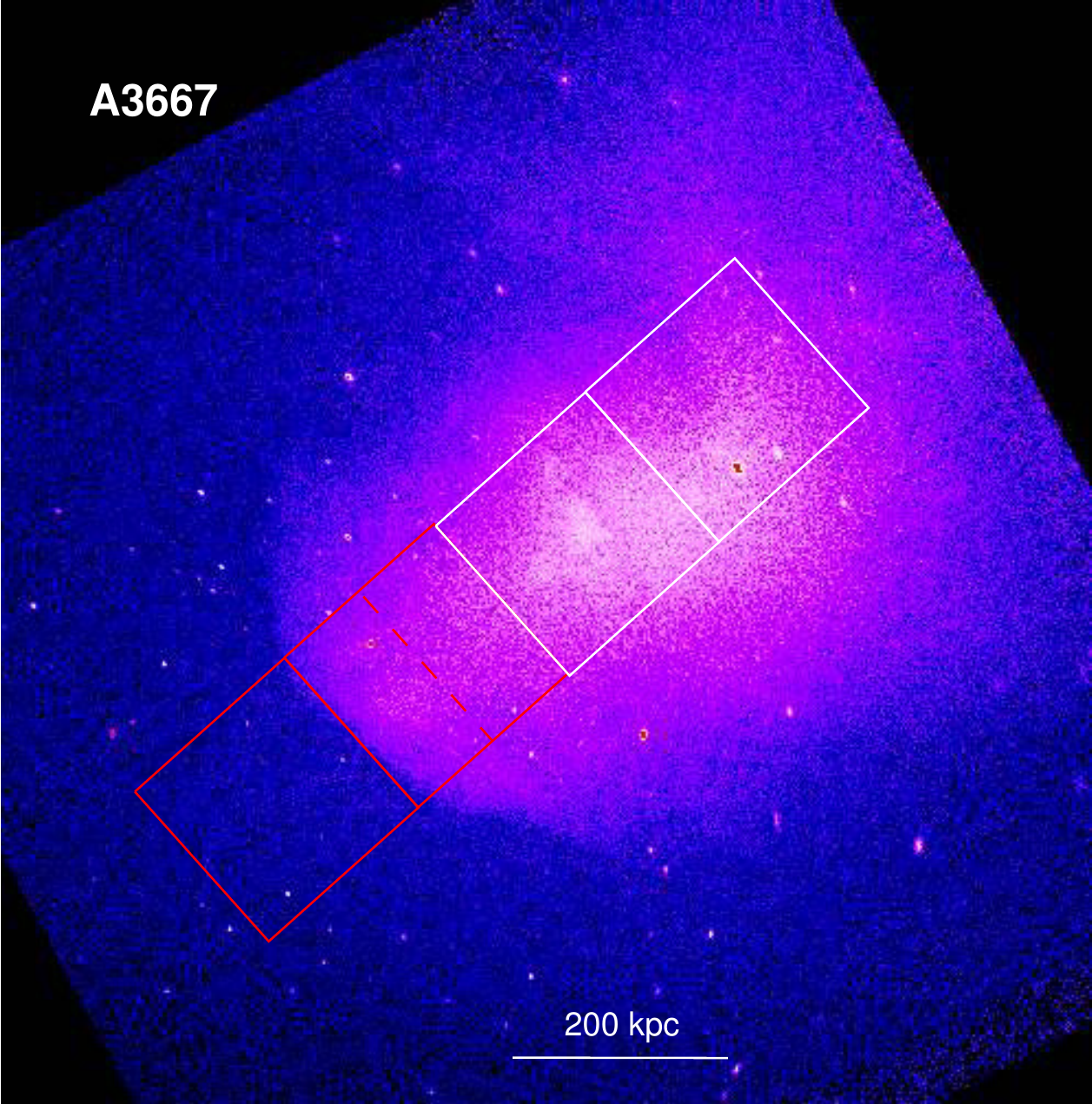}
\includegraphics[width=0.469\textwidth]{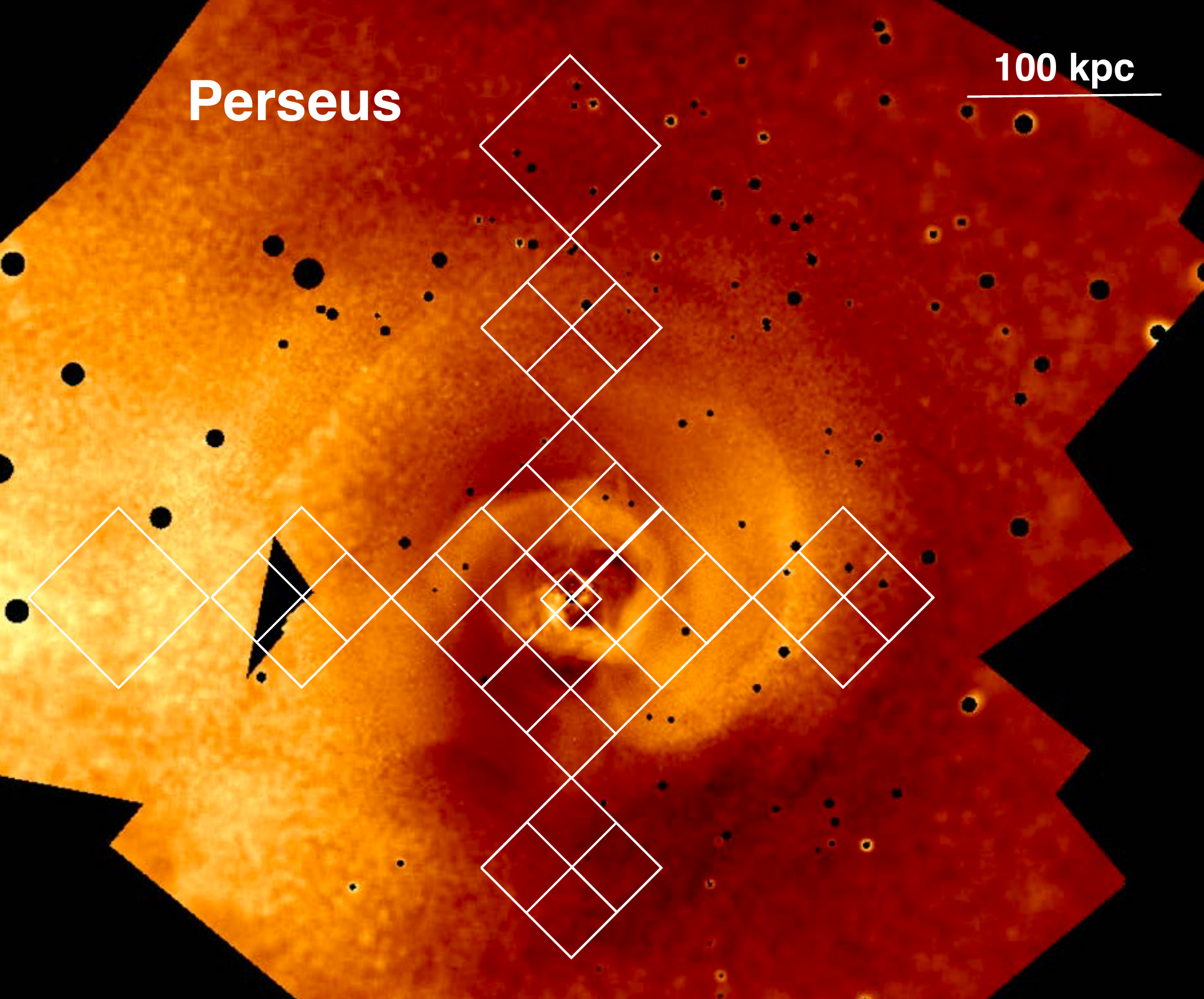}
\caption{{\it Chandra} images of nearby clusters with cold fronts. 3'$\times$3' squares are shown to indicate positions of SXS pointings which could potentially yield interesting results based on the findings of this work. All panels except the top-left (A2029) are taken from \citet{kit14}.}
\label{fig:nearby_clusters}
\end{figure*}

\section{Discussion}\label{sec:disc}

\subsection{Implications for Estimating the ICM Viscosity}\label{sec:estimating_viscosity}

Our results show that the line shapes produced by the sloshing motions are very similiar, regardless of whether or not the ICM is inviscid or very viscous (see Section \ref{sec:vel_dist_viscous} and Figure \ref{fig:compare_sim_lines}). This is not unexpected, as the largest velocities along the line of sight will be associated with the driving scale of the sloshing ($\sim$100~kpc), which is much larger than the viscous dissipation scale. \citet{ino03} used a study of ICM turbulence to show that the fast gas motions at the driving scale will be mostly responsible for the shape of the spectral lines. Sloshing represents an analogous situation, producing smaller-scale turbulent motions with smaller velocities which will have less of an effect on the shape of the spectral line \citep[see][]{vaz12,zuh13}.

Even in nearby clusters, the viscous dissipation scale will be unresolved by {\it Astro-H}, unless this scale is implausibly large. For example, \citet{zuh15} showed using simulations of turbulent velocity fields in a Coma-like cluster that SXS's spatial resolution will be too coarse to constrain the dissipation scale of the turbulent cascade. Definitive constraints on viscosity from gas motions in clusters will likely have to wait for missions with similar spectral resolution and better spatial resolution, such as {\it Athena} and {\it X-ray Surveyor}.

\subsection{Application to Possible {\it Astro-H} Targets: Where Should We Point?}\label{sec:targets}

The results from our simulation analysis of sloshing cold fronts may be relevant for a number of nearby clusters with similar features. The cold gas component of sloshing, located underneath the front surface, is likely to be a region where emission lines will be significantly broadened. If our line of sight is within the plane of the sloshing motions, we also expect to measure a significant line shift. Here we identify a few nearby systems with cold fronts which could be observed by {\it Astro-H}, to determine if the line shifting and broadening predicted in this work can be detected.

Two nearby clusters which have clear-cut examples of sloshing cold fronts are Abell 2029 and Abell 2319. Abell 2029 is at a redshift of 0.07728, and in this case the cold fronts can be entirely confined within a single SXS pointing of 3'$\times$3' (see the top-left panel of Figure \ref{fig:nearby_clusters}). For this cluster, the SXS pointing should be placed at the cluster center, and the different portions of the cold spiral can be resolved at a scale of $\sim$1', though accurate modeling of the effects of the PSF will be required. In Abell 2319, at a redshift of 0.0557, the cold fronts are larger in the sky plane, and can be split across a few SXS pointings (an example pointing is shown in the top-right panel of Figure \ref{fig:nearby_clusters}). In any case, each pointing will need to have long enough exposure to ensure that enough counts will be within the Fe-K complexes to measure the line shift and width accurately \citep{ota15}.

Another interesting system is Abell 3667, which possesses a large cold front to the southeast (see the bottom-left panel of Figure \ref{fig:nearby_clusters}). Kinematic measurements in the X-ray band have the potential in this case to distinguish between different merging scenarios for this cluster. Abell 3667 is assumed to be an ongoing major merger nearly in the plane of the sky \citep{rot97,vik01,joh08,owe09,dat14}. \citet{vik01} used hydrodynamic arguments to estimate the velocity of the cold front to be transsonic or supersonic, supporting this interpretation, though a recent analysis \citet{dat14} suggests the velocity is subsonic. However, if there is a significant line-of-sight velocity shift of the gas underneath the cold front surface (the pointing outlined in red with a dashed line through the center in Figure \ref{fig:nearby_clusters}), this may indicate that we are observing a sloshing cold front with our line of sight at least partially aligned with the plane of the gas motions, or at least that the merger is not strictly in the plane of the sky. Simulations indicate that sloshing cold fronts can also appear in major mergers with large impact parameters \citep{ric01,poo06,zuh11b}.

Finally, an extensive {\it Astro-H} study is already planned for the Perseus Cluster ($z$ = 0.0179), the brightest X-ray cluster in the sky. Due to its proximity of $\sim$68~Mpc, {\it Astro-H} will be able to map the core of Perseus with excellent resolution, with 1' $\sim$ 20~kpc. Using a combination of {\it ROSAT}, {\it XMM-Newton}, and {\it Suzaku} observations, \citet{sim12} demonstrated evidence of large-scale sloshing motions extending from the cluster core to the virial radius. The bottom-right panel of Figure \ref{fig:nearby_clusters} shows the residual image of the cluster core, with planned {\it Astro-H}/SXS pointings overlaid \citep{kit14}, in which the large-scale sloshing spiral can be readily seen. It appears that the plane of the sloshing motions is oriented nearly perpendicular to our line of sight, so on the basis of this work we expect to find no significant line shifts associated with the cold gas underneath the front surfaces, but may find significant line broadening from the expansion of the cold fronts in our line of sight. However, Perseus is also a site of significant AGN activity, which is likely driving significant turbulence, which will also produce significant line broadening. Teasing the effects of sloshing and AGN-driven turbulence apart will be an interesting challenge for future simulations.

\section{Summary}\label{sec:summary}

We have carried out a detailed investigation of gas motions in simulations of a relaxed, cool-core cluster, specifically with regard to how these motions may be observed by the {\it Astro-H} X-ray observatory. Our main results are as follows:

\begin{itemize}
\item {\it Astro-H} will be able to detect the large-scale bulk motions associated with sloshing gas in cool-core clusters. These motions will produce significant shifting and broadening of spectral lines that will be easily measured by {\it Astro-H}'s high-resolution calorimeter. The regions with the most detectable velocities are located underneath the cold front surfaces, associated with the cold, dense phase.
\item The line shifts produced by cold fronts will be most observable when our line of sight is directly within the plane of the sloshing motions. This is also the direction where the evidence of sloshing from imaging studies will be the least obvious; cold fronts will still be seen, but without an obvious spiral pattern. Line widths seen in this orientation will be produced predominately by the variation of the sloshing motion along the line of sight. In cases where a clear sloshing spiral is seen, our line of sight is nearly perpendicular to the sloshing plane, and line shifts from the sloshing motions should be insignificant. However, significant line broadening may be produced by sloshing-driven turbulence or by the radial expansion of the cold fronts perpendicular to the sloshing plane. In either case, sloshing motions provide evidence of potentially significant non-thermal pressure support.
\item Sloshing motions produce line shapes that are well-represented by one or more Gaussian components. One-component Gaussian velocity models provide good fits to our synthetic spectra, regardless of the underlying velocity distribution. However, depending on the degree of non-Gaussianity of the underlying distribution, line shifts and widths under this assumption may be biased. For long exposures, two-component models may be used to fit complex velocity distributions with well-separated components.
\item The line shapes produced in our inviscid and highly viscous simulations are very similar, and for practical purposes will be indistinguishable. This reflects the fact that the medium-to-large spatial scale motions have the strongest effect on the line shape, which are relatively unaffected by viscous dissipation, even if the ICM is very viscous. For this reason, constraints on viscosity from line shapes produced by sloshing motions are likely to be limited.
\item Measured line shifts and widths from sloshing cold fronts will be affected by systematic effects, such as uncertainty in the line spread function and gain. In particular, the systematic error on the line shift is likely to dominate over other sources of error discussed in this work, including statistical uncertainties, errors arising from fitting Gaussian models to non-Gaussian velocity distributions, and PSF scattering from other regions.
\item Though a smaller effect, measured line shifts and widths will also be affected by PSF scattering of photons from the bright cluster core. In our case, we find that the photons scattered from the nearby core into the field of view centered on the sloshing cold fronts can bias line shifts and widths by several tens of km/s, though the bias on line widths is typically smaller. This indicates that spectra from cold front regions in clusters should be modeled in combination with that from the core region to mitigate this effect, as we showed in Section \ref{sec:joint_modeling} \citep[see also][]{kit14}.
\end{itemize}

There are a number of unresolved questions that our work does not address, and a number of avenues for future investigation are suggested by our results.
\begin{itemize}
\item Though our idealized setup provides the opportunity to determine what effects sloshing motions have on spectral lines, in real clusters the situation is more complicated. For example, the dominant driver of all gas motion, including turbulence, in our simulation is the sloshing motions themselves. However, even in relaxed systems, there will be other sources exciting plasma motions, such as AGN and substructure, so the gas motion produced in our simulations is likely a lower limit on what can be expected. Analysis of a more complicated system, such as a cluster extracted from a cosmological simulation, would provide a fuller picture of the combined effect of turbulence and bulk motions on observations of spectral lines.
\item For this work, we chose to analyze a single epoch shortly after the core passage of the subcluster, which is late enough so that cold fronts have had time to develop, but early enough so that they are still bright and prominent. A detailed analysis of the later epochs of our simulation are needed to determine if our conclusions hold as the cold fronts expand. We have had a cursory look at the data from later times, which suggests that similar results will indeed hold.
\item We predict that line shapes produced by sloshing may have a detectable non-Gaussianity. This may be modeled with a superposition of Gaussians, as suggested by \citet{sha12}. Alternatively, new models for spectral fitting could be developed that incorporate physically motivated non-Gaussian velocity distributions. However, our ability to constrain these models will be primarily limited by the SXS gain uncertainty.
\item A separate analysis of our data using model responses for the upcoming {\it Athena} and {\it X-ray Surveyor} missions would be useful to characterize the advantages of larger effective area and smaller PSF. This is also possible using the \code{SIMX} package.
\item It is not clear to what extent the results of our analysis would apply to cold fronts produced in major mergers. In these systems, shocks are also present, the bulk motions are significantly larger, and the turbulence produced can be significant. A similar analysis of a major merger simulation is necessary.
\end{itemize}

\acknowledgments
JAZ thanks Maxim Markevitch and Randall Smith for useful discussions, and JAZ and AS would especially like to thank Daisuke Nagai and Erwin Lau for a stimulating discussion over dinner at the Snowcluster 2015 meeting. JAZ acknowledges support from NASA though subcontract SV2-8203 to MIT from the Smithsonian Astrophysical Observatory. This work required the use and integration of several Python packages for astronomy, including yt \citep[\code{\url{http://yt-project.org}},][]{tur11}, AstroPy \citep[\code{\url{http://astropy.org}},][]{ast13}, APLpy (\code{\url{http://aplpy.github.com}}), and pyregion (\code{\url{http://pyregion.readthedocs.org/}}).

\appendix

\section{Spectral Modeling Verification Test}

We performed a verification test of our procedure for generating and fitting synthetic {\it Astro-H} spectra. In this test, we ensure we can recover the input plasma temperature and metallicity, as well as the shift and width of spectral lines, for a simplified cluster model. For our test, we set up an isothermal, kT = 6~keV cluster with a density profile given by a $\beta$-model \citep{cav76,cav78}, with a core radius $r_c$ = 50~kpc, $\beta$ = 1, and a core electron density $n_c$ = 0.035~cm$^{-3}$. We then add a velocity field to the cluster model using a single Gaussian random field with $\mu_z$ = 400~km/s and $\sigma_z$ = 400~km/s. We perform 200 realizations of this velocity field, and from each one we compute a realization of the spectrum taken from an entire SXS pointing located at the center of the model cluster. We use the 200 different spectral realizations to compute the mean and 1-$\sigma$ errors for each parameter.

In the case of the normalization parameter $\eta$, we must take into account that a fraction of the photons within the region covered by the SXS pointing will be scattered out of the region due to PSF scattering and vignetting effects. To quantify this effect, we perform the same simulation of the synthetic observation with these effects turned off. We determine that $\sim7\%$ of the photons have been removed from the region by these effects, and have adjusted the expected value of $\eta$ accordingly. The results of the spectral fitting test are shown in Table \ref{tab:beta_model_test}. We find that we are able to recover the values of all of the parameters within the 1-$\sigma$ errors.

\renewcommand{\arraystretch}{2.0}
\begin{table}[thdp]
\tabletypesize{\scriptsize}
\caption{Results of Spectral-Fitting Verification Test\label{tab:beta_model_test}}
\begin{center}
\begin{tabular}{lll}
\hline
\hline
Parameter & True Value & Fitted Value \\
\hline
kT (keV) & 6 & 5.97 $\substack{+0.03 \\ -0.04}$ \\
Z (Z$_\odot$) & 0.3 & 0.299 $\substack{+0.006 \\ -0.005}$ \\   
$\mu_z$ (km/s) & 400 & 420 $\substack{+28 \\ -31}$ \\ 
$\sigma_z$ (km/s) & 400 & 398 $\substack{+29 \\ -22}$ \\
$\eta$ (10$^{-2}$ cm$^{-5}$) & 1.522 & 1.520 $\substack{+0.003 \\ -0.004}$ \\ 
\hline
\end{tabular}
\end{center}
\end{table}

\section{Viscous Simulation Figures}\label{sec:visc_figures}

In what follows, we present the rest of the slice, phase space, and line shape plots for the viscous simulation. We refer back to Figures \ref{fig:map_z_visc}-\ref{fig:map_y_visc} for the locations of the spectral extracting regions and slice planes.

\begin{figure*}[p]
\begin{center}
\begin{minipage}[b]{0.49\linewidth}
\includegraphics[width=\textwidth]{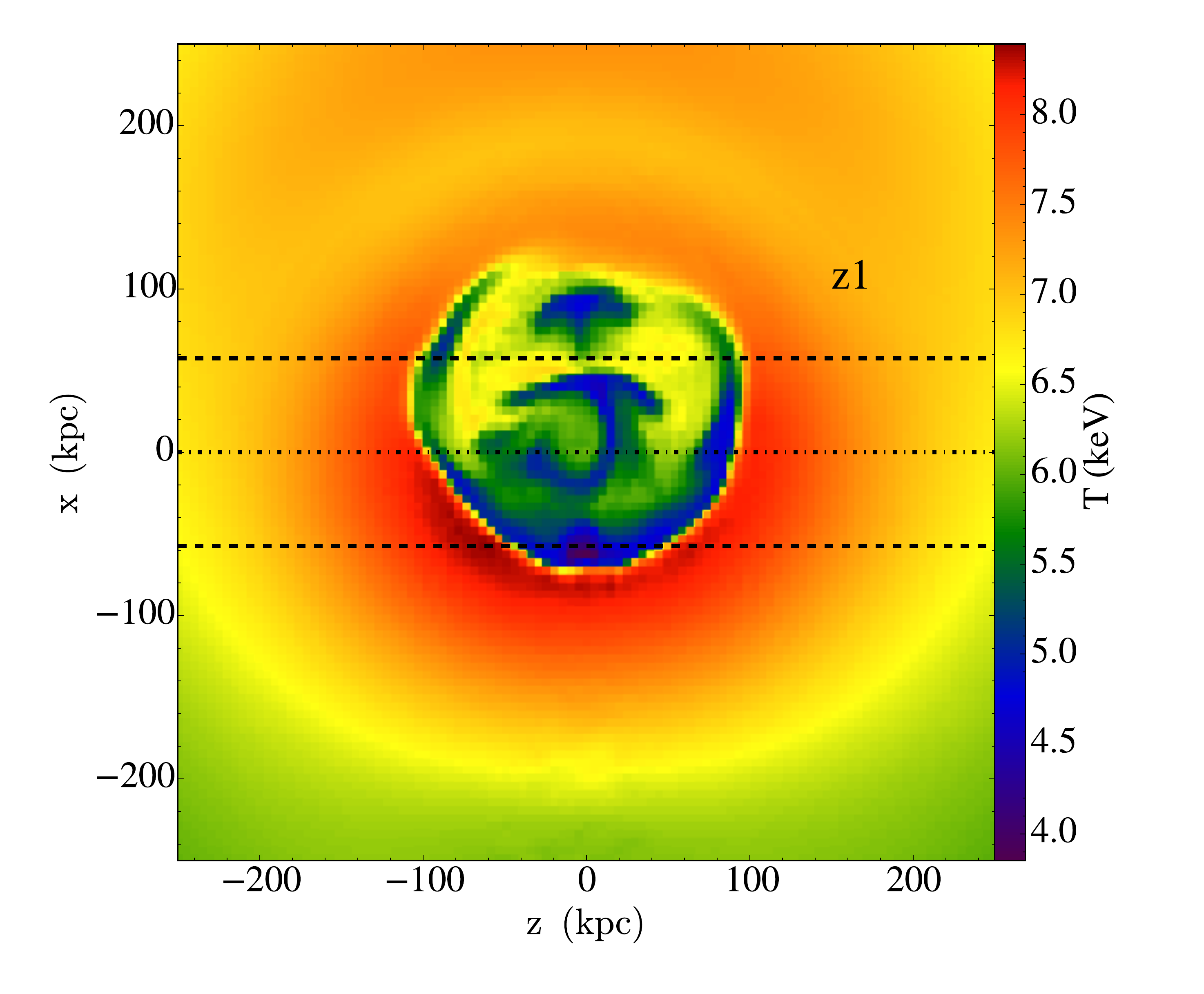}
\end{minipage}
\begin{minipage}[b]{0.49\linewidth}
\includegraphics[width=\textwidth]{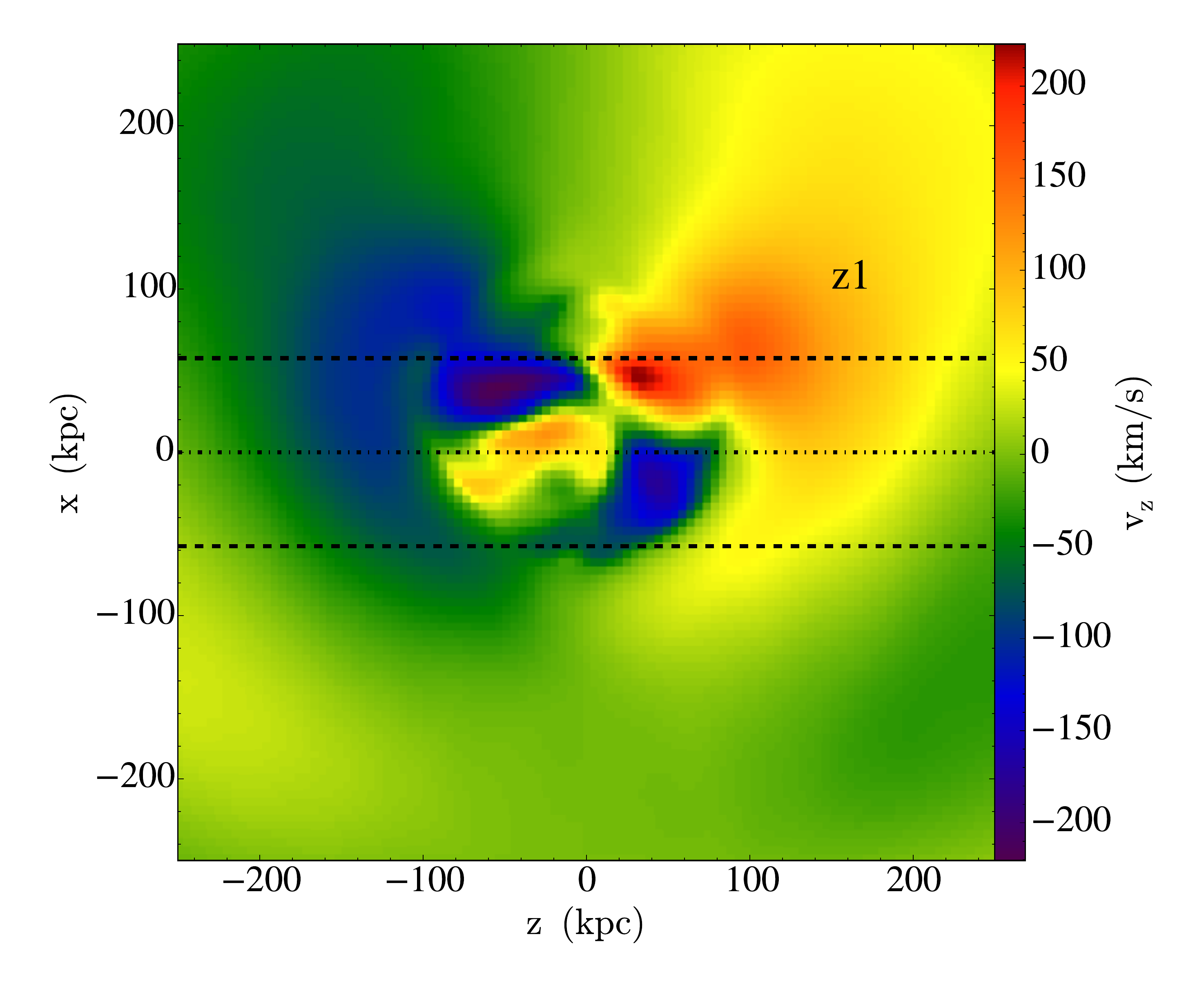}
\end{minipage}
\begin{minipage}[b]{0.51\linewidth}
\includegraphics[width=0.97\textwidth]{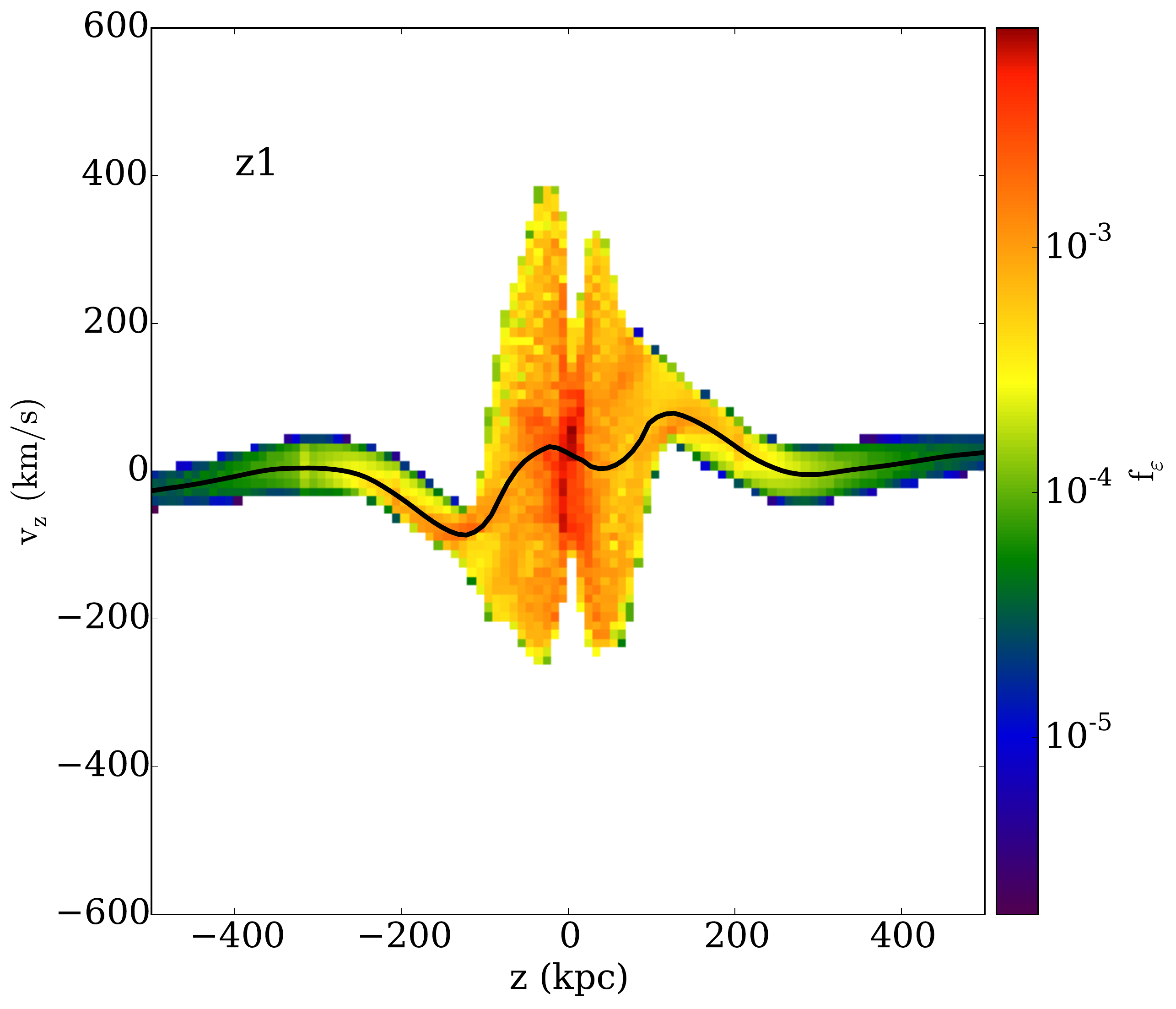}
\end{minipage}
\begin{minipage}[b]{0.47\linewidth}
\includegraphics[width=0.92\textwidth]{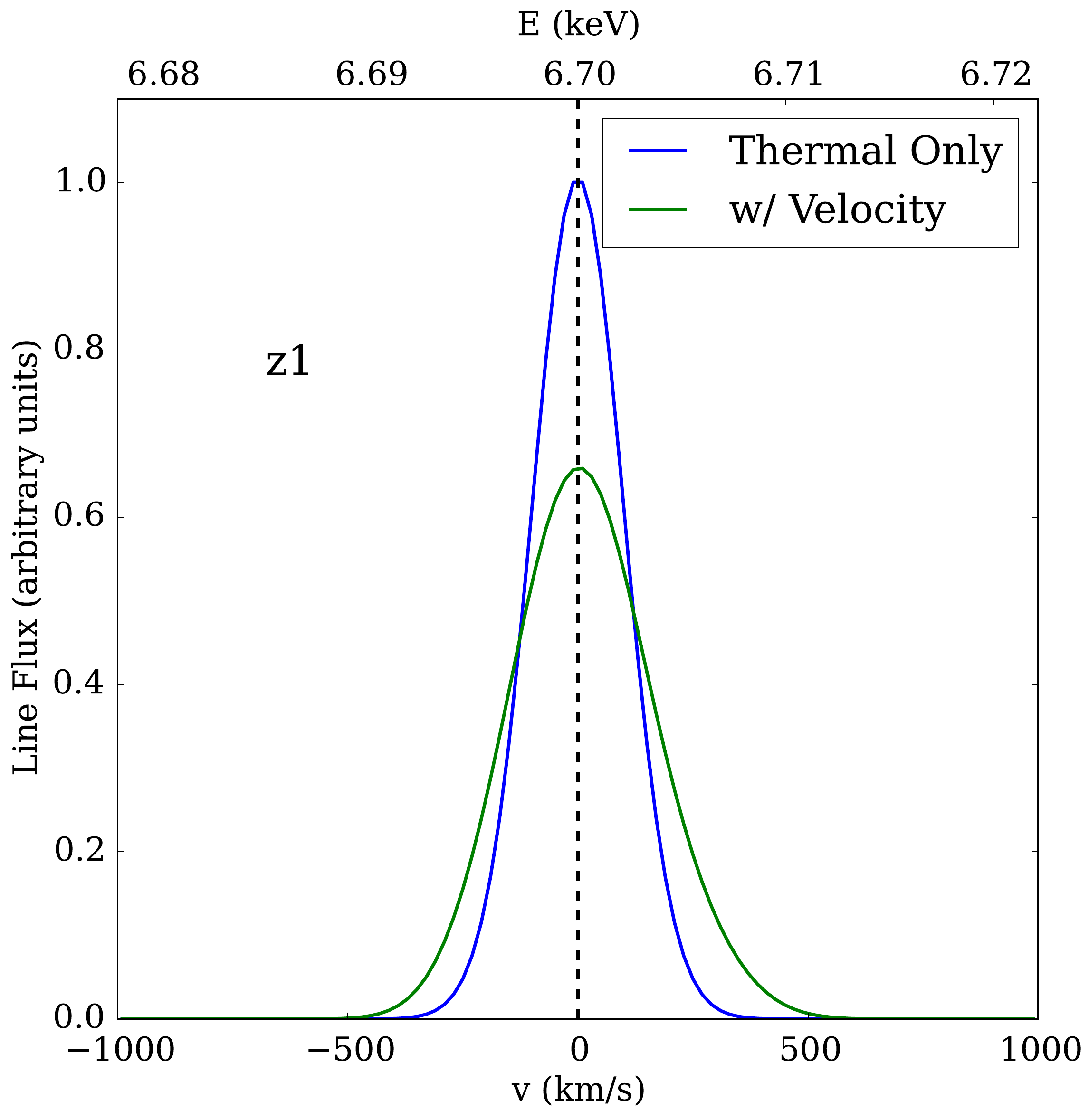}
\end{minipage}
\caption{Characteristics of the velocity field along the $z$-axis of the viscous simulation, for region ``z1''. Upper panels: slices through the $x-z$-plane at the center of region ``z1'', of temperature (left) and the $z$-component of the velocity (right). Black lines indicate the center and edges of the elliptical cylinder corresponding to the region in Figure \ref{fig:map_z_visc}. Lower-left panel: Phase space plot showing the fraction of emission as a function of position and velocity within the cylinder. The black line indicates the emission-weighted average value. Lower-right panel: Effect of plasma motion on a ``toy'' He-like iron line for the emission with the region.\label{fig:vz_visc_dist1}}
\end{center}
\end{figure*}

\begin{figure*}[p]
\begin{center}
\begin{minipage}[b]{0.49\linewidth}
\includegraphics[width=\textwidth]{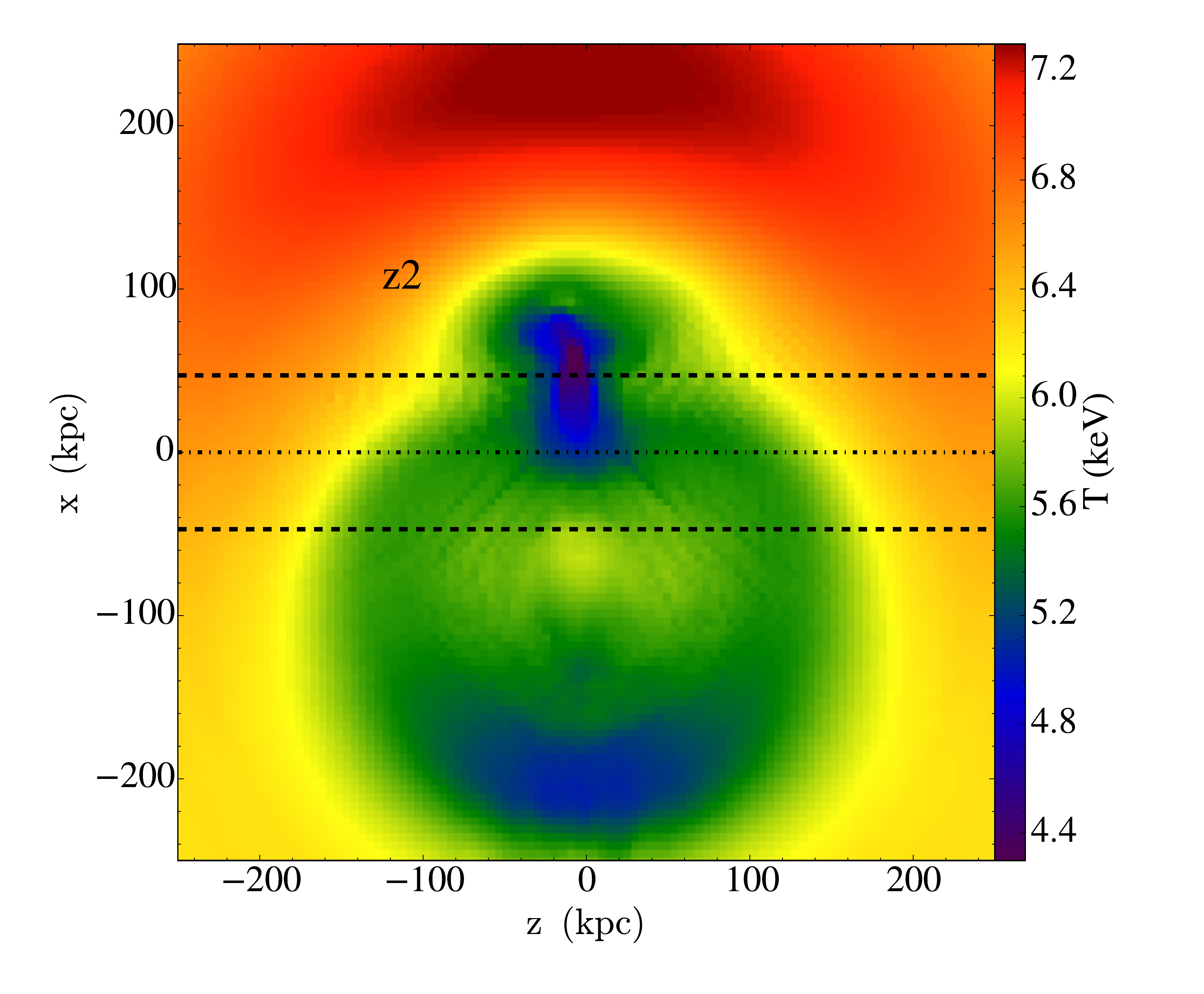}
\end{minipage}
\begin{minipage}[b]{0.49\linewidth}
\includegraphics[width=\textwidth]{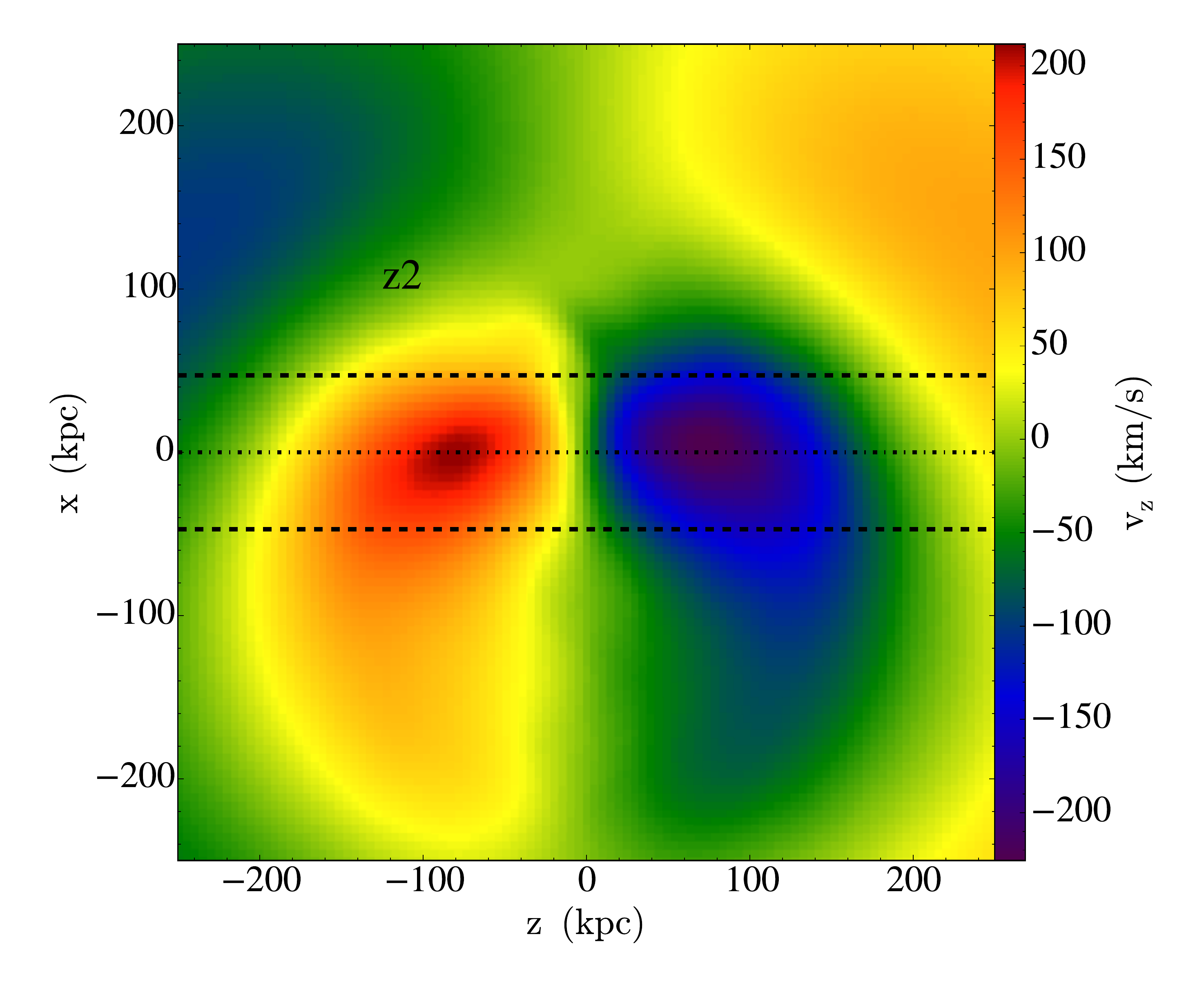}
\end{minipage}
\begin{minipage}[b]{0.51\linewidth}
\includegraphics[width=0.97\textwidth]{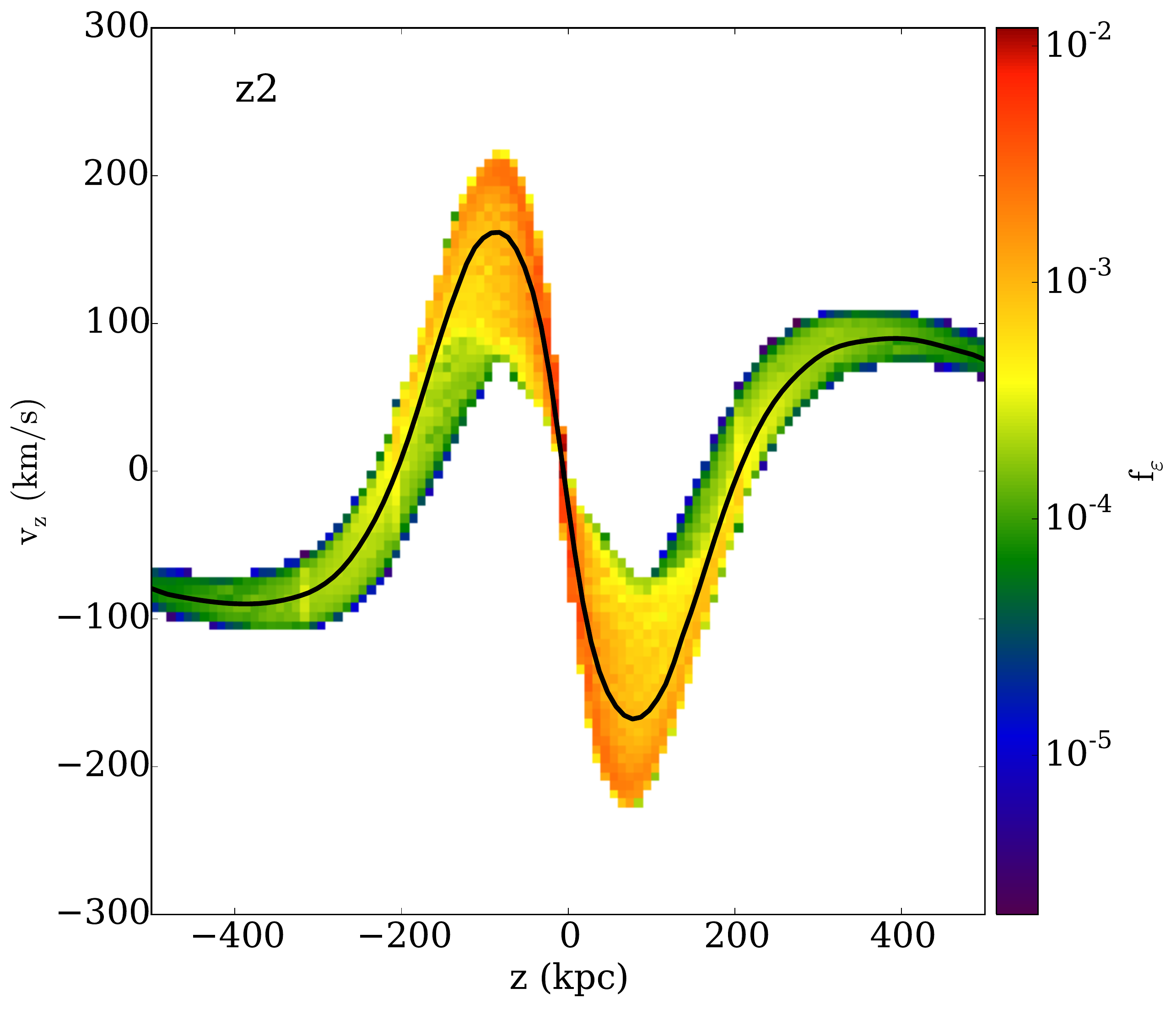}
\end{minipage}
\begin{minipage}[b]{0.47\linewidth}
\includegraphics[width=0.92\textwidth]{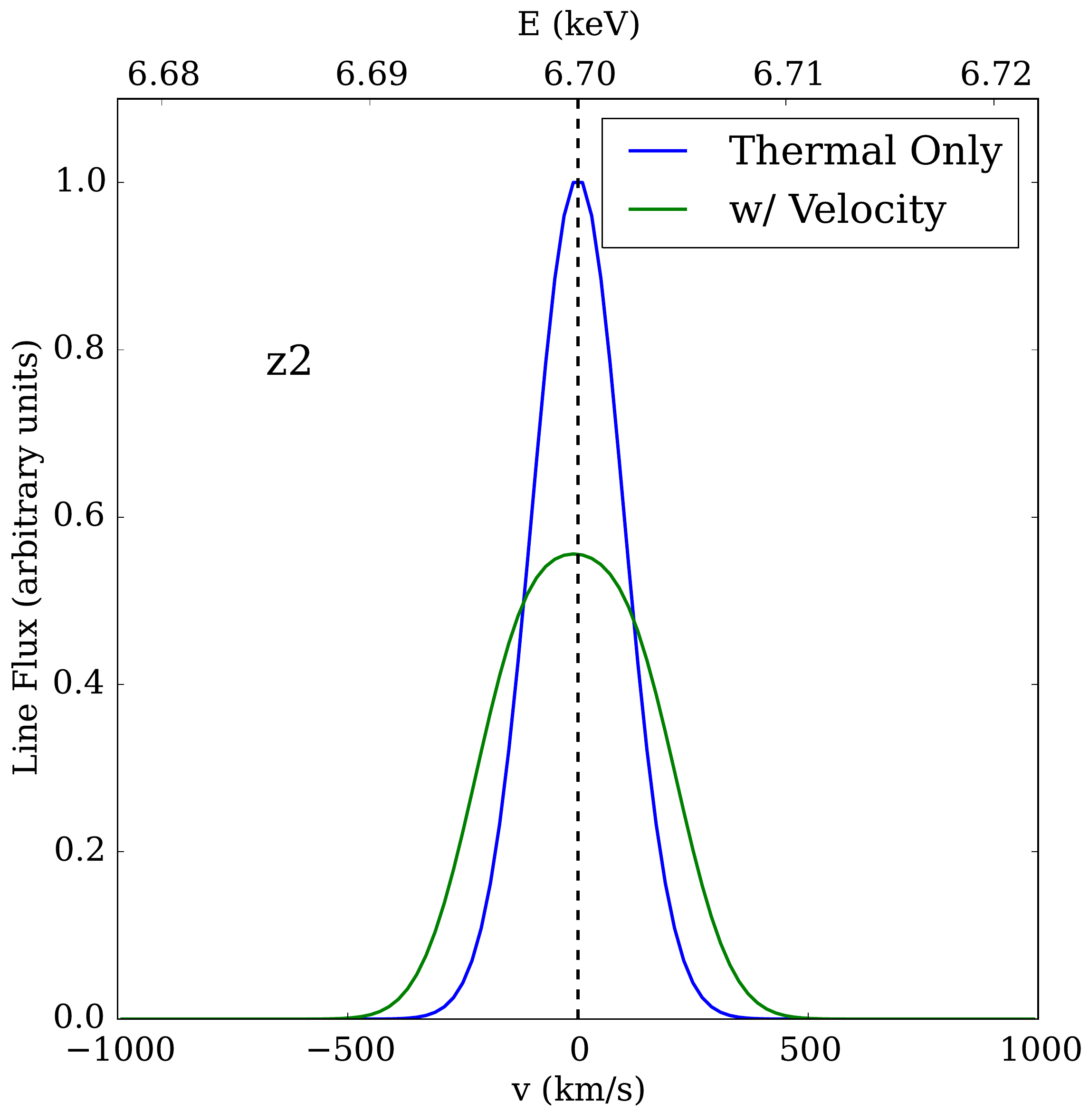}
\end{minipage}
\caption{Characteristics of the velocity field along the $z$-axis of the viscous simulation, for region ``z2''. Upper panels: slices through the $x-z$-plane at the center of region ``z2'', of temperature (left) and the $z$-component of the velocity (right). Black lines indicate the center and edges of the elliptical cylinder corresponding to the region in Figure \ref{fig:map_z_visc}. Lower-left panel: Phase space plot showing the fraction of emission as a function of position and velocity within the cylinder. The black line indicates the emission-weighted average value. Lower-right panel: Effect of plasma motion on a ``toy'' He-like iron line for the emission with the region.\label{fig:vz_visc_dist2}}
\end{center}
\end{figure*}

\begin{figure*}[p]
\begin{center}
\begin{minipage}[b]{0.49\linewidth}
\includegraphics[width=\textwidth]{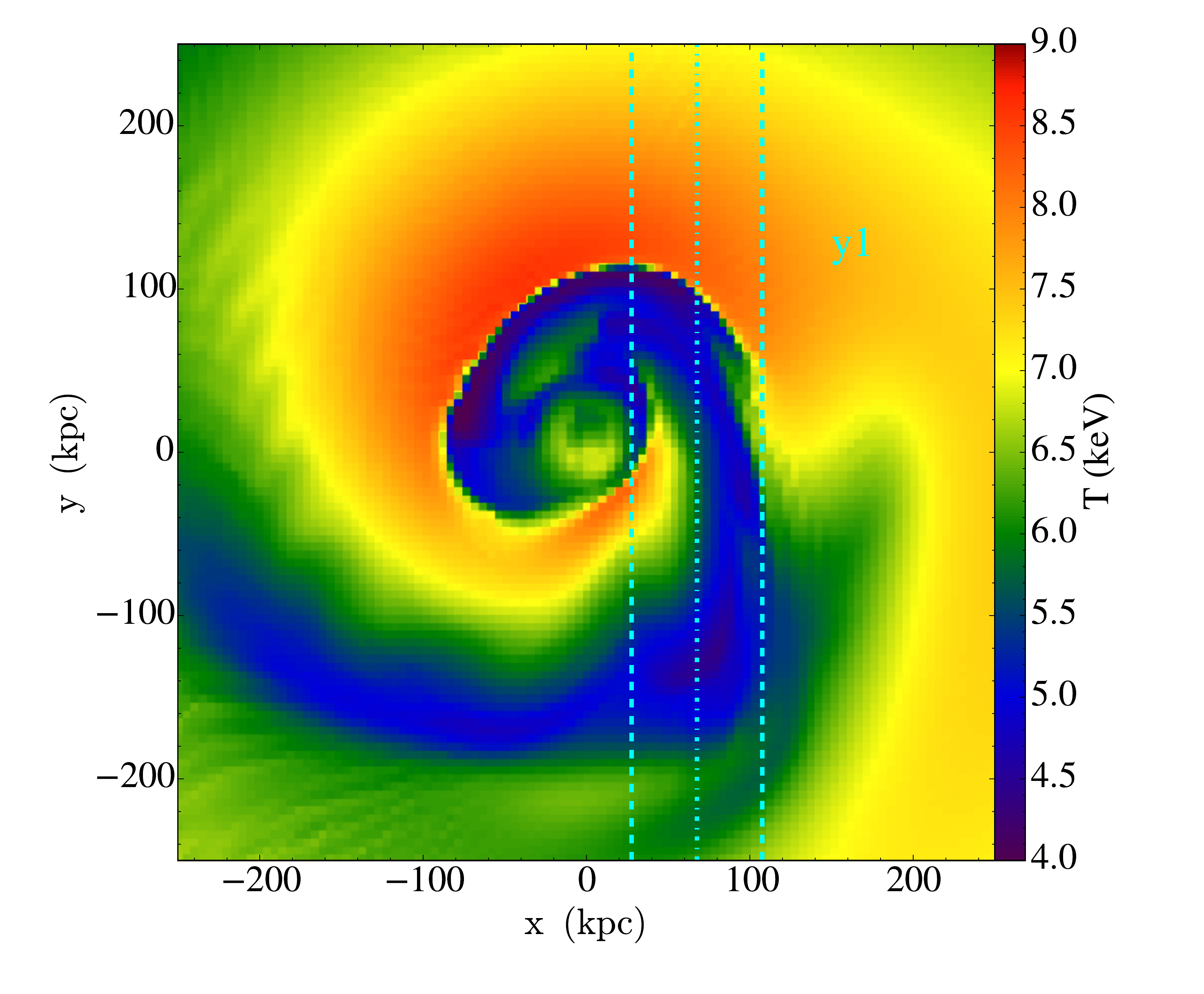}
\end{minipage}
\begin{minipage}[b]{0.49\linewidth}
\includegraphics[width=\textwidth]{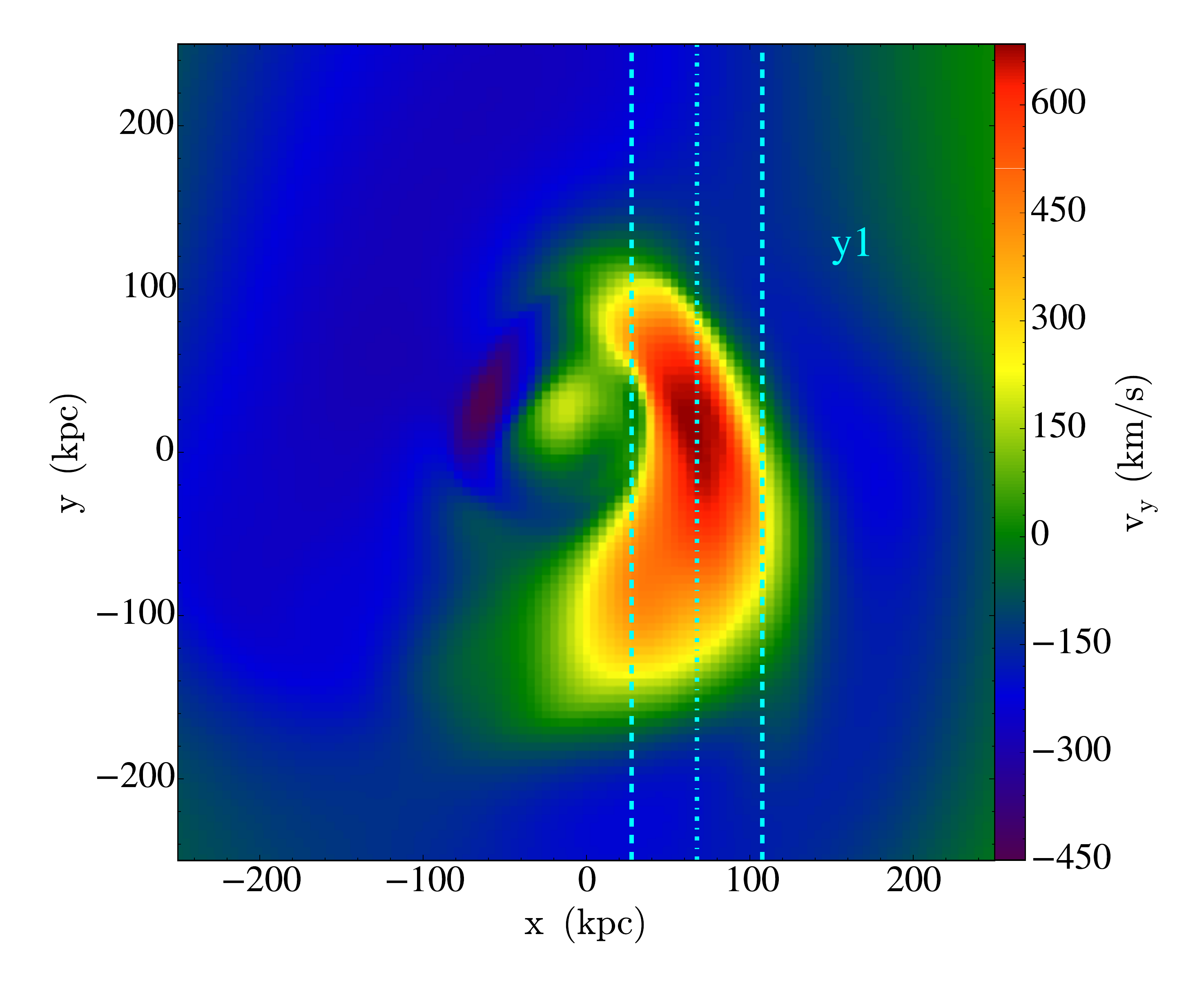}
\end{minipage}
\begin{minipage}[b]{0.51\linewidth}
\includegraphics[width=0.97\textwidth]{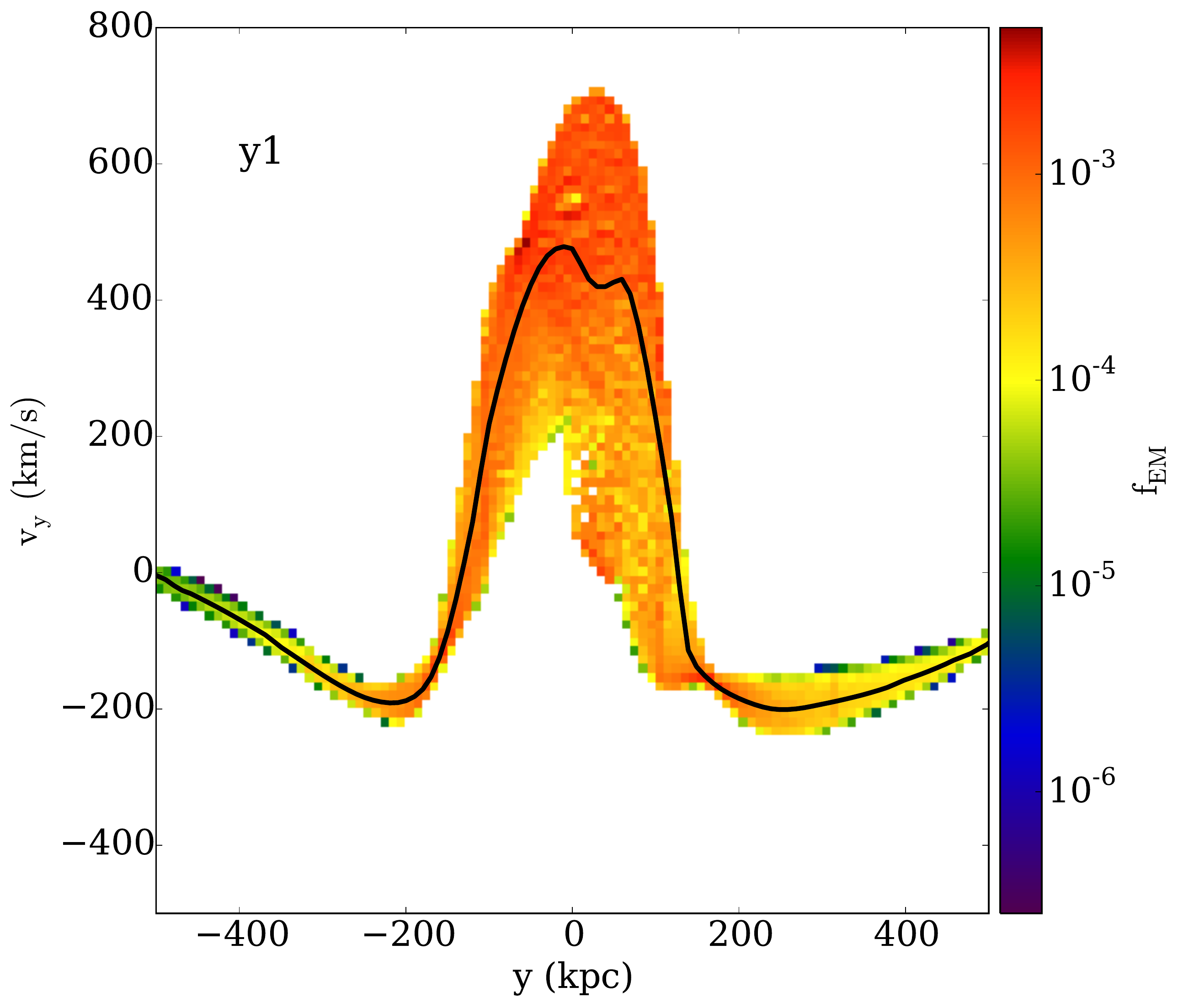}
\end{minipage}
\begin{minipage}[b]{0.47\linewidth}
\includegraphics[width=0.92\textwidth]{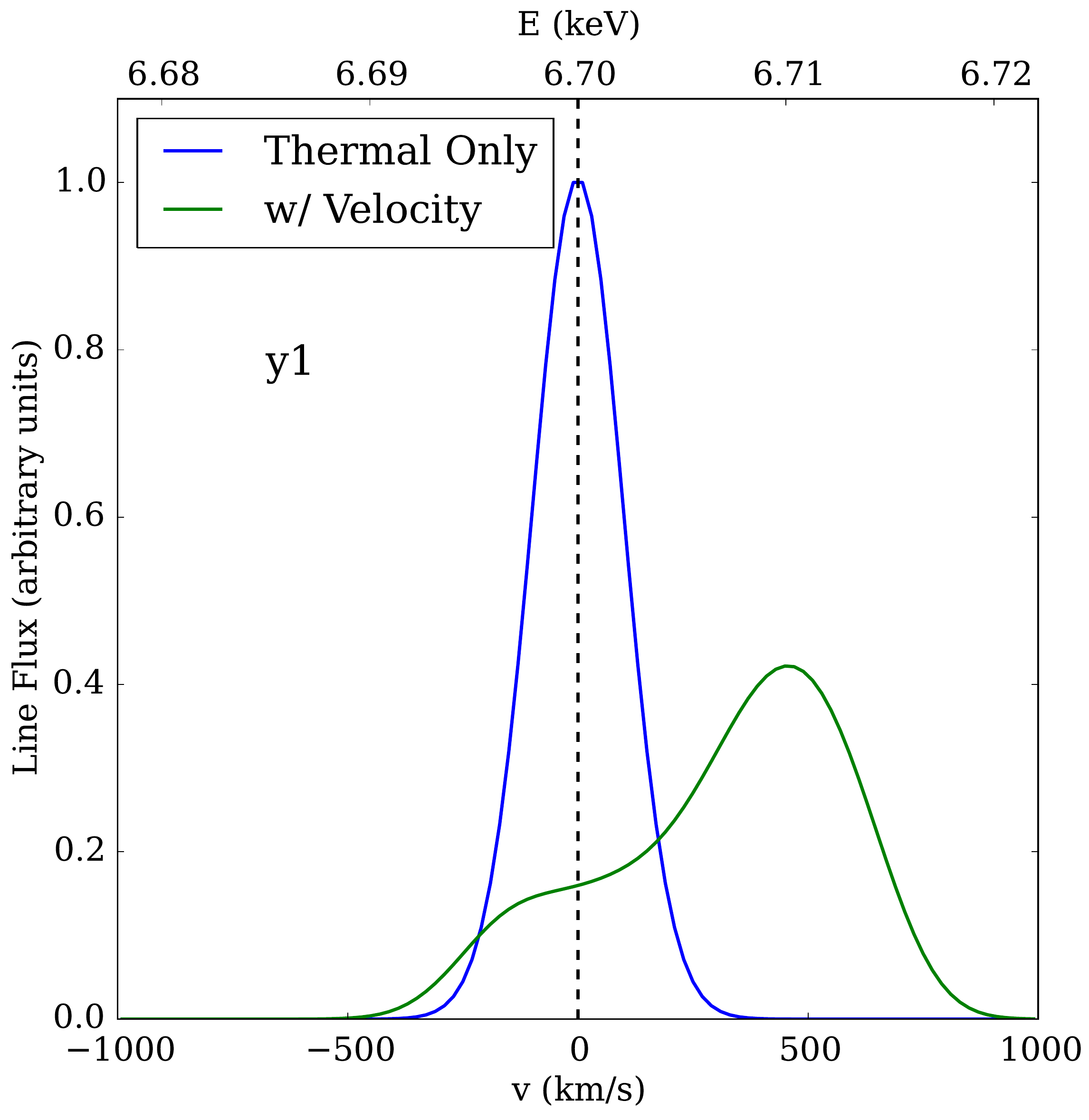}
\end{minipage}
\caption{Characteristics of the velocity field along the $y$-axis of the viscous simulation. Upper panels: slices through the $x-y$-plane at $z = 0$, of temperature (left) and the $y$-component of the velocity (right). Cyan lines indicate the center and edges of the elliptical cylinder corresponding to the region in Figure \ref{fig:map_y_visc}. Lower-left panel: Phase space plot showing the fraction of emission as a function of position and velocity within the cylinder. The black line indicates the emission-weighted average value. Lower-right panel: Effect of plasma motion on a ``toy'' He-like iron line for the emission with the region.\label{fig:vy_visc_dist}}
\end{center}
\end{figure*}


\begin{thebibliography}{}
\bibitem[Anders \& Grevesse(1989)]{and89} Anders, E., \& Grevesse, N.\ 1989, \gca, 53, 197
\bibitem[Applegate et al.(2014)]{app14} Applegate, D.~E., von der Linden, A., Kelly, P.~L., et al.\ 2014, \mnras, 439, 48
\bibitem[Ascasibar \& Markevitch(2006)]{AM06} Ascasibar, Y., \& Markevitch, M. 2006, \apj, 650, 102
\bibitem[Astropy Collaboration et al.(2013)]{ast13} Astropy Collaboration, Robitaille, T.~P., Tollerud, E.~J., et al.\ 2013, \aap, 558, A33
\bibitem[Bautz et al.(2009)]{bau09} Bautz, M.~W., Miller, E.~D., Sanders, J.~S., et al.\ 2009, \pasj, 61, 1117
\bibitem[Biffi et al.(2012)]{bif12} Biffi, V., Dolag, K., B{\"o}hringer, H., \& Lemson, G.\ 2012, \mnras, 420, 3545
\bibitem[Biffi et al.(2013)]{bif13} Biffi, V., Dolag, K., B{\"o}hringer, H.\ 2013, \mnras, 428, 1395
\bibitem[Braginskii(1965)]{bra65} Braginskii, S.~I.\ 1965, Reviews of Plasma Physics, 1, 205
\bibitem[Brunetti \& Lazarian(2007)]{bru07} Brunetti, G., \& Lazarian, A.\ 2007, \mnras, 378, 245
\bibitem[Cash(1979)]{cas79} Cash, W.\ 1979, \apj, 228, 939
\bibitem[Cavaliere \& Fusco-Femiano(1976)]{cav76} Cavaliere, A., \& Fusco-Femiano, R.\ 1976, \aap, 49, 137
\bibitem[Cavaliere \& Fusco-Femiano(1978)]{cav78} Cavaliere, A., \& Fusco-Femiano, R.\ 1978, \aap, 70, 677
\bibitem[Churazov et al.(2003)]{chu03} Churazov, E., Forman, W., Jones, C., B{\"o}hringer, H.\ 2003, \apj, 590, 225
\bibitem[Colella \& Woodward(1984)]{col84} Colella, P., \& Woodward, P.~R.\ 1984, Journal of Computational Physics, 54, 174
\bibitem[Datta et al.(2014)]{dat14} Datta, A., Schenck, D.~E., Burns, J.~O., Skillman, S.~W., \& Hallman, E.~J.\ 2014, \apj, 793, 80
\bibitem[Dennis \& Chandran(2005)]{den05} Dennis, T.~J., \& Chandran, B.~D.~G.\ 2005, \apj, 622, 205
\bibitem[Donnert et al.(2013)]{don13} Donnert, J., Dolag, K., Brunetti, G., \& Cassano, R.\ 2013, \mnras, 429, 3564
\bibitem[{Dubey} et~al.(2009)]{dub09} {Dubey}, A., {Antypas}, K.,
  {Ganapathy}, M.~K., {Reid}, L.~B., {Riley}, K.~M., {Sheeler}, D.,
  {Siegel}, A., {Weide}, K. Extensible component based architecture
  for FLASH, a massively parallel, multiphysics simulation
  code. Parallel Computing 35~(10-11), 512--522.
\bibitem[En{\ss}lin et al.(2011)]{ens11} En{\ss}lin, T., Pfrommer, C., Miniati, F., \& Subramanian, K.\ 2011, \aap, 527, A99
\bibitem[Evrard et al.(1996)]{evr96} Evrard, A.~E., Metzler, C.~A., \& Navarro, J.~F.\ 1996, \apj, 469, 494
\bibitem[Fujita et al.(2004)]{fuj04} Fujita, Y., Matsumoto, T., \& Wada, K.\ 2004, \apjl, 612, L9
\bibitem[Fujita et al.(2005)]{fuj05} Fujita, Y., Matsumoto, T., Wada, K., \& Furusho, T.\ 2005, \apjl, 619, L139
\bibitem[Ghizzardi et al.(2010)]{ghi10} Ghizzardi, S., Rossetti, M., \& Molendi, S.\ 2010, \aap, 516, A32
\bibitem[Inogamov \& Sunyaev(2003)]{ino03} Inogamov, N.~A., \& Sunyaev, R.~A.\ 2003, Astronomy Letters, 29, 791
\bibitem[Johnston-Hollitt et al.(2008)]{joh08} Johnston-Hollitt, M., Hunstead, R.~W., \& Corbett, E.\ 2008, \aap, 479, 1
\bibitem[Keshet(2012)]{kes11} Keshet, U.\ 2012, \apj, 753, 120
\bibitem[Kitayama et al.(2014)]{kit14} Kitayama, T., Bautz, M., Markevitch, M., et al.\ 2014, arXiv:1412.1176
\bibitem[Kunz et al.(2014)]{kun14} Kunz, M.~W., Schekochihin, A.~A., \& Stone, J.~M.\ 2014, Physical Review Letters, 112, 205003
\bibitem[Mahdavi et al.(2013)]{mah13} Mahdavi, A., Hoekstra, H., Babul, A., et al.\ 2013, \apj, 767, 116
\bibitem[Markevitch \& Vikhlinin(2007)]{MV07} Markevitch, M., \& Vikhlinin, A.\ 2007, \physrep, 443, 1
\bibitem[Mazzotta et al.(2004)]{maz04} Mazzotta, P., Rasia, E., Moscardini, L., \& Tormen, G.\ 2004, \mnras, 354, 10
\bibitem[Mitsuda et al.(2014)]{mit14} Mitsuda, K., Kelley, R.~L., Akamatsu, H., et al.\ 2014, \procspie, 9144, 91442A
\bibitem[Nagai et al.(2007)]{nag07} Nagai, D., Vikhlinin, A., \& Kravtsov, A.~V.\ 2007, \apj, 655, 98
\bibitem[Nagai et al.(2013)]{nag13} Nagai, D., Lau, E.~T., Avestruz, C., Nelson, K., \& Rudd, D.~H.\ 2013, \apj, 777, 137
\bibitem[Nelson et al.(2014)]{nel14} Nelson, K., Lau, E.~T., \& Nagai, D.\ 2014, \apj, 792, 25
\bibitem[Owers et al.(2009)]{owe09} Owers, M.~S., Couch, W.~J., \& Nulsen, P.~E.~J.\ 2009, \apj, 693, 901
\bibitem[Ota et al.(2015)]{ota15} Ota, N., Nagai, D., \& Lau, E.~T.\ 2015, arXiv:1507.02730
\bibitem[Piffaretti \& Valdarnini(2008)]{pif08} Piffaretti, R., \& Valdarnini, R.\ 2008, \aap, 491, 71
\bibitem[Pinto et al.(2015)]{pin15} Pinto, C., Sanders, J.~S., Werner, N., et al.\ 2015, \aap, 575, A38
\bibitem[Poole et al.(2006)]{poo06} Poole, G.~B., Fardal, M.~A., Babul, A., et al.\ 2006, \mnras, 373, 881
\bibitem[Rasia et al.(2006)]{ras06} Rasia, E., Ettori, S., Moscardini, L., et al.\ 2006, \mnras, 369, 2013
\bibitem[Rebusco et al.(2006)]{reb06} Rebusco, P., Churazov, E., B{\"o}hringer, H., \& Forman, W.\ 2006, \mnras, 372, 1840
\bibitem[Rebusco et al.(2008)]{reb08} Rebusco, P., Churazov, E., Sunyaev, R., B{\"o}hringer, H., \& Forman, W.\ 2008, \mnras, 384, 1511
\bibitem[Ricker \& Sarazin(2001)]{ric01} Ricker, P.~M., \& Sarazin, C.~L.\ 2001, \apj, 561, 621
\bibitem[Ricker(2008)]{ric08} Ricker, P.~M.\ 2008, ApJS, 176, 293
\bibitem[Roediger et al.(2011)]{rod11} Roediger, E., Br{\"u}ggen, M., Simionescu, A., et al.\ 2011, \mnras, 413, 2057
\bibitem[Roediger et al.(2012)]{rod12} Roediger, E., Lovisari, L., Dupke, R., et al.\ 2012, \mnras, 420, 3632
\bibitem[Roediger et al.(2013)]{rod13} Roediger, E., Kraft, R.~P., Forman, W.~R., et al.\ 2013, \apj, 764, 60
\bibitem[Rossetti et al.(2013)]{ros13} Rossetti, M., Eckert, D., De Grandi, S., et al.\ 2013, \aap, 556, A44
\bibitem[Rottgering et al.(1997)]{rot97} Rottgering, H.~J.~A., Wieringa, M.~H., Hunstead, R.~W., \& Ekers, R.~D.\ 1997, \mnras, 290, 577
\bibitem[Sanders et al.(2011)]{san11} Sanders, J.~S., Fabian, A.~C., \& Smith, R.~K.\ 2011, \mnras, 410, 1797
\bibitem[Sanders \& Fabian(2013)]{san13} Sanders, J.~S., \& Fabian, A.~C.\ 2013, \mnras, 429, 2727
\bibitem[Sarazin(1988)]{sar88} Sarazin, C.~L.\ 1988, X-Ray Emission from Clusters of Galaxies (Cambridge: Cambridge Univ. Press)
\bibitem[Shang \& Oh(2012)]{sha12} Shang, C., \& Oh, S.~P.\ 2012, \mnras, 426, 3435
\bibitem[Simionescu et al.(2012)]{sim12} Simionescu, A., Werner, N., Urban, O., et al.\ 2012, \apj, 757, 182
\bibitem[Smith et al.(2001)]{smi01} Smith, R.~K., Brickhouse, N.~S., Liedahl, D.~A., \& Raymond, J.~C.\ 2001, \apjl, 556, L91
\bibitem[Spitzer(1962)]{spi62} Spitzer, L.\ 1962, Physics of Fully Ionized Gases, New York: Interscience (2nd edition), 1962
\bibitem[Takahashi et al.(2014)]{tak14} Takahashi, T., Mitsuda, K., Kelley, R., et al.\ 2014, \procspie, 9144, 914425
\bibitem[Tittley \& Henriksen(2005)]{tit05} Tittley, E.~R., \& Henriksen, M.\ 2005, \apj, 618, 227
\bibitem[Turk et al.(2011)]{tur11} Turk, M.~J., Smith, B.~D., Oishi,
  J.~S., Skory, S., Skillman, S.~W., Abel, T., \& Norman, M.~L.\ 2011,
  \apjs, 192, 9
\bibitem[Vazza et al.(2010)]{vaz10} Vazza, F., Gheller, C., \& Brunetti, G.\ 2010, \aap, 513, A32
\bibitem[Vazza et al.(2012)]{vaz12} Vazza, F., Roediger, E., \& Br{\"u}ggen, M.\ 2012, \aap, 544, A103
\bibitem[Vikhlinin et al.(2001)]{vik01} Vikhlinin, A., Markevitch, M., \& Murray, S.~S.\ 2001, \apj, 551, 160
\bibitem[Vikhlinin et al.(2005)]{vik05} Vikhlinin, A., Markevitch, M., Murray, S.~S., Jones, C., Forman, W., \& Van Speybroeck, L.\ 2005, \apj, 628, 655
\bibitem[von der Linden et al.(2014)]{vdl14} von der Linden, A., Mantz, A., Allen, S.~W., et al.\ 2014, \mnras, 443, 1973
\bibitem[Walker et al.(2014)]{wal14} Walker, S.~A., Fabian, A.~C., \& Sanders, J.~S.\ 2014, \mnras, 441, L31
\bibitem[Weisskopf et al.(2015)]{wei15} Weisskopf, M.~C., Gaskin, J., Tananbaum, H., \& Vikhlinin, A.\ 2015, arXiv:1505.00814
\bibitem[Wilms et al.(2000)]{wil00} Wilms, J., Allen, A., \& McCray, R.\ 2000, \apj, 542, 914
\bibitem[Zhang et al.(2010)]{zha10} Zhang, Y.-Y., Okabe, N., Finoguenov, A., et al.\ 2010, \apj, 711, 1033
\bibitem[ZuHone et al.(2010)]{zuh10} ZuHone, J.~A., Markevitch, M., \& Johnson, R.~E.\ 2010, \apj, 717, 908 (ZMJ10)
\bibitem[ZuHone et al.(2011)]{zuh11a} ZuHone, J.~A., Markevitch, M., \& Lee, D.\ 2011, \apj, 743, 16
\bibitem[ZuHone(2011)]{zuh11b} ZuHone, J.~A.\ 2011, \apj, 728, 54
\bibitem[ZuHone et al.(2013)]{zuh13} ZuHone, J.~A., Markevitch, M., Brunetti, G., \& Giacintucci, S.\ 2013, \apj, 762, 78
\bibitem[ZuHone et al.(2014)]{zuh14} ZuHone, J.~A., Biffi, V., Hallman, E.~J., et al.\ 2014, arXiv:1407.1783
\bibitem[ZuHone et al.(2015)]{zuh15} ZuHone, J., Markevitch,
M., \& Zhuravleva, I.\ 2015, arXiv:1505.07848
\end{thebibliography}
\end{document}